\documentclass[a4paper,onecolumn,10pt,accepted=2026-10-10]{quantumarticle}
\pdfoutput=1
\usepackage[utf8]{inputenc}
\usepackage[english]{babel}
\usepackage[T1]{fontenc}
\usepackage{csquotes}
\usepackage{multicol}
\usepackage{caption}
\usepackage{comment}
\usepackage{amsmath}
\usepackage{tabularx}
\usepackage{bm}
\usepackage[caption=false]{subfig}

\usepackage{amsmath,amsfonts,amsthm,amssymb,graphics,graphicx,mathtools,epstopdf,etoolbox,indentfirst,enumitem,mleftright,xcolor,booktabs,footnote,relsize,microtype}

\usepackage[scr=euler]{mathalpha} 
\usepackage{accents,yhmath} 

\let\originalleft\left
\let\originalright\right
\renewcommand{\left}{\mathopen{}\mathclose\bgroup\originalleft}
\renewcommand{\right}{\aftergroup\egroup\originalright}

\usepackage[
backend=bibtex,
style=alphabetic,
giveninits=true, 
maxbibnames=99,
minalphanames=3,
date=year
]{biblatex}
\makeatletter
\let\AdditivityDefaultBblFile\blx@bblfile
\def\blx@bblfile{%
  \IfFileExists{\jobname.bbl}
    {\AdditivityDefaultBblFile}
    {\IfFileExists{additivity_q.bbl}
       {\begingroup
          \def\jobname{additivity_q}%
          \AdditivityDefaultBblFile
        \endgroup}
       {\AdditivityDefaultBblFile}}%
}
\makeatother

\usepackage{hyperref} 
\usepackage{doi} 

\renewbibmacro{in:}{}

\AtEveryBibitem{\clearfield{issue}} 

\newbibmacro{string+doiurl}[1]{%
	\iffieldundef{doi}{%
		\iffieldundef{url}{
			#1%
		}{%
			\href{\thefield{url}}{#1}%
		}%
	}{%
		\href{https://doi.org/\thefield{doi}}{#1}%
	}%
}
\DeclareFieldFormat{title}{\usebibmacro{string+doiurl}{\mkbibemph{#1}}}
\DeclareFieldFormat[article,thesis,incollection,inproceedings,manual]{title}%
{\usebibmacro{string+doiurl}
	{#1} 
}
\usepackage{algorithm} 
\usepackage[noend]{algpseudocode}
\floatname{algorithm}{Protocol} 
\algrenewcommand\alglinenumber[1]{\normalsize #1.} 

\newcounter{algsubstate}




\definecolor{darkmagenta}{rgb}{0.85, 0, 0.45}
\hypersetup{
breaklinks=true,
colorlinks = true,
citecolor= blue,
urlcolor= blue,
linkcolor = blue
}

\newcommand{\ket}[1]{\left| #1 \right>}
\newcommand{\bra}[1]{\left< #1 \right|}
\newcommand{\ketbra}[2]{\ket{#1} \! \bra{#2}}
\newcommand{\pure}[1]{\ketbra{#1}{#1}}
\newcommand{\tr}[2][]{\operatorname{Tr}_{#1}\left[#2\right]} 
\newcommand{\Tr}{\operatorname{Tr}} 
\newcommand{\reg}{\mathrm{reg}} 

\newcommand{\defvar}{\coloneqq} 
\newcommand{\dop}[1]{\operatorname{S}_{#1}} 
\newcommand{\eps}{\varepsilon}
\newcommand{\freq}{\operatorname{freq}}
\newcommand{\Hmin}{H_\mathrm{min}}

\newcommand{\id}{\mathbb{I}} 
\newcommand{\idnorm}{\mathbb{U}} 
\newcommand{\idmap}{\operatorname{id}} 

\newcommand{\norm}[1]{\left\lVert#1\right\rVert} 
\newcommand{\Pos}{\operatorname{Pos}} 
\newcommand{\CPTP}{\operatorname{CPTP}} 
\newcommand{\CPo}{\operatorname{CP}} 
\newcommand{\suchthat}{\text{ s.t.}} 
\newcommand{\supp}{\operatorname{supp}} 
\newcommand{\term}[1]{\textup{\textbf{#1}}}

\newcommand{\Renyi}{R\'{e}nyi}
\newcommand{\mbf}[1]{\mathbf{#1}} 
\newcommand{\bsym}[1]{\boldsymbol{#1}} 

\newcommand{\EATchann}{\mathcal{M}}
\newcommand{\CS}{\overline{C}} 

\newcommand{\cS}{\bar{c}} 
\newcommand{\alphCS}{\overline{\mathcal{C}}}
\newcommand{\CP}{\widehat{C}} 

\newcommand{\cP}{\hat{c}} 
\newcommand{\alphCP}{\widehat{\mathcal{C}}}

\newcommand{\pf}{\operatorname{\mathtt{Pur}}} 
\newcommand{\inQ}{\widetilde{Q}} 

\newcommand{\downsymb}{{}}
\newcommand{\downsymbcomma}{{}}

\DeclareRobustCommand{\rchi}{{\mathpalette\irchi\relax}}
\newcommand{\irchi}[2]{\raisebox{\depth}{$#1\chi$}} 

\newtheorem{remark}{Remark}[section]
\newtheorem{theorem}{Theorem}[section]
\newtheorem{lemma}{Lemma}[section]
\newtheorem{corollary}{Corollary}[section]

\newtheorem{fact}{Fact}[section]

\newtheorem{manualtheoreminner}{Theorem}
\newenvironment{manualtheorem}[1]{%
  \IfBlankTF{#1}
    {\renewcommand{\themanualtheoreminner}{\unskip}}
    {\renewcommand\themanualtheoreminner{#1}}%
  \manualtheoreminner
}{\endmanualtheoreminner}

\theoremstyle{definition} 
\newtheorem{definition}{Definition}[section]

\newtheorem{prot}{Protocol}

\newcommand{\markup}[1]{#1} 

\begin{document}

\title{Marginal-constrained entropy accumulation theorem}

\author{Amir Arqand}
\author{Ernest Y.-Z.\ Tan}
\affiliation{Institute for Quantum Computing and Department
of Physics and Astronomy, University of Waterloo, Waterloo, Ontario N2L 3G1, Canada.}

\date{}

\maketitle

\begin{abstract}
We derive a novel chain rule for a family of channel conditional entropies, covering von Neumann and sandwiched {\Renyi} entropies. In the process, we show that these channel conditional entropies are equal to their regularized version, and more generally, additive across tensor products of channels. For the purposes of cryptography, applying our chain rule to sequences of channels yields a new variant of {\Renyi} entropy accumulation, in which we can impose some specific forms of marginal-state constraint on the input states to each individual channel. This generalizes a recently introduced security proof technique that was developed to analyze prepare-and-measure QKD with no limitations on the repetition rate. In particular, our generalization yields ``fully adaptive'' protocols that can in principle update the entropy estimation procedure during the protocol itself, similar to the quantum probability estimation framework.
\end{abstract}

\section{Introduction}
Additivity questions are one of the fundamental questions in quantum information theory, concerning whether certain entropic quantities associated with quantum channels are additive when multiple channels are used together. At its core, the question seeks to understand whether the optimal performance of quantum channels, measured under these quantities, can be achieved independently for each channel or if there is an advantage to joint usage. This problem arises in various areas of quantum information theory, for example in the context of channel capacities~\cite{AC97,Shor02,King03}.
In this work, 
we analyze this question for a certain entropic quantity that is relevant in the context of cryptography.

We begin by presenting a family of channel conditional entropies defined based on sandwiched {\Renyi} divergences~\cite{MDS+13}, which basically quantify the minimum output conditional entropy of a given quantum channel. In other words, given a channel $\mathcal{M}\in\CPTP(A\widetilde{A},BC)$ and (optionally) some state $\psi_A$ on register $A$, we denote the channel conditional entropy by $H_\alpha^\uparrow\left(\mathcal{M},B,[\psi_A]\right)$ and define it as
\begin{align}\label{eq:informal_channel_ent}
	H_\alpha^\uparrow\left(\mathcal{M},B,[\psi_A]\right)\defvar\inf_{\substack{\rho_{A \widetilde{A}
	\widetilde{R}
	}\\
		\rho_A=\psi_A}}H_\alpha^\uparrow(B|C\widetilde{R})_{\mathcal{M}[\rho]},
\end{align}
where the optimization is over all density operators $\rho_{A\widetilde{A}
\widetilde{R}
}$ with a marginal constraint on $A$, and $\widetilde{R}$ is a stabilizing register.
This family can be viewed as a generalization of the channel entropies studied in~\cite{GW21,Yuan19}, in that we can recover the definitions in those works by setting the $C$ register to be trivial and omitting the marginal constraint (equivalently, setting $A$ to be trivial) in the above definition.\footnote{A different notion of channel entropy was studied in~\cite{DGP24}; apart from the marginal constraint, it also differs from our definition in terms of some technical points involving the conditioning and stabilizing registers. However, in special cases the definitions coincide, for instance as described in~\cite[Theorem~5]{DGP24} for the case of von Neumann entropy.}

We extend some results from those works (see also~\cite{KMK05,DKR06,WWY14,GW15,SPSD24}) 
by demonstrating that for $\alpha\geq 1$, the above version of channel conditional entropy (Eq.~\eqref{eq:informal_channel_ent}) is also strongly additive across tensor products of channels. More broadly, we prove a chain rule that is a generalization of this property, by showing that this quantity is superadditive under channel composition in a particular form. 
Our proofs are based on techniques recently developed in~\cite{FFF24}, which used a variational characterization of measured divergences to derive various chain rules, followed by a regularization argument to convert the measured divergences to sandwiched divergences. 

We highlight that the result of strong additivity across tensor products has been independently obtained in a concurrent work~\cite{FHKR25},
using different proof techniques. That work also derives a number of other chain rules; however, those results are ``incomparable to'' (i.e.~neither strictly stronger nor weaker than) our chain rule for channel composition. This is mainly because our chain rule incorporates marginal constraints, making it more suitable for prepare-and-measure QKD protocols --- for such protocols, it yields a more general form of ``adaptive'' protocol as compared to the results in~\cite{FHKR25}; we discuss this in more detail in Sec.~\ref{subsec:simplified}. If the chain rules in that work could be modified to incorporate marginal constraints, then they would yield similar results, with some potential advantages such as not requiring a stabilizing register.

We then apply our results to QKD scenarios. It is often the case that quantum cryptography protocols consist of performing $n$ rounds of some operations, producing registers that will be processed into the final secret key produced by the protocol. Usually, the amount of extractable secret key in the protocol is measured by some entropic quantity such as smooth min-entropy or {\Renyi} entropy 
of the final overall state. It is crucial to relate this entropic quantity to some simpler quantity that can be computed by only analyzing single rounds of protocols. A tool developed to this end is the \term{entropy accumulation theorem} (EAT)~\cite{DFR20,DF19}. Informally speaking, the EAT gives a relation that can be described as follows. Suppose the state $\rho$ in the protocol can be produced by $n$ channels acting ``sequentially'' in some sense, and denote the resulting ``secret'' raw data as $S_1^n$ and some final side-information register as $E_n$.
Then letting $\Omega$ denote the event that the protocol accepts (based on some ``test data'' computed over the $n$ rounds), the EAT states that as long as $\rho$ satisfies some Markov conditions, its smooth min-entropy conditioned on $\Omega$ satisfies a bound with the following form:
\begin{align}\label{eq:EATsketch}
	\text{(informal summary)}\quad \Hmin^\eps(S_1^n | E_n)_{\rho_{|\Omega}} \geq n h_\mathrm{vN} - O(\sqrt{n}),
\end{align}
where loosely speaking, $h_\mathrm{vN}$ is a value slightly smaller than the minimum von Neumann entropy over all single-round states that are ``compatible with'' the accept condition, which suffices to yield a security proof. Subsequently, a \term{generalized entropy accumulation theorem} (GEAT) was developed~\cite{MFSR22,MFSR24}, which relaxed the Markov conditions to a less restrictive non-signalling condition, allowing a broader range of applications. 

However, the EAT and GEAT results had their own limitations. Firstly, the key lengths calculated through either of them were somewhat suboptimal. For example, the key lengths calculated through those frameworks were outperformed by methods such as entropic uncertainty relations (EURs) for smooth entropies~\cite{TL17,LXP+21}. Secondly, they were not applicable to prepare-and-measure (PM) QKD protocols in their full generality. In particular, the EAT was mostly suitable for device-dependent entanglement-based (EB) QKD and device-independent (DI) protocols. While the GEAT was applicable for both PM and EB protocols, it was restrictive in that the adversary was restricted to only interact with a single quantum register sent by the sender at a time, which would in practice limit the repetition rate of the protocol. 

The first limitation was addressed in~\cite{inprep_weightentropy,AHT24} by exploiting methods from the \term{quantum probability estimation} (QPE) framework~\cite{ZFK20}, and working entirely with {\Renyi} entropies. 
As for the second limitation (the repetition-rate restriction), while a solution was given in~\cite{inprep_weightentropy}, it is somewhat restrictive in other senses. Essentially, the idea in that work is as follows. Through the source-replacement technique~\cite{CLL04,FL12}, we can analyze a PM protocol 
as an EB protocol, by supposing that rather than having the sender (Alice) prepare some ensemble of states, she instead prepares a bipartite entangled state and sends only one part of that through a quantum channel (controlled by an adversary Eve) to the receiver (Bob). 
Therefore, by the source-replacement technique, without loss of generality, one can assume that the adversary is instead preparing the joint state of Alice and Bob, but with a constraint on Alice's marginal state that reflects the fact that Alice does not send that part of the state through the adversarial channel. The solution proposed in~\cite{inprep_weightentropy} is restricted in the following sense. Firstly, 
it requires Alice's single-round marginal state to be the same in each round. 
Secondly, 
its proof approach involves modelling the protocol via a specific tensor product of channels, in a manner that imposes some technical restrictions in terms of how the side-information registers can be updated.

In this work, we address those limitations simultaneously. In particular, we derive a new variant of {\Renyi} entropy accumulation, which not only allows imposing different marginal constraints for each channel, but also allows updating the side-information registers in a manner similar to the GEAT. The latter property also allows ``fully adaptive'' updating of the ``statistical estimation'' procedure over the course of the protocol, in a manner described in~\cite{ZFK20} (see also~\cite{AHT24}). We obtain this version of entropy accumulation, which we term the \term{marginal-constrained entropy accumulation theorem} (MEAT), by combining our  result on superadditivity of channel conditional entropy with the QPE-inspired method developed in~\cite{inprep_weightentropy,AHT24} to handle statistical analysis. This allows our new variant of entropy accumulation to be applicable directly in PM-QKD protocols, while benefiting from the tight analysis in~\cite{inprep_weightentropy,AHT24}. We also develop a slight generalization of the notion of \term{$f$-weighted {\Renyi} entropies} from~\cite{inprep_weightentropy}, which may be useful for some more general protocols.

\markup{
We view the main practical advantage of this result over the GEAT as the fact that it removes the repetition-rate restriction when applying the latter to PM-QKD. However, we note that for protocols that do not require a secret memory register (see Sec~\ref{sec:conclusion}), the present result also yields a slightly tighter keyrate bound. 
This improvement follows directly from the fact that our analysis is based on the tighter conditional {\Renyi} entropy  $H^\uparrow_\alpha$, rather than the conditional {\Renyi} entropy $H_\alpha$ used in~\cite{AHT24}.
}
We note however that more generally (outside the context of PM-QKD), our result is neither strictly stronger nor weaker than the GEAT. This is because while we allow for marginal constraints (which the GEAT did not accommodate), we do not allow for ``secret'' internal memory registers that were present in the GEAT. We discuss this in more detail later in Sec~\ref{sec:conclusion}, after having established the relevant notation.

\subsection{Outline of main results}
Our first technical result establishes a superadditivity property of the channel conditional entropy defined in Eq.~\eqref{eq:informal_channel_ent}. In particular, for two channels $\mathcal{E}_1\in\CPTP(A_0Y_0,X_1Y_1)$, and $\mathcal{E}_2\in\CPTP(A_1Y_1,X_2Y_2)$, we have:
\begin{align}
H_\alpha^{\uparrow}\left(\mathcal{E}_2\circ\mathcal{E}_1,X_1X_2,[\psi_{A_0}\otimes\phi_{A_1}]\right)\geq
		H_\alpha^{\uparrow}\left(\mathcal{E}_2,X_2,[\phi_{A_1}]\right)
		+H_\alpha^{\uparrow}\left(\mathcal{E}_1,X_1,[\psi_{A_0}]\right).
\end{align}
As a special case of the above result, for $\mathcal{E}_1\in\CPTP(A_0B_0,X_0Y_0)$, and $\mathcal{E}_2\in\CPTP(A_1B_1,X_1Y_1)$ we get strong additivity under the tensor product of channels:
\begin{align}
    H_\alpha^{\uparrow}&\left(\mathcal{E}_2\otimes\mathcal{E}_1,X_0X_1,[\psi_{A_0}\otimes\phi_{A_1}]\right)=H_\alpha^{\uparrow}\left(\mathcal{E}_2,X_1,[\phi_{A_1}]\right)
				+H_\alpha^{\uparrow}\left(\mathcal{E}_1,X_0,[\psi_{A_0}]\right).
\end{align}
To prove the above results, we first note that it satisfies \term{weak additivity} (i.e.~additivity across a tensor product of IID copies of a channel) by Lemma~\ref{lemma:channel iid additivity}. Then, to prove strong additivity under tensor product and superadditivity under channel composition, we begin by relating the channel conditional entropy to the minimized channel divergence (defined in~\cite{FFF24}) via Lemma~\ref{lemma: channel duality}. We then prove the same statement for their corresponding measured {\Renyi} entropies in Lemma~\ref{lemma:measured chain rule}, and then convert the result into a relation for sandwiched {\Renyi} conditional entropies by regularizing the measured entropies and using the result in~\cite{FFF24}.

Our next technical result is a new variant of the entropy accumulation theorem that preserves a form of marginal constraint and is directly applicable to PM-QKD protocols.
We begin by recalling the source-replacement technique, which states that the state generated in a PM-QKD protocol  can be reformulated as an equivalent state in an EB protocol. More specifically, let $\mathcal{M}_j\in\CPTP(A_{j-1}E_{j-1},S_j\CP_jE_j)$ 
represent some ``effective actions'' (details in Sec.~\ref{sec:security}) performed by Alice, Bob and Eve in the $j$-th round of the protocol,
where $A_{j-1}$ is Alice's marginal register, $E_{j-1}$ and $E_j$ are the adversary's side information registers before/after that channel, $S_j$ is the raw key register prior to the privacy amplification step, and $\CP_j$ is the register that will be used to decide by Alice and Bob to decide whether to abort. Then the global state 
after Bob has completed all his measurements
is given by:
\begin{align}\label{eq:intro global state}
    \rho_{S_1^n\CP_1^nE_n}=\mathcal{M}_n\circ\cdots\circ\mathcal{M}_1[\omega_{A_0^{n-1}E_0}].
\end{align}
Note that since we invoked the source-replacement technique, the marginal on $A_0^{n-1}$ registers must be inaccessible to the adversary. This implies that, assuming no correlation in the source preparation step by Alice, the global state on $A_0^{n-1}$ must be of the form
\begin{align}\label{eq:intro MC}
    \omega_{A_0^{n-1}}=\sigma_{A_0}^{(0)}\otimes\cdots\otimes\sigma_{A_{n-1}}^{(n-1)},
\end{align}
where $\sigma_{A_j}^{(j)}$ is the trusted source-replaced state induced by Alice's state preparation process (see Sec.~\ref{sec:security} for more detail).
For simplicity, we focus on fixed-length protocols (those that always output a key of fixed length whenever they accept). Let Alice and Bob store the information announced for the purpose of the acceptance test\footnote{This step is sometimes referred to as ``parameter estimation''.} in registers $\CP_1^n$, and they accept the protocol if the observed frequency distribution on those registers lies within some predetermined set $S_\Omega$; let this event be denoted as $\Omega$. 
Our result can essentially be summarized as the following bound (see Theorems~\ref{thrm:MEAT_simp},\ref{thrm:MEAT} for formal statements):
\begin{align}
H^\uparrow_\alpha(S_1^n | \CP_1^n E_n)_{\rho_{|\Omega}} \geq  n  h^\uparrow_{\alpha}
- \frac{\alpha}{\alpha-1} \log\frac{1}{p_\Omega}, 
\end{align}
where 
(letting $\bsym{\nu}_{\CP}$ denote the distribution on $\CP$ induced by any state $\nu_{\CP}$)
\begin{align}
h^\uparrow_{\alpha} &= 
\inf_{\mbf{q} \in S_\Omega} \inf_{\nu\in\Sigma_{S\CP E\widetilde{E}}} \left( \frac{\alpha}{{\alpha}-1}D\left(\mbf{q} \middle\Vert \bsym{\nu}_{\CP}\right)+\sum_{\cP\in\supp(\bsym{\nu}_{\CP})}q(\cP)H^\uparrow_{{\alpha}}(S|E\widetilde{E})_{\nu_{|\cP}}  \right) 
,
\end{align}
with $\Sigma_{S\CP E \widetilde{E}}$ informally being the convex hull of the union of outputs of channels $\EATchann_j \otimes \idmap_{\widetilde{E}}$ such that the marginal on $A_j$ is preserved (see Definition~\ref{def:range} for details), where $\widetilde{E}$ is a register of large enough dimension to serve as a purifying register for any of the $A_{j-1}E_{j-1}$ registers. 
As outlined in the introduction of~\cite{AHT24}, the quantity $h^\uparrow_{\alpha}$ can be qualitatively viewed as a relaxation of the minimum value of the single-round {\Renyi} entropy $H^\uparrow_{{\alpha}}(S|E\widetilde{E})$ over all states ``compatible with'' $S_\Omega$; the important aspect here is that it has already accounted for all finite-size and non-IID effects in the protocol.

\subsection{Paper structure}
The structure of the paper is as follows: 
\begin{itemize}
\item In Sec~\ref{sec:notation}, we lay out preliminary notations and concepts. 

\item In Sec~\ref{sec:additivity}, we provide the additivity and chain rule results that serve as a building block of this new variant of entropy accumulation. In particular, Theorem~\ref{thrm:channel cond ent chain rule} is our chain rule for channel composition, and it yields Corollary~\ref{cor:channel cond ent additivity} for strong additivity across tensor products.

\item In Sec~\ref{sec:MEAT}, we introduce and generalize the notion of {$f$-weighted {\Renyi} entropies} from~\cite{inprep_weightentropy}, which we then use to prove the main result of this work, the MEAT (Theorem~\ref{thrm:MEAT}, or Theorem~\ref{thrm:MEAT_simp} for a slightly simpler special case). 

\item In Sec~\ref{sec:security}, we present an outline of how this can provide a security proof for PM-QKD protocols.

\item In Sec~\ref{sec:conclusion} we discuss prospects for future work.
\end{itemize}

\section{Preliminaries}
\label{sec:notation}

\begin{table}[ht!]
\caption{List of notation}\label{tab:notation}
\def\arraystretch{1.5} 
\setlength\tabcolsep{.28cm}
\begin{tabular}{c l}
\toprule
\textit{Symbol} & \textit{Definition} \\
\toprule
$\log$ & Base-$2$ logarithm \\
\hline
$H$ & Base-$2$ von Neumann entropy \\
\hline
$\mathbb{R}\cup\{-\infty,+\infty\}$ & Extended real line \\
\hline
$\norm{\cdot}_p$ & Schatten $p$-norm \\
\hline
$\left|\cdot\right|$ & Absolute value of operator; $\left|M\right| \defvar \sqrt{M^\dagger M}$ \\
\hline
$A\odot B$ & Entrywise multiplication of $A$ and $B$ \\
\hline
$A\perp B$ & $A$ and $B$ are orthogonal; $AB=BA=0$ \\
\hline
$X\geq Y$ (resp.~$X>Y$) & $X-Y$ is positive semidefinite (resp.~positive definite)\\
\hline
$\Pos(A)$ 
& Set of positive semidefinite operators on register $A$\\
\hline
$\operatorname{Herm}(A)$ 
& Set of hermitian operators on register $A$\\
\hline
$\CPTP(A,B)$
& Set of completely positive and trace preserving maps from $A$ to $B$.\\
\hline
$\CPo(A,B)$
& Set of completely positive maps from $A$ to $B$.\\
\hline
$\dop{=}(A)$ (resp.~$\dop{\leq}(A)$) & Set of normalized (resp.~subnormalized) states on register $A$ \\
\hline
$\idnorm_A$ & Maximally mixed state on register $A$ \\
\hline
$A_j^k$ & Registers $A_j \dots A_k$ \\
\toprule
\end{tabular}
\def\arraystretch{1}
\end{table}

We list some basic notation in Table~\ref{tab:notation}.
Apart from the notation in 
that table,
we will also need to use some other concepts, which we shall define below, and briefly elaborate on in some cases. 
In this work, we will assume that all systems are finite-dimensional, but we will not impose any bounds on the system dimensions unless otherwise specified. 
All entropies are defined in base~$2$.
Throughout this work we will often leave tensor products with identity channels implicit; e.g.~given a channel $\mathcal{E}\in\CPTP(Q,Q')$, we often use the compact notation
\begin{align}
\mathcal{E}[\rho_{QR}] \defvar (\mathcal{E} \otimes \idmap_R)[\rho_{QR}].
\end{align}
The same convention holds for completely positive maps as well.
\begin{definition}\label{def:freq}
(Frequency distributions) For a string $z_1^n\in\mathcal{Z}^n$ on some alphabet $\mathcal{Z}$, $\freq_{z_1^n}$ denotes the following probability distribution on $\mathcal{Z}$:
\begin{align}
\freq_{z_1^n}(z) \defvar \frac{\text{number of occurrences of $z$ in $z_1^n$}}{n} .
\end{align}
\end{definition}

\begin{definition}
A state $\rho \in \dop{\leq}(CQ)$ is said to be \term{classical on $C$} (with respect to a specified basis on $C$) if it is in the form 
\begin{align}
\rho_{CQ} = \sum_c \lambda_c \pure{c} \otimes \sigma_c,
\label{eq:cq}
\end{align}
for some normalized states $\sigma_c \in \dop{=}(Q) $ and weights $\lambda_c \geq 0$, with $\ket{c}$ being the specified basis states on $C$. In most circumstances, we will not explicitly specify this ``classical basis'' of $C$, leaving it to be implicitly defined by context.
It may be convenient to absorb the weights $\lambda_c$ into the states $\sigma_c$, writing them as subnormalized states $\omega_c = \lambda_c\sigma_c \in \dop{\leq}(Q)$ instead. 
\end{definition}

\begin{definition}\label{def:cond}
(Conditioning on classical events) For a state $\rho \in \dop{\leq}(CQ)$ classical on $C$, written in the form
$\rho_{CQ} = \sum_c \pure{c} \otimes \omega_c$ 
for some $\omega_c \in \dop{\leq}(Q)$,
and an event $\Omega$ defined on the register $C$, we will define a corresponding \term{partial state} and \term{conditional state} as, respectively,
\begin{align}
\rho_{\land\Omega} \defvar \sum_{c\in\Omega} \pure{c} \otimes \omega_c, \qquad\qquad \rho_{|\Omega} \defvar \frac{\tr{\rho}}{\tr{\rho_{\land\Omega}}} \rho_{\land\Omega} = \frac{
\sum_{c} \tr{\omega_c}
}{\sum_{c\in\Omega} \tr{\omega_c}} \rho_{\land\Omega} .
\end{align}
Strictly speaking, the latter is not well-defined when $\tr{\rho_{\land\Omega}}=0$; in this work we shall simply interpret $\rho_{|\Omega}$ to be defined by setting it to some arbitrary state in such circumstances (our results are valid under this interpretation).
The process of taking partial states is commutative and ``associative'', in the sense that for any events $\Omega,\Omega'$ we have $(\rho_{\land\Omega})_{\land\Omega'} = (\rho_{\land\Omega'})_{\land\Omega} = \rho_{\land(\Omega\land\Omega')}$; hence for brevity we will denote all of these expressions as
\begin{align}
\rho_{\land\Omega\land\Omega'} \defvar (\rho_{\land\Omega})_{\land\Omega'} = (\rho_{\land\Omega'})_{\land\Omega} = \rho_{\land(\Omega\land\Omega')}.
\end{align}
On the other hand, some disambiguating parentheses are needed when combined with taking conditional states (due to the normalization factors).
\end{definition}

In light of the preceding two definitions, for a normalized state $\rho \in \dop{=}(CQ)$ that is classical on $C$, it is reasonable to write it in the form
\begin{align}
\rho_{CQ} = \sum_c \rho(c) \pure{c} \otimes \rho_{Q|c},
\label{eq:cqstateprobs}
\end{align}
where $\rho(c)$ denotes the probability of $C=c$ according to $\rho$, and $\rho_{Q|c}$ can indeed be interpreted as the conditional state on $Q$ corresponding to $C=c$, i.e.~$\rho_{Q|c} = \tr[C]{\rho_{|\Omega}}$ where $\Omega$ is the event $C=c$. 
We may sometimes denote the distribution on $C$ induced by $\rho$ as the tuple
\begin{align}\label{eq:stateprobvec}
\bsym{\rho}_C \defvar \left(\rho(1),\rho(2),\dots\right).
\end{align}
Further extending this notation, if there are two classical registers $C$ and $C'$, we write $\rho(c|c')$ to denote the conditional probabilities induced by a state $\rho$, i.e.
\begin{align}
\rho(c|c') \defvar \frac{\rho(cc')}{\rho(c')} \quad \text{for } \rho(c') \neq 0.
\end{align}

\begin{definition}
(Measure-and-prepare or read-and-prepare channels)
A (projective) \term{measure-and-prepare channel} is a channel $\mathcal{E} \in \CPTP(Q,QQ')$ of the form
\begin{align}
\mathcal{E}[\rho_Q] = \sum_j (P_j \rho_Q P_j) \otimes \sigma_{Q'|j},
\end{align}
for some projective measurement $\{P_j\}$ on $Q$ and some normalized states $\sigma_{Q'|j}$.
If $Q$ is classical and the measurement is a projective measurement in its classical basis, we shall refer to it as a \term{read-and-prepare channel}.
Note that a read-and-prepare channel always simply extends the state ``without disturbing it'', i.e. tracing out $Q'$ results in the original state again.
\end{definition}

The following definitions of {\Renyi} divergences and entropies are reproduced from~\cite{Tom16}, and coincide with those in~\cite{DFR20,MFSR24} for normalized states. 

\begin{definition}\label{def:sandwiched divergence}
({\Renyi} divergence)
For any $\rho,\sigma\in\Pos(A)$ with $\tr{\rho}\neq0$, and $\alpha\in(0,1)\cup (1,\infty)$, the (sandwiched) \Renyi\ divergence between $\rho$, $\sigma$ is defined as:
\begin{align}
    \label{eq:sand_renyi_div}
    D_\alpha(\rho\Vert\sigma)=\begin{cases}
    \frac{1}{\alpha-1}\log\frac{\tr{ \left(\sigma^{\frac{1-\alpha}{2\alpha}}\rho\sigma^{\frac{1-\alpha}{2\alpha}}\right)^\alpha}}{\tr{\rho}} &\left(\alpha < 1\ \land\ \rho\not\perp\sigma\right)\vee \left(\supp(\rho)\subseteq\supp(\sigma)\right) \\ 
    +\infty & \text{otherwise},
    \end{cases}  
\end{align}
where for $\alpha>1$ the $\sigma^{\frac{1-\alpha}{2\alpha}}$ terms are defined via the Moore-Penrose pseudoinverse if $\sigma$ is not full-support~\cite{Tom16}.
The above definition is extended to $\alpha \in \{0,1,\infty\}$ by taking the respective limits.
For the $\alpha=1$ case, it reduces to the Umegaki divergence:
\begin{align}
    \label{eq:umegaki_div}
    D(\rho\Vert\sigma)=\begin{cases}
        \frac{\tr{\rho\log\rho-\rho\log\sigma}}{\tr{\rho}} & \supp(\rho)\subseteq\supp(\sigma)\\
    +\infty & \text{otherwise}. 
    \end{cases}
\end{align}
For any two classical probability distributions $\mbf{p},\mbf{q}$ on a common alphabet, we also define the {\Renyi} divergence $D_\alpha(\mbf{p}\Vert\mbf{q})$ analogously, e.g.~by viewing the distributions as diagonal density matrices in the above formulas; in the $\alpha=1$ case this gives the Kullback–Leibler (KL) divergence.
\end{definition}

\begin{definition}\label{def:sandwiched entropy}
({\Renyi} entropies)
For any bipartite state $\rho\in
\dop{=}(AB)
$, and $\alpha\in[0,\infty]$, we define the following two (sandwiched) {\Renyi} conditional entropies:
\begin{align}
    \label{eq:cond_renyi}
    &H_\alpha(A|B)_\rho=-D_\alpha(\rho_{AB}\Vert\id_A\otimes\rho_B)\notag\\
    &H_\alpha^\uparrow(A|B)_\rho=\sup_{\sigma_B\in\dop{=}(B)}-D_\alpha(\rho_{AB}\Vert\id_A\otimes\sigma_B).
\end{align}
For $\alpha=1$, both the above values coincide and are equal to the von Neumann entropy.
\end{definition}

\begin{definition}\label{def:measured divergence}
	(Measured {\Renyi} divergence)
	For any state $\rho,\sigma\in\Pos(A)$, with $\tr{\rho}\neq0$, and $\alpha\in [0,\infty]$, the measured {\Renyi} divergence between $\rho, \sigma$ is defined as:
	\begin{align}
		\label{eq:measured divergence}
		D_{\mathbb{M},\alpha}(\rho\Vert\sigma)=\sup_{\mathcal{X},M}D_\alpha(P_{\rho,M}\Vert P_{\sigma,M}),
	\end{align}
	where the optimization is over finite sets $\mathcal{X}$, and POVMs $M$ on it, with $P_{\rho,M}(x)=\tr{M(x)\rho}$, and $D_\alpha(P_{\rho,M}\Vert P_{\sigma,M})$ is the classical {\Renyi} divergence between the two distributions.
\end{definition}

\begin{remark}
In the above definitions of divergences, for generality we allowed arbitrary nonzero positive semidefinite operators in the first argument; however, throughout the rest of this work we will exclusively be focusing on only the case where those operators are normalized. Hence we can apply various results from~\cite{FFF24} that were derived under that restriction.
\end{remark}

The work~\cite{FFF24} studied \term{minimized channel divergences}, which differ from the more ``standard'' definition of channel divergences by minimizing over input states, rather than maximizing. We now present a slight generalization of that concept, incorporating some marginal-state constraints. Note that there is no ``stabilizing register'' (i.e.~purification of the input states) when defining minimized channel divergences --- since it involves minimization instead of maximization, including such a register would make no difference whenever the underlying operator divergence satisfies data-processing.
\begin{definition}\label{def:minimized channel divergence}
	(Minimized channel divergence)
	For $\mathcal{M}\in\CPTP(A\widetilde{A},B)$, $\mathcal{N}\in\CPo(A'\widetilde{A}',B)$, and $\psi_A \in \dop{=}(A)$, $\phi_{A'} \in \dop{=}(A')$, the (marginal-constrained) minimized channel divergence between $\mathcal{M},\mathcal{N}$ is defined as:
	\begin{align}
		\label{eq: constrained minimized channel divergence}
		\mathbb{D}^{\inf}(\mathcal{M},[\psi_A]\Vert\mathcal{N},[\phi_{A'}])=\inf_{\substack{\rho\in\dop{=}(A\widetilde{A})\ \suchthat\ \rho_A=\psi_A\\ \sigma\in\dop{=}(A'\widetilde{A}')\ \suchthat\ \sigma_{A'}=\phi_{A'}}}\mathbb{D}(\mathcal{M}[\rho]\Vert\mathcal{N}[\sigma]),
	\end{align}
	where $\mathbb{D}$ can be any of the divergences defined in this work. Its regularized version is defined as
	\begin{align}
		\label{eq:constrained regularized min channel div}
		\mathbb{D}^{\inf,\reg}(\mathcal{M},[\psi_A]\Vert\mathcal{N},[\phi_{A'}])=\lim_{m\rightarrow\infty}\frac{1}{m}\mathbb{D}^{\inf}(\mathcal{M}^{\otimes m},[\psi_A^{\otimes m}]\Vert\mathcal{N}^{\otimes m},[\phi_{A'}^{\otimes m}]),
	\end{align}
	if the limit exists. We view the marginal-state specifications in the above notation as ``optional arguments'', in that if we do not wish to impose a marginal constraint on the registers $A$, $A'$, or both, then we denote the corresponding minimized divergences as $\mathbb{D}^{\inf}(\mathcal{M}\Vert\mathcal{N},[\phi_{A'}])$, $\mathbb{D}^{\inf}(\mathcal{M},[\psi_A]\Vert\mathcal{N})$, and $\mathbb{D}^{\inf}(\mathcal{M}\Vert\mathcal{N})$, respectively, e.g.
	\begin{align}
		\mathbb{D}^{\inf}(\mathcal{M},[\psi_A]\Vert\mathcal{N})=\inf_{\substack{\rho\in\dop{=}(A\widetilde{A})\ \suchthat\ \rho_A=\psi_A\\ \sigma\in\dop{=}(A'\widetilde{A}')}}\mathbb{D}(\mathcal{M}[\rho]\Vert\mathcal{N}[\sigma]),
	\end{align}
	and so on. 
	Moreover, the regularized versions for these cases are defined and represented similarly, if the corresponding limits exist.
\end{definition}

We also define channel entropies with marginal constraints, generalizing the definitions from~\cite{GW21,Yuan19}. In this case, there is indeed a nontrivial stabilizing register in the definition --- this is essentially because it is defined via a minimization, which corresponds more closely to the standard version of channel divergences (since by Definition~\ref{def:sandwiched entropy}, minimizing over entropies is basically maximizing over divergences). However, we show later in Lemma~\ref{lemma: channel duality} that it also has a ``dual'' relation with minimized channel divergences.

\begin{definition}\label{def:channel conditional entropy}
	({\Renyi} channel conditional entropy)
	For any channel $\mathcal{M}\in\CPTP(A\widetilde{A},BC)$, any $\psi_A\in\dop{=}(A)$, and $\alpha\in [0,\infty]$, the (marginal-constrained) {\Renyi} channel conditional entropy is defined as\footnote{As we have assumed all registers are finite-dimensional in this work, this infimum is always finite.}
	\begin{align}
		\label{eq:channel conditional entropy}
		H_\alpha^{\uparrow}(\mathcal{M},B,[\psi_A])=\inf_{\rho\in\dop{=}(A\widetilde{A}\widetilde{R})\ \suchthat\ \rho_A=\psi_A}H_\alpha^\uparrow(B|C\widetilde{R})_{\mathcal{M}[\rho]},
	\end{align}
	where $\widetilde{R}$ is isomorphic to $A\widetilde{A}$. Note that the infimum in the above optimization is always attained for finite-dimensional registers (by compactness), and furthermore can be restricted to the set of pure states, since by data-processing, purifying the input state only decreases the conditional entropy of the states in the above optimization. Its regularized version is defined as
	\begin{align}
		\label{eq:reg channel conditional entropy}
		H_\alpha^{\uparrow,\reg}(\mathcal{M},B,[\psi_A])=\lim_{m\rightarrow\infty}\frac{1}{m}H_\alpha^{\uparrow}(\mathcal{M}^{\otimes m},B^m,[\psi_A^{\otimes m}]),
	\end{align}
	noting that the limit always exists by Fekete's lemma~\cite{Fekete,Davis07}.
\end{definition}

Note that the regularized {\Renyi} channel conditional entropy is additive across IID copies of the channel, i.e.~weakly additive:
\begin{align}
	\label{eq:additivity channel cond ent}
	H_\alpha^{\uparrow,\reg}(\mathcal{M}^{\otimes n},B^n,[\psi_A^{\otimes n}])&=\lim_{m\rightarrow\infty}\frac{1}{m}H_\alpha^{\uparrow}(\mathcal{M}^{\otimes nm},B^{nm},[\psi_A^{\otimes nm}])\notag\\
	&=n\lim_{m'\rightarrow\infty}\frac{1}{m'}H_\alpha^{\uparrow}(\mathcal{M}^{\otimes m'},B^{m'},[\psi_A^{\otimes m'}])\notag\\
	&=nH_\alpha^{\uparrow,\reg}(\mathcal{M},B,[\psi_A]),
\end{align}
where the second line holds via a change of indices (and noting that since the $m'\to\infty$ limit exists, any subsequence of it converges to the same limit).

In some parts of this work, we will need the notion of taking some arbitrary purification of a state. For such situations, it is convenient to introduce the concept of a purifying function: 
\begin{definition}\label{def:purify}
	For registers $Q,Q'$ with $\dim(Q) \leq \dim(Q')$, a \term{purifying function for $Q$ onto $Q'$} is a function $\pf: \dop{\leq}(Q) \to \dop{\leq}(QQ')$ such that for any state $\rho_Q$, the state $\pf(\rho_Q)$ is a purification of $\rho_Q$ onto the register $Q'$, i.e.~a (possibly subnormalized) rank-$1$ operator such that $\tr[Q']{\pf(\rho_Q)} = \rho_Q$.
\end{definition}
Note that a purifying function is \emph{not} a channel (i.e.~CPTP map), for instance because it is necessarily nonlinear.

We now also list some useful properties we will use throughout our work.

\begin{fact} \label{fact:DPI}
(Data-processing~\cite[Theorem~1]{FL13}; see also~\cite{MDS+13,Beigi13,MO14,Tom16}) For any $\alpha\in[\tfrac{1}{2},\infty]$, any $\rho,\sigma\in\Pos(Q)$ with $\tr{\rho}\neq0$, and any channel $\mathcal{E}\in\CPTP(Q,Q')$, we have:
\begin{align}
D_\alpha(\rho\Vert\sigma) \geq D_\alpha(\mathcal{E}[\rho]\Vert\mathcal{E}[\sigma]),
\end{align}
and thus also for any $\rho\in\dop{=}(Q''Q)$,
\begin{align}
H_\alpha(Q''|Q)_{\rho} \leq H_\alpha(Q''|Q')_{\mathcal{E}[\rho]}, \quad H^\uparrow_\alpha(Q''|Q)_{\rho} \leq H^\uparrow_\alpha(Q''|Q')_{\mathcal{E}[\rho]}.
\end{align}
If $\mathcal{E}$ is an isometry, all the above bounds hold with equality.
\end{fact}

\begin{fact} \label{fact:classmix}
(Conditioning on classical registers; see~\cite[Sec.~III.B.2 and Proposition~9]{MDS+13} or~\cite[Eq.~(5.32) and Proposition~5.1]{Tom16}) Let $\rho,\sigma \in \dop{=}(C Q)$ be states classical on $C$. Then 
\begin{align}
    \label{eq:classmixD}
    \forall \alpha\in(0,1)\cup (1,\infty), \qquad D_\alpha(\rho\Vert\sigma)=\frac{1}{\alpha-1}\log\left(\sum_{c}\rho(c)^\alpha\sigma(c)^{1-\alpha}2^{(\alpha-1)D_\alpha(\rho_{Q|c}\Vert\sigma_{Q|c})}\right),
\end{align}
where any terms with $\rho(c)=0$ are interpreted as having zero contribution to the sum, even if the corresponding value of $\sigma(c)^{1-\alpha}2^{(\alpha-1)D_\alpha(\rho_{Q|c}\Vert\sigma_{Q|c})}$ is infinite.
Hence for a state $\rho \in \dop{=}(C Q Q')$ classical on $C$ (under the convention that $\frac{\alpha}{\alpha-1}=1$ when $\alpha=\infty$):
\begin{align}
\label{eq:classmixHdown}
\forall \alpha\in(0,1)\cup (1,\infty), \qquad &H_\alpha(Q|CQ')=\frac{1}{1-\alpha}\log\left(\sum_{c}\rho(c)2^{(1-\alpha)H_\alpha(Q|Q')_{\rho|c}}\right),\\
\label{eq:classmixHup}
\forall \alpha\in[\tfrac{1}{2},1)\cup (1,\infty], \qquad &H^\uparrow_\alpha(Q|CQ')=\frac{\alpha}{1-\alpha}\log\left(\sum_{c}\rho(c)2^{\frac{1-\alpha}{\alpha}H^\uparrow_\alpha(Q|Q')_{\rho|c}}\right).
\end{align}

\begin{remark}
\label{rem:classmixlimits}
While the above relations are sometimes only stated for $\alpha\neq\infty$, 
we have presented~\eqref{eq:classmixHup} in particular as explicitly including $\alpha=\infty$. This is because it is usually implicit that the $\alpha\to\infty$ limits yield these additional results:
\begin{align}
\label{eq:classmixHmindown}
&H_\infty(Q|CQ')=\min_{c\in\supp(\bsym{\rho}_{C})} H_\infty(Q|Q')_{\rho|c},\\
\label{eq:classmixHminup}
&H^\uparrow_\infty(Q|CQ')=\log\left(\sum_{c}\rho(c)2^{H^\uparrow_\infty(Q|Q')_{\rho|c}}\right),
\end{align} 
and~\eqref{eq:classmixHminup} is equivalent to setting $\alpha=\infty$ in \eqref{eq:classmixHup}, under the stated convention that $\frac{\alpha}{\alpha-1}=1$ when $\alpha=\infty$. Note that it is less straightforward to claim a similar such correspondence between~\eqref{eq:classmixHmindown} and~\eqref{eq:classmixHdown}, because ``naively'' setting $\frac{1}{\alpha-1}=0$ and $2^{-\alpha}=0$ in that case yields an underspecified $0\log0$ expression.\footnote{It is technically still valid under the interpretation that all expressions for $\alpha=\infty$ in the extended reals are interpreted by taking the limit ``as a whole'' rather than ``term-by-term''. In some sense, the correspondence between~\eqref{eq:classmixHminup} and \eqref{eq:classmixHup} happens to work because taking the limit of the $\frac{\alpha}{\alpha-1}$ terms already yields a non-singular expression that commutes with taking this limit. 
However, we will not further discuss this nuance within this work.} (We have also addressed a minor oversight in~\cite[Proposition~5.1]{Tom16} by slightly restricting the range of $\alpha$ in~\eqref{eq:classmixHminup} to exclude $\alpha<\frac{1}{2}$, as the proof of that relation uses a data-processing argument, at least for pinching channels.) 
\end{remark}
\end{fact}
\begin{fact}\label{fact:duality}
	(Duality relation~\cite[Proposition~5.3]{Tom16}; see also~\cite[Theorem 10]{MDS+13})
	For any pure state $\rho\in\dop{=}(ABC)$, and any $\alpha\in [\frac{1}{2},\infty]$, we have
	\begin{align}
		\label{eq:duality}
		H^\uparrow_\alpha(A|B)_\rho + H^\uparrow_\beta(A|C)_\rho=0,\quad \text{for } \frac{1}{\alpha}+\frac{1}{\beta}=2.
	\end{align}
\end{fact}

\section{Chain rule and additivity}
\label{sec:additivity}
In this section, we prove a chain rule for the (marginal-constrained) {\Renyi} channel conditional entropy when channels are composed; put another way, it is superadditive under channel composition. As a special case of this result, we establish the strong additivity of this quantity under tensor products. To achieve this, we initially show that the {\Renyi} channel conditional entropy exhibits weak additivity, i.e.~additivity under IID copies of channels.
\begin{lemma}\label{lemma:channel iid additivity}
	(Weak additivity) Consider a channel $\mathcal{E}\in\CPTP(AY,X\widehat{Y})$ and a state $\psi_A \in \dop{=}(A)$. For any $\alpha\in[1,\infty]$, 
	and any $m\in\mathbb{N}$, we have
	\begin{align}
		\label{eq:channel iid additivity}
		H_\alpha^{\uparrow}\left(\mathcal{E},X,[\psi_A]\right)=H_\alpha^{\uparrow,\reg}\left(\mathcal{E},X,[\psi_A]\right)=\frac{1}{m}H_\alpha^{\uparrow}\left(\mathcal{E}^{\otimes m},X^m,[\psi_A^{\otimes m}]\right).
		\end{align}
	\begin{proof}
		This result was previously proven in~\cite{inprep_weightentropy}, but for completeness we include a proof in this work.
		First suppose $\alpha>1$. The first equality in~\eqref{eq:channel iid additivity} is established using nearly the same argument as in the proof of~\cite[Lemma~3.5]{MFSR24}\footnote{A similar approach has been used in previous work, for instance~\cite{GW21}.
		}, just replacing the de Finetti theorem used in that argument with a generalized version described in~\cite[Corollary~3.2]{FR15}. Subsequently, the second equality holds by the weak additivity property of the regularized channel conditional entropy in~\eqref{eq:additivity channel cond ent}.
		
		To prove the case where $\alpha=1$, we first note that {\Renyi} conditional entropy $H_\alpha^\uparrow(Q|Q')$ converges to von Neumann conditional entropy $H(Q|Q')$ as $\alpha\to 1^+$, and the convergence is uniform (in the sense of being independent of the state, apart from the system dimension). More specifically, 
		by~\cite[Lemma~B.9]{DFR20} we have the following continuity bound:
		\begin{align}
			\label{eq:uniform_cont}
            H_\alpha^\uparrow(Q|Q') \geq H(Q|Q') - (\alpha-1) K_{\dim(Q)}             
            \quad\qquad\forall\alpha \in \left(1,1+\frac{1}{\log(1+2\dim(Q))}\right),
		\end{align}
		where $K_{\dim(Q)}$ is a positive constant depending only on the system dimensions.
		Hence, we have the following chain of relations: 
		\begin{align}
			\label{eq:von-nuemann add}
			&\;\forall\alpha>1\qquad H_\alpha^{\uparrow}\left(\mathcal{E},X,[\psi_A]\right)=\frac{1}{m}H_\alpha^{\uparrow}\left(\mathcal{E}^{\otimes m},X^m,[\psi_A^{\otimes m}]\right)\notag\\
			\iff
			&\;
			\forall\alpha>1\qquad\inf_{\substack{\rho\in\dop{=}(AY\widetilde{R})\\ \suchthat\ \rho_A=\psi_A}} 	H_\alpha^\uparrow(X|\widehat{Y}\widetilde{R})_{\mathcal{E}[\rho]}=\inf_{\substack{\rho\in\dop{=}(A^mY^m\widetilde{R})\\ \suchthat\ \rho_{A^m}=\psi_A^{\otimes m}}}\frac{1}{m} 	H_\alpha^\uparrow(X^m|\widehat{Y}^m\widetilde{R})_{\mathcal{E}^{\otimes m}[\rho]}\notag\\
			\implies& 
			\lim_{\alpha\to 1^+}\inf_{\substack{\rho\in\dop{=}(AY\widetilde{R})\\ \suchthat\ \rho_A=\psi_A}} 	H_\alpha^\uparrow(X|\widehat{Y}\widetilde{R})_{\mathcal{E}[\rho]}=\lim_{\alpha\to 1^+}\inf_{\substack{\rho\in\dop{=}(A^mY^m\widetilde{R})\\ \suchthat\ \rho_{A^m}=\psi_A^{\otimes m}}}\frac{1}{m} 	H_\alpha^\uparrow(X^m|\widehat{Y}^m\widetilde{R})_{\mathcal{E}^{\otimes m}[\rho]}\notag\\
			\iff&\inf_{\substack{\rho\in\dop{=}(AY\widetilde{R})\\ \suchthat\ \rho_A=\psi_A}}\lim_{\alpha\to 1^+} 	H_\alpha^\uparrow(X|\widehat{Y}\widetilde{R})_{\mathcal{E}[\rho]}=\inf_{\substack{\rho\in\dop{=}(A^mY^m\widetilde{R})\\ \suchthat\ \rho_{A^m}=\psi_A^{\otimes m}}}\lim_{\alpha\to 1^+}\frac{1}{m} 	H_\alpha^\uparrow(X^m|\widehat{Y}^m\widetilde{R})_{\mathcal{E}^{\otimes m}[\rho]}\notag\\
			\iff& \inf_{\substack{\rho\in\dop{=}(AY\widetilde{R})\\ \suchthat\ \rho_A=\psi_A}} 	H(X|\widehat{Y}\widetilde{R})_{\mathcal{E}[\rho]}=\inf_{\substack{\rho\in\dop{=}(A^mY^m\widetilde{R})\\ \suchthat\ \rho_{A^m}=\psi_A^{\otimes m}}}\frac{1}{m} 	H(X^m|\widehat{Y}^m\widetilde{R})_{\mathcal{E}^{\otimes m}[\rho]}
			\notag\\
			\iff&\qquad\qquad H\left(\mathcal{E},X,[\psi_A]\right)=\frac{1}{m}H\left(\mathcal{E}^{\otimes m},X^m,[\psi_A^{\otimes m}]\right)
,
		\end{align}
		where the fourth line holds due to
		Eq.~\eqref{eq:uniform_cont}, which allows us to swap the infimum and one-sided limit on both sides, recalling that the relevant dimensions are all constants for the purposes of this analysis, and that the $\alpha \to 1^+$ limit of $H_\alpha^\uparrow$ exists. (Explicitly: given a function $f: (1,A) \times \mathcal{D} \to (B,\infty)$ for some set $\mathcal{D}$ and constants $A>1$, $B>-\infty$ , if we have that $\lim_{x\to 1^+} f(x,y) \eqqcolon f_1(y)$ exists for every $y$, and there exists a constant $K$ such that $f(x,y) \geq f_1(y) - (x-1)K$ for all $x$ and $y$, then
        \begin{align}
        \lim_{x\to 1^+}\inf_{y\in\mathcal{D}} f(x,y) &\geq \lim_{x\to 1^+}\inf_{y\in\mathcal{D}} \left(f_1(y) - (x-1)K\right) \nonumber\\
        &= \inf_{y\in\mathcal{D}} f_1(y) - \lim_{x\to 1^+}(x-1)K \nonumber\\
        &= \inf_{y\in\mathcal{D}} f_1(y) 
        =\inf_{y\in\mathcal{D}} \lim_{x\to 1^+} f(x,y), 
        \end{align}
        and conversely we have the ``generic'' bound
        \begin{align}
        \inf_{y\in\mathcal{D}} \lim_{x\to 1^+} f(x,y) &\geq \inf_{y\in\mathcal{D}} \lim_{x\to 1^+} \inf_{y'\in\mathcal{D}} f(x,y') \nonumber\\
        &= \lim_{x\to 1^+} \inf_{y'\in\mathcal{D}} f(x,y') ,
        \end{align}
        so the two orderings of the limit and infimum coincide.)

        Finally, to obtain the first equality in~\eqref{eq:channel iid additivity} for $\alpha=1$, we take the limit of $m\to\infty$ from the last line in~\eqref{eq:von-nuemann add}.
	\end{proof}
\end{lemma}

The following lemma 
uses the duality relation in Fact~\ref{fact:duality} to establish a connection between channel conditional entropy and minimized channel divergence. Through the rest of this section, we will frequently be using the notation $\mathbb{I}_{Q'} \otimes \idmap_{Q} \in \CPo(Q,QQ')$ to denote the CP map that simply outputs the tensor product of the input state on $Q$ with the identity operator $\mathbb{I}_{Q'}$, i.e.
\begin{align}
(\mathbb{I}_{Q'} \otimes \idmap_{Q}) [\rho_{Q}] \defvar \mathbb{I}_{Q'} \otimes \rho_{Q}.
\end{align}

\begin{lemma}\label{lemma: channel duality}
	Consider a channel $\mathcal{E}\in\CPTP(AY,X\widehat{Y})$ and a state $\psi_A \in \dop{=}(A)$. Let $V\in\CPTP(AY,X\widehat{Y}Z)$ be any Stinespring dilation of $\mathcal{E}$. Then, for any $\alpha\in [\frac{1}{2},\infty]$ we have:
	\begin{align}
		\label{eq:channel duality}
		H_\alpha^{\uparrow}\left(\mathcal{E},X,[\psi_A]\right)=D_\beta^{\inf}\left(\Tr_{\widehat{Y}}\circ V,[\psi_A]\Vert \id_X \otimes \idmap_Z \right),
	\end{align}
	where $\beta=\frac{\alpha}{2\alpha -1} \in [\frac{1}{2},\infty]$.
	\begin{proof}
		For any pure state $\rho\in\dop{=}(AY\widetilde{R})$, $V[\rho]$ is also a pure state. Thus, we have:        
		\begin{align}
			\label{eq:channel duality proof}
			H_\alpha^\uparrow(X|\widehat{Y}\widetilde{R})_{\mathcal{E}[\rho]}&=-H_\beta^\uparrow(X|Z)_{V[\rho]}\notag\\
            &=-H_\beta^\uparrow(X|Z)_{\Tr_{\widehat{Y}\widetilde{R}}\circ V[\rho]}\notag\\
            &=-H_\beta^\uparrow(X|Z)_{\Tr_{\widehat{Y}}\circ V\circ\Tr_{\widetilde{R}}[\rho]}\notag\\       
			&=-H_\beta^\uparrow(X|Z)_{\Tr_{\widehat{Y}}\circ V[\rho_{AY}]}\notag\\
			&=\inf_{\sigma_{Z}\in\dop{=}(Z)}D_\beta\left(\Tr_{\widehat{Y}}\circ V[\rho_{AY}]\Vert\id_{X}\otimes\sigma_Z\right),
		\end{align}         
        where the first line follows from Fact.~\ref{fact:duality}, the third line holds since $V$ does not act on $\widetilde{R}$, and the last line holds by the definition of {\Renyi} conditional entropies in Definition~\ref{def:sandwiched entropy}. Taking the infimum over all
		pure
		$\rho\in\dop{=}(AY\widetilde{R})$ such that $\rho_A=\psi_A$, and noting that the last line is independent of $\widetilde{R}$, proves the claim.
	\end{proof}
\end{lemma}
With the above two lemmas established, we can now derive several further relationships that link channel conditional entropy to minimized channel divergence and their regularized counterparts.
\begin{corollary}\label{cor:equivalence channel ent and min div}
	Let $\mathcal{E}\in\CPTP(AY,X\widehat{Y})$ be a channel, $\psi_A \in \dop{=}(A)$ be a state, and $V\in\CPTP(AY,X\widehat{Y}Z)$ be any Stinespring dilation of $\mathcal{E}$. Then, for any $\alpha\in[1,\infty]$, we have
	\begin{align}
		\label{eq:equivalence channel ent and min div}
			&H_\alpha^{\uparrow}\left(\mathcal{E},X,[\psi_A]\right)=D_\beta^{\inf}\left(\Tr_{\widehat{Y}}\circ V,[\psi_A]\Vert \id_X \otimes \idmap_Z \right)\\
			=&\,\frac{1}{m}H_\alpha^{\uparrow}\left(\mathcal{E}^{\otimes m},X^m,[\psi_A^{\otimes m}]\right)=\frac{1}{m}D_\beta^{\inf}\left(\Tr^{\otimes m}_{\widehat{Y}}\circ V^{\otimes m},[\psi_A^{\otimes m}]\Vert \id^{\otimes m}_X\otimes \idmap_{Z^m} \right)\\
			=&\,H_\alpha^{\uparrow,\reg}\left(\mathcal{E},X,[\psi_A]\right)=D_\beta^{\inf,\reg}\left(\Tr_{\widehat{Y}}\circ V,[\psi_A]\Vert \id_X \otimes \idmap_Z \right),
	\end{align}
	where $\beta=\frac{\alpha}{2\alpha -1} \in [\frac{1}{2},1]$. In particular, this implies $D_\beta^{\inf,\reg}\left(\Tr_{\widehat{Y}}\circ V,[\psi_A]\Vert \id_X \otimes \idmap_Z \right)$ exists, in that the limit in its definition indeed converges.
	\begin{proof}
        The equality between the leftmost expressions in each line is just Lemma~\ref{lemma:channel iid additivity}.
		The middle equalities in the first two lines follow from Lemma~\ref{lemma: channel duality} (noting $V^{\otimes m}$ is a valid Stinespring dilation of $\mathcal{E}^{\otimes m}$),
		and the middle equality in the third line then follows by taking the $m\to\infty$ limit on both expressions in the second line.
	\end{proof}
\end{corollary}
Before demonstrating the superadditivity of {\Renyi} channel conditional entropy for composed channels, we will first present some preliminary results. We begin by establishing a similar superadditivity property for measured
{\Renyi} divergence. As the proof of this superadditivity property relies on Lemmas 22 and 24 found in \cite{FFF24}, we will restate their conclusions here for the sake of completeness.
\begin{fact}(Lemma~22 and~24 in~\cite{FFF24})
	\label{fact:set measured chain rule}
	Consider any sets $\mathbf{D}_1(A_1)\subseteq\dop{=}(A_1)$, $\mathbf{D}_2(A_2)\subseteq\dop{=}(A_2)$, and $\mathbf{D}_3(A_1A_2)\subseteq\dop{=}(A_1A_2)$ and also $\mathbf{D}'_1(A_1)\subseteq\Pos(A_1)$, $\mathbf{D}'_2(A_2)\subseteq\Pos(A_2)$, and $\mathbf{D}'_3(A_1A_2)\subseteq\Pos(A_1A_2)$. 
	Suppose furthermore that all those sets are convex and compact, and the following inclusions hold:
	\begin{align}
		\label{eq:inclusion polar set 1}
		&\Big\{\Gamma_{A_1}\otimes\Gamma_{A_2} \;\Big|\;
		\Gamma_{A_j} \in \Pos(A_j) \land
		\sup_{\nu\in\mathbf{D}_j(A_j)} \Tr[\Gamma_{A_j}\nu_{A_j}]\leq 1 \;\textnormal{ for } j\in\{1,2\}
		\Big\}\notag\\
		&\subseteq\Big\{\Gamma_{A_1A_2} \;\Big|\;
		\Gamma_{A_1A_2} \in \Pos(A_1A_2) \land
		\sup_{\nu\in\mathbf{D}_3(A_1A_2)} \Tr[\Gamma_{A_1A_2}\nu_{A_1A_2}]
		\leq 1 \Big\},
	\end{align}
	\begin{align}
		\label{eq:inclusion polar set 2}
		&\Big\{\Gamma_{A_1}\otimes\Gamma_{A_2} \;\Big|\; 
		\Gamma_{A_j} \in \Pos(A_j) \land
		\sup_{\rchi\in\mathbf{D}'_j(A_j)} \Tr[\Gamma_{A_j}\rchi_{A_j}]\leq 1
		\;\textnormal{ for } j\in\{1,2\} 
		\Big\}\notag\\
		&\subseteq\Big\{\Gamma_{A_1A_2} \;\Big|\;
		\Gamma_{A_1A_2} \in \Pos(A_1A_2) \land
		\sup_{\rchi\in\mathbf{D}'_3(A_1A_2)} \Tr[\Gamma_{A_1A_2}\rchi_{A_1A_2}]
		\leq 1  
		\Big\}.		
	\end{align}
	Then for any $\alpha\in[\frac{1}{2},1]$, we have the following:
	\begin{align}
		\label{eq:set measured chain rule}
		\inf_{\substack{\nu\in\mathbf{D}_3(A_1A_2)\\ \rchi\in\mathbf{D}'_3(A_1A_2)}}D_{\mathbb{M},\alpha}(\nu\Vert\rchi)\geq \inf_{\substack{\nu\in\mathbf{D}_1(A_1)\\ \rchi\in\mathbf{D}'_1(A_1)}}D_{\mathbb{M},\alpha}(\nu\Vert\rchi)
		+
		\inf_{\substack{\nu\in\mathbf{D}_2(A_2)\\ \rchi\in\mathbf{D}'_2(A_2)}}D_{\mathbb{M},\alpha}(\nu\Vert\rchi).
	\end{align}
\end{fact}

With this, we show the following:

\begin{lemma}\label{lemma:measured chain rule}
	Let $\mathcal{F}_1\in\CPTP(A_0Y_0,X_1Y_1Z_1)$, and $\mathcal{F}_2\in\CPTP(A_1Y_1,X_2Y_2Z_2)$ be quantum channels. Consider states $\psi_{A_0} \in \dop{=}(A_0)$ and $\phi_{A_1}\in\dop{=}(A_1)$.
	Then, for any $\alpha\in[\frac{1}{2},1]$, we have
	\begin{align}
		\label{eq:measured chain rule}
		D_{\mathbb{M},\alpha}^{\inf}&\left(\Tr_{Y_2}\circ\mathcal{F}_2\circ\mathcal{F}_1,[\psi_{A_0}\otimes\phi_{A_1}]\Vert \id_{X_1X_2} \otimes \idmap_{Z_1 Z_2} \right) \nonumber\\
		&\geq
		D_{\mathbb{M},\alpha}^{\inf}\left(\Tr_{Y_2}\circ\mathcal{F}_2,[\phi_{A_1}]\Vert \id_{X_2} \otimes \idmap_{Z_2} \right)
		+D_{\mathbb{M},\alpha}^{\inf}\left(\Tr_{Y_1}\circ\mathcal{F}_1,[\psi_{A_0}]\Vert \id_{X_1} \otimes \idmap_{Z_1} \right).
	\end{align}
\end{lemma}
\begin{proof}
	Consider Fact~\ref{fact:set measured chain rule} with the following choices of sets:
	let $\mathbf{D}_1(X_1Z_1),\allowbreak \mathbf{D}_2(X_2Z_2)$ and $\mathbf{D}_3(X_1Z_1X_2Z_2)$ be
	\begin{align}
		\label{eq:polar set 1}
		&\mathbf{D}_1(X_1Z_1)\defvar\left\{\Tr_{Y_1}\circ\mathcal{F}_1[\rho_{A_0Y_0}]|\ \rho_{A_0Y_0}\in\dop{=}(A_0Y_0)\land\rho_{A_0}=\psi_{A_0} \right\}\notag\\
		&\mathbf{D}_2(X_2Z_2)\defvar\left\{\Tr_{Y_2}\circ\mathcal{F}_2[\rho_{A_1Y_1}]|\ \rho_{A_1Y_1}\in\dop{=}(A_1Y_1)\land\rho_{A_1}=\phi_{A_1} \right\}\notag\\
		&\mathbf{D}_3(X_1Z_1X_2Z_2)\defvar\left\{\Tr_{Y_2}\circ\mathcal{F}_2\circ\mathcal{F}_1[\rho_{A_0A_1Y_0}]|\ \rho\in\dop{=}(A_0A_1Y_0)\land\rho_{A_0A_1}=\psi_{A_0}\otimes\phi_{A_1} \right\},
	\end{align}
	and furthermore, let $\mathbf{D}'_1(X_1Z_1), \mathbf{D}'_2(X_2Z_2)$ and $\mathbf{D}'_3(X_1Z_1X_2Z_2)$ be
\begin{align}
		\label{eq:polar set 2}
		&\mathbf{D}'_1(X_1 Z_1)\defvar\left\{\id_{X_1}\otimes\sigma_{Z_1}|\ \sigma_{Z_1}\in\dop{=}(Z_1)\right\}\notag\\
		&\mathbf{D}'_2(X_2Z_2)\defvar\left\{\id_{X_2}\otimes\sigma_{Z_2}|\ \sigma_{Z_2}\in\dop{=}(Z_2)\right\}\notag\\
		&\mathbf{D}'_3(X_1Z_1X_2Z_2)\defvar\left\{\id_{X_1X_2}\otimes\sigma_{Z_1Z_2}|\ \sigma_{Z_1Z_2}\in\dop{=}(Z_1Z_2)\right\}.
	\end{align}
	With this, the bound~\eqref{eq:set measured chain rule} in Fact~\ref{fact:set measured chain rule} is equivalent to our desired bound~\eqref{eq:measured chain rule}, so it suffices to show that the conditions in Fact~\ref{fact:set measured chain rule} are satisfied.\footnote{We find that some of the optimizations and SDPs in the following proof do not seem to have a very intuitive interpretation. It may be helpful to look at the proof of Corollary~\ref{cor:channel cond ent additivity} (strong additivity) below, where the SDPs seem to have a slightly more intuitive tensor-product structure.}
	
	First, we note that 
	the sets $\mathbf{D}_1(X_1Z_1), \mathbf{D}_2(X_2Z_2)$, and $\mathbf{D}_3(X_1Z_1X_2Z_2)$ are convex, bounded, and closed; therefore they are compact. Similarly, $\mathbf{D}'_1(X_1Z_1), \mathbf{D}'_2(X_2Z_2)$ and $\mathbf{D}'_3(X_1Z_1X_2Z_2)$ are also convex and compact.
	Therefore, it now suffices to show the inclusion relations in Eqs.~\eqref{eq:inclusion polar set 1} and~\eqref{eq:inclusion polar set 2}, for the sets defined in~\eqref{eq:polar set 1} and~\eqref{eq:polar set 2}, respectively. 
	
	We first consider Eq.~\eqref{eq:inclusion polar set 2}, as it is more straightforward. We need to show that for any $\Gamma_{X_1Z_1}\in\Pos(X_1Z_1)$ and $\Gamma_{X_2Z_2} \in \Pos(X_2Z_2)$ such that 
        \begin{align}
			\sup_{\sigma\in\dop{=}(Z_1)}
			\Tr\left[\Gamma_{X_1Z_1}\left(\id_{X_1}\otimes\sigma_{Z_1}\right)\right]\leq 1 \quad\text{and}\quad
			\sup_{\sigma\in\dop{=}(Z_2)}
			\Tr\left[\Gamma_{X_2Z_2}\left(\id_{X_2}\otimes\sigma_{Z_2}\right)\right]\leq 1,
		\end{align}
		we have
		\begin{align}
			\sup_{\sigma\in\dop{=}(Z_1Z_2)}&\Tr\left[\left(\Gamma_{X_1Z_1}\otimes\Gamma_{X_2Z_2}\right)\left(\id_{X_1X_2}\otimes\sigma_{Z_1Z_2}\right) \right]\leq 1.
		\end{align}
		The initial two conditions are equivalent to $\Gamma_{Z_1}\leq \id_{Z_1}$ and $\Gamma_{Z_2}\leq \id_{Z_2}$, respectively. Therefore we have
		\begin{align}
			&\sup_{\sigma\in\dop{=}(Z_1Z_2)}\Tr\left[\left(\Gamma_{X_1Z_1}\otimes\Gamma_{X_2Z_2}\right)\left(\id_{X_1X_2}\otimes\sigma_{Z_1Z_2}\right) \right]\notag\\
			&=\sup_{\sigma\in\dop{=}(Z_1Z_2)}\Tr\left[\left(\Gamma_{Z_1}\otimes\Gamma_{Z_2}\right)\left(\sigma_{Z_1Z_2}\right) \right]\notag\\
			&\leq \sup_{\sigma\in\dop{=}(Z_1Z_2)}\Tr\left[\sigma_{Z_1Z_2}\right]\notag\\
			&\leq 1,
		\end{align}
	which is the desired result.
	
	Let us now prove the inclusion in~\eqref{eq:inclusion polar set 1}.
	In other words, we need to show that for any $\Gamma_{X_1Z_1} \in \Pos(X_1Z_1)$ and $\Gamma_{X_2Z_2} \in \Pos(X_2Z_2)$ such that
	\begin{align}
	\sup_{\nu\in\mathbf{D}_1(X_1Z_1)} \Tr\left[\Gamma_{X_1Z_1}\nu_{X_1Z_1}\right]\leq 1 \quad\text{and}\quad \sup_{\nu\in\mathbf{D}_2(X_2Z_2)} \Tr\left[\Gamma_{X_2Z_2}\nu_{X_2Z_2}\right]\leq 1,
	\end{align} 
	we have
	\begin{align}
		\label{eq:joint result}
		\sup_{\nu\in\mathbf{D}_3(X_1Z_1X_2Z_2)} \Tr\left[\left(\Gamma_{X_1Z_1}\otimes\Gamma_{X_2Z_2}\right)\nu_{X_1Z_1X_2Z_2}\right]\leq 1 .
	\end{align}
	
	The initial two conditions may be equivalently expressed as upper bounding the subsequent SDPs by 1 (note that since $\psi_{A_0} , \phi_{A_1}$ are normalized states by definition, we do not need to separately impose a normalization constraint):
\begin{multicols}{2}
	\noindent
	\begin{align}\label{eq:SDP individual 1}
		\sup_{\rho\in\Pos(A_0Y_0)}
		&\Tr\left[\rho_{A_0Y_0}\mathcal{F}_1^\dagger\circ\Tr_{Y_1}^\dagger[\Gamma_{X_1Z_1}]\right] \notag\\
		&\qquad\qquad \suchthat
		\notag\\
		&\quad\qquad\rho_{A_0}=\psi_{A_0} 
	\end{align}
	\begin{align}\label{eq:SDP individual 2}
		\sup_{\rho\in\Pos(A_1Y_1)}&\Tr\left[\rho_{A_1Y_1}\mathcal{F}_2^\dagger\circ\Tr_{Y_2}^\dagger[\Gamma_{X_2Z_2}]\right] \notag,\\
		&\qquad\qquad\suchthat\notag\\
		&\qquad\quad\rho_{A_1}=\phi_{A_1}.
	\end{align}
\end{multicols}
\noindent Additionally, Eq.~\eqref{eq:joint result} can be equivalently expressed as upper bounding the following SDP by 1:
\begin{align}\label{eq: SDP joint}
	\sup_{\rho\in\Pos(A_0A_1Y_0)}&\Tr\left[\rho_{A_0A_1Y_0}\mathcal{F}_1^\dagger\circ\mathcal{F}_2^\dagger\circ\Tr_{Y_2}^\dagger\left[\Gamma_{X_0Z_0}\otimes\Gamma_{X_1Z_1}\right]\right]\notag\\
	&\qquad\qquad\qquad\suchthat\notag\\
	&\quad\qquad\rho_{A_0A_1}=\psi_{A_0}\otimes\phi_{A_1}.
\end{align}
Without loss of generality, we shall suppose that $\psi_{A_0}$ and $\phi_{A_1}$ have full support.\footnote{In case $\psi_{A_0}$ or $\phi_{A_1}$ do not have full support, we can instead replace all of these optimizations with restricted versions where the Hilbert spaces on $A_0$ and $A_1$ are replaced by $\supp(\psi_{A_0})$ and $\supp(\phi_{A_1})$ (including when they appear in the domains of the channels $\mathcal{F}_1,\mathcal{F}_2$). Observe that this restriction does not change the optimal values since any feasible point in the original space corresponds to a feasible solution in the reduced space, and vice versa.}

The dual problems of the SDPs in Eq.~\eqref{eq:SDP individual 1}, and \eqref{eq:SDP individual 2} are as follows:
	\begin{multicols}{2}
		\noindent
		\begin{align}\label{eq:SDP dual individual 1}
			&\inf_{\Lambda\in\operatorname{Herm}(A_0)}\Tr\left[\psi_{A_0}\Lambda_{A_0}\right] \notag\\
			&\qquad\qquad\qquad\suchthat\notag\\
			&\Lambda_{A_0}\otimes\id_{Y_0}\geq\mathcal{F}_1^\dagger[\Gamma_{X_1Z_1}\otimes\id_{Y_1}] 
		\end{align}
		\begin{align}\label{eq:SDP dual individual 2}
			&\inf_{\Lambda\in\operatorname{Herm}(A_1)}\Tr\left[\phi_{A_1}\Lambda_{A_1}\right] \notag\\
			&\qquad\qquad\qquad\suchthat \notag\\
			&\Lambda_{A_1}\otimes\id_{Y_1}\geq\mathcal{F}_2^\dagger[\Gamma_{X_2Z_2}\otimes\id_{Y_2}],
		\end{align}
	\end{multicols}
	\noindent and the dual problem to the SDP in Eq.~\eqref{eq: SDP joint} is given by
	\begin{align}
		\label{eq:SDP dual joint}
		&\inf_{\Lambda\in\operatorname{Herm}(A_0A_1)}\Tr\left[\left(\psi_{A_0}\otimes\phi_{A_1}\right)\Lambda_{A_0A_1}\right]\qquad\qquad\qquad\qquad\notag\\
		&\qquad\qquad\qquad\suchthat\notag\\
		&\Lambda_{A_0A_1}\otimes\id_{Y_0}\geq\mathcal{F}_1^\dagger\circ\mathcal{F}_2^\dagger\left[\Gamma_{X_1Z_1}\otimes\Gamma_{X_2Z_2}\otimes\id_{Y_2}\right].
	\end{align}
    Note that since $\Gamma_{X_1Z_1}\geq 0$, $\Gamma_{X_2Z_2}\geq 0$, and the adjoint of a CP map is also CP, the constraints in Eqs.~\eqref{eq:SDP dual individual 1},~\eqref{eq:SDP dual individual 2}, and~\eqref{eq:SDP dual joint} imply that $\Lambda_{A_0}\geq 0$, $\Lambda_{A_1}\geq 0$, and $\Lambda_{A_0A_1}\geq 0$. Hence without loss of generality, the optimizations in Eqs.~\eqref{eq:SDP dual individual 1},~\eqref{eq:SDP dual individual 2}, and~\eqref{eq:SDP dual joint} can be restricted to $\Lambda\in\Pos(A_0)$, $\Lambda\in\Pos(A_1)$, and $\Lambda\in\Pos(A_0A_1)$, respectively.
    
	For these SDPs, 
	strong duality holds by the Slater condition~\cite{BV04v8}, recalling we supposed that $\psi_{A_0}$ and $\phi_{A_1}$ have full support.  Specifically, $\rho_{A_0Y_0}=\psi_{A_0}\otimes\idnorm_{Y_0}$, $\rho_{A_1Y_1}=\phi_{A_1}\otimes\idnorm_{Y_1}$, and $\rho_{A_0A_1Y_0}=\psi_{A_0}\otimes\phi_{A_1}\otimes\idnorm_{Y_0}$ are strict feasible points for the SDPs in Eqs.~\eqref{eq:SDP individual 1}, \eqref{eq:SDP individual 2}, and~\eqref{eq: SDP joint}, respectively. Moreover, $\Lambda_{A_0}=c_0\id_{A_0}$, $\Lambda_{A_1}=c_1\id_{A_1}$, and $\Lambda_{A_0A_1}=c_{01}\id_{A_0A_1}$, for large enough values of $c_0,\ c_1,\ c_{01}$, are strict feasible points for the SDPs in Eqs.~\eqref{eq:SDP dual individual 1}, \eqref{eq:SDP dual individual 2}, and \eqref{eq:SDP dual joint}.
 Therefore, we see
 the Slater condition is satisfied, and we have strong duality.
	
	Since strong duality holds, we instead analyze the dual problems.
	Observe that it
	suffices to show that the tensor product of 
	any feasible points of
	the SDPs in Eq.~\eqref{eq:SDP dual individual 1} 
	and \eqref{eq:SDP dual individual 2} is a feasible point for the SDP in Eq.~\eqref{eq:SDP dual joint}.\footnote{Explicitly: by fundamental infimum properties, each of the optimizations in Eq.~\eqref{eq:SDP dual individual 1}--\eqref{eq:SDP dual individual 2} has a sequence of feasible points converging to the optimal values, which are at most $1$ by hypothesis. If the tensor products of such feasible points are always feasible in the Eq.~\eqref{eq:SDP dual joint} optimization, this yields a sequence of feasible points in the latter that converge to a value of at most $1$, as desired. (Alternatively, one can use the fact that by the Slater condition, the optimal values in Eq.~\eqref{eq:SDP dual individual 1}--\eqref{eq:SDP dual individual 2} are indeed attained, and apply the subsequent arguments specifically to some \emph{optimal} solutions for those optimizations.)} Denote some arbitrary feasible points of the SDPs in~\eqref{eq:SDP dual individual 1} and~\eqref{eq:SDP dual individual 2} by $\Lambda^*_{A_0}$ and $\Lambda^*_{A_1}$, respectively. We have 
\begin{align}\label{eq:feasability}
	&\Lambda_{A_1}^* \otimes \id_{Y_1} 
	\geq \mathcal{F}_2^\dagger[\Gamma_{X_2Z_2} \otimes \id_{Y_2}] \notag\\
	\iff\ 
	&\Gamma_{X_1Z_1}\otimes\Lambda_{A_1}^*\otimes\id_{Y_1}
	\geq
	\Gamma_{X_1Z_1}\otimes\mathcal{F}_2^\dagger[\Gamma_{X_2Z_2}\otimes\id_{Y_2}]\notag\\
	\implies\ 
	&
	\mathcal{F}_1^\dagger\left[\Gamma_{X_1Z_1}\otimes\id_{Y_1}\right]\otimes\Lambda_{A_1}^*
	\geq
	\mathcal{F}_1^\dagger\left[\Gamma_{X_1Z_1}\otimes\mathcal{F}_2^\dagger[\Gamma_{X_2Z_2}\otimes\id_{Y_2}]\right]\notag\\
	\implies\ 
	&\Lambda_{A_0}^*\otimes\Lambda_{A_1}^*\otimes\id_{Y_0}
	\geq
	\mathcal{F}_1^\dagger\left[\Gamma_{X_1Z_1}\otimes\mathcal{F}_2^\dagger[\Gamma_{X_2Z_2}\otimes\id_{Y_2}]\right],		
\end{align}
where the first line is the constraint of the SDP in Eq.~\eqref{eq:SDP dual individual 2}, the second step holds since $\Gamma_{X_1Z_1} \geq 0$,
the third line holds because the adjoint of a CP map is also CP,
and the last line follows from applying the constraint in Eq.~\eqref{eq:SDP dual individual 1} to the LHS of the penultimate line, and noting that we have
$\Lambda_{A_1}^*\geq 0$ as mentioned above.
The desired claim then follows by noting that the RHS of the last line of Eq.~\eqref{eq:feasability} is equal to the RHS of the constraint in Eq.~\eqref{eq:SDP dual joint}.
\end{proof}

Next, we show that superaddivity holds for the regularization of minimized {\Renyi} divergence.
\begin{corollary}\label{cor:reg renyi chain rule}
	For any channels $\mathcal{F}_1\in\CPTP(A_0Y_0,X_1Y_1Z_1)$, and $\mathcal{F}_2\in\CPTP(A_1Y_1,X_2Y_2Z_2)$, let $\psi_{A_0} \in \dop{=}(A_0)$ and $\phi_{A_1}\in\dop{=}(A_1)$ be states. Then, for any $\alpha\in[\frac{1}{2},1]$, we have
	\begin{align}
		\label{eq:reg renyi chain rule}
		D_{\alpha}^{\inf,\reg}&\left(\Tr_{Y_2}\circ\mathcal{F}_2\circ\mathcal{F}_1,[\psi_{A_0}\otimes\phi_{A_1}]\Vert \id_{X_1X_2} \otimes \idmap_{Z_1 Z_2} \right) \nonumber\\
		&\geq
		D_{\alpha}^{\inf,\reg}\left(\Tr_{Y_2}\circ\mathcal{F}_2,[\phi_{A_1}]\Vert \id_{X_2} \otimes \idmap_{Z_2} \right)
		+D_{\alpha}^{\inf,\reg}\left(\Tr_{Y_1}\circ\mathcal{F}_1,[\psi_{A_0}]\Vert \id_{X_1} \otimes \idmap_{Z_1} \right).
	\end{align}
	\begin{proof}
		Applying Lemma~\ref{lemma:measured chain rule} to $\mathcal{F}_1^{\otimes m}\in\allowbreak\CPTP(A_0^mY_0^m,X_1^mY_1^mZ_1^m)$, and\newline $\mathcal{F}_2^{\otimes m}\in\allowbreak\CPTP(A_1^mY_1^m,X_2^mY_2^mZ_2^m)$, and dividing both sides by $m$, we have
		\begin{align}
			\label{eq:reg measured chain rule}
				\frac{1}{m}D_{\mathbb{M},\alpha}^{\inf}&\left(\Tr_{Y_2}^{\otimes m}\circ\mathcal{F}_2^{\otimes m}\circ\mathcal{F}_1^{\otimes m},[\psi_{A_0}^{\otimes m}\otimes\phi_{A_1}^{\otimes m}]
				\Vert\id_{X_1X_2}^{\otimes m} \otimes \idmap_{Z_1 Z_2}^{\otimes m}\right) \notag\\
			&\geq
			\frac{1}{m}D_{\mathbb{M},\alpha}^{\inf}\left(\Tr_{Y_2}^{\otimes m}\circ\mathcal{F}_2^{\otimes m},[\phi_{A_1}^{\otimes m}]\Vert\id_{X_2}^{\otimes m} \otimes \idmap_{Z_2}^{\otimes m}\right)
			\notag\\
			&\quad+\frac{1}{m}D_{\mathbb{M},\alpha}^{\inf}\left(\Tr_{Y_1}^{\otimes m}\circ\mathcal{F}_1^{\otimes m},[\psi_{A_0}^{\otimes m}]\Vert \id_{X_1}^{\otimes m} \otimes \idmap_{Z_1}^{\otimes m}\right).
		\end{align}
		The proof follows by taking the limit as $m\rightarrow\infty$ of the above relation, and using the result of Lemma~29, Eq.~(96) in~\cite{FFF24}, 
		noting that the sets of states being optimized over in these terms indeed satisfy the required conditions in that lemma --- namely, some convexity, compactness and permutation-invariance conditions, along with $O(m)$ max-divergence bounds such as
		\begin{align}
		D_{\infty}^{\inf}\left(\Tr_{Y_1}^{\otimes m}\circ\mathcal{F}_1^{\otimes m},[\psi_{A_0}^{\otimes m}]\Vert \id_{X_1}^{\otimes m} \otimes \idmap_{Z_1}^{\otimes m} \right)
		&\leq 
		D_{\infty}\left( \rho^*_{X_1^m Z_1^m} \Vert\id_{X_1^m}\otimes \rho^*_{Z_1^m} \right) \nonumber\\
		&=-H_\infty(X_1^m|Z_1^m)_{\rho^*}\nonumber\\
		&\leq m \log\dim(X) \quad\text{(see e.g.~\cite[Lemma~5.2]{Tom16})}
		,
		\end{align}
		where $\rho^*$ is just any state obtained by picking some arbitrary feasible point in the first argument of the $D_{\infty}^{\inf}$ optimization,
		and similarly for the other two terms.
	\end{proof}
\end{corollary}
We can now demonstrate our main result in this section, the superadditivity of {\Renyi} channel conditional entropy under channel composition.
\begin{theorem}\label{thrm:channel cond ent chain rule}
	Let $\mathcal{E}_1\in\CPTP(A_0Y_0,X_1Y_1)$, and $\mathcal{E}_2\in\CPTP(A_1Y_1,X_2Y_2)$ be quantum channels. Furthermore, consider states $\psi_{A_0} \in \dop{=}(A_0)$ and $\phi_{A_1}\in\dop{=}(A_1)$. Then, for any $\alpha\in [1,\infty]$ we have
	\begin{align}
		\label{eq:channel cond ent chain rule}
		H_\alpha^{\uparrow}\left(\mathcal{E}_2\circ\mathcal{E}_1,X_1X_2,[\psi_{A_0}\otimes\phi_{A_1}]\right)\geq
		H_\alpha^{\uparrow}\left(\mathcal{E}_2,X_2,[\phi_{A_1}]\right)
		+H_\alpha^{\uparrow}\left(\mathcal{E}_1,X_1,[\psi_{A_0}]\right).
	\end{align}
	\begin{proof}
		Let $V_1\in\CPTP(A_0Y_0,X_1Y_1Z_1)$, and $V_2\in\CPTP(A_1Y_1,X_2Y_2Z_2)$ be any Stinespring dilations of $\mathcal{E}_1$, and $\mathcal{E}_2$, respectively. Note that $V_2\circ V_1\in\CPTP(A_0A_1Y_0,X_1X_2Y_2Z_1Z_2)$ is a valid Stinespring dilation for $\mathcal{E}_2\circ\mathcal{E}_1$. Therefore, 
		\begin{align}
				H_\alpha^{\uparrow}&\left(\mathcal{E}_2\circ\mathcal{E}_1,X_1X_2,[\psi_{A_0}\otimes\phi_{A_1}]\right)\notag\\
				&=D_{\beta}^{\inf,\reg}\left(\Tr_{Y_2}\circ V_2\circ V_1,[\psi_{A_0}\otimes\phi_{A_1}]\Vert \id_{X_1X_2} \otimes \idmap_{Z_1Z_2} \right)\notag\\
				&\geq
				D_{\beta}^{\inf,\reg}\left(\Tr_{Y_2}\circ V_2,[\phi_{A_1}]\Vert\id_{X_2} \otimes \idmap_{Z_2} \right)
				+D_{\beta}^{\inf,\reg}\left(\Tr_{Y_1}\circ V_1,[\psi_{A_0}]\Vert\id_{X_1} \otimes \idmap_{Z_1} \right)\notag\\
				&=H_\alpha^{\uparrow}\left(\mathcal{E}_2,X_2,[\phi_{A_1}]\right)
				+H_\alpha^{\uparrow}\left(\mathcal{E}_1,X_1,[\psi_{A_0}]\right),
		\end{align}  
		where the second and fourth lines follow from Corollary~\ref{cor:equivalence channel ent and min div}, and the third line is the result of Corollary~\ref{cor:reg renyi chain rule}.
	\end{proof}
\end{theorem}
As a corollary, we can now prove the strong additivity of {\Renyi} conditional entropy under tensor products.
\begin{corollary}\label{cor:channel cond ent additivity} (Strong additivity)
	Let $\mathcal{E}_1\in\CPTP(A_0B_0,X_0Y_0)$, and $\mathcal{E}_2\in\CPTP(A_1B_1,X_1Y_1)$ be quantum channels, and $\psi_{A_0} \in \dop{=}(A_0)$, and $\phi_{A_1}\in\dop{=}(A_1)$ be states. Then, for any $\alpha\in [1,\infty]$ we have
	\begin{align}
		\label{eq:channel cond ent additivity}
		H_\alpha^\uparrow\left(\mathcal{E}_1\otimes\mathcal{E}_2,X_0X_1,[\psi_{A_0}\otimes\phi_{A_1}]\right)=H_\alpha^\uparrow\left(\mathcal{E}_1,X_0,[\psi_{A_0}]\right)+H_\alpha^\uparrow\left(\mathcal{E}_2,X_1,[\phi_{A_1}]\right).
	\end{align}
	\begin{proof}
		The $\leq$ direction holds by simply noting that the LHS contains an infimum over inputs, as defined in Eq.~\eqref{eq:channel conditional entropy}, and hence restricting the infimum to the tensor product of the optimal inputs for the individual channels yields an upper bound, with value $H_\alpha^\uparrow\left(\mathcal{E}_1,X_0,[\psi_{A_0}]\right)+H_\alpha^\uparrow\left(\mathcal{E}_2,X_1,[\phi_{A_1}]\right)$.
        
        For the $\geq$ direction, it can be obtained as a corollary of Theorem~\ref{thrm:channel cond ent chain rule} by simply taking a suitable choice of channels in that theorem (basically the same as in the Corollary~\ref{cor:tensorcase} proof we present later). 
		However, we find it may be instructive to see how an argument analogous to the Theorem~\ref{thrm:channel cond ent chain rule} proof would proceed in this scenario (to prove the $\geq$ direction), and hence we present this in some detail here. Let $V_1\in\CPTP(A_0B_0,X_0Y_0Z_0)$, and $V_2\in\CPTP(A_1B_1,X_1Y_1Z_1)$ be any Stinespring dilations of $\mathcal{E}_1$ and $\mathcal{E}_2$, respectively. Then we only need to prove the following holds:
		\begin{align}\label{eq:measured additivity}
			D_{\mathbb{M},\alpha}^{\inf}&\left(\Tr_{Y_0Y_1}\circ\left(V_2\otimes V_1\right),[\psi_{A_0}\otimes\phi_{A_1}]\Vert\id_{X_0X_1}\otimes\idmap_{Z_0Z_1}\right) \nonumber\\
			&\geq
			D_{\mathbb{M},\alpha}^{\inf}\left(\Tr_{Y_1}\circ V_2,[\phi_{A_1}]\Vert \id_{X_1}\otimes\idmap_{Z_1} \right)
			+D_{\mathbb{M},\alpha}^{\inf}\left(\Tr_{Y_0}\circ V_1,[\psi_{A_0}]\Vert \id_{X_0}\otimes\idmap_{Z_0} \right).
		\end{align}
		The reason for this is that	once we prove
		Eq.~\eqref{eq:measured additivity}, we can derive an analogous result to Corollary~\ref{cor:reg renyi chain rule}. Then by employing an argument similar to that in Theorem~\ref{thrm:channel cond ent chain rule} the claim is proven. Therefore, we focus only on proving Eq.~\eqref{eq:measured additivity}. 

		We utilize an argument similar to that of Lemma~\ref{lemma:measured chain rule}. First note that the proof of the inclusion relation in Eq.~\eqref{eq:inclusion polar set 2} is identical to that of Lemma~\ref{lemma:measured chain rule}. 
		Then, we only need to prove the inclusion relation in Eq.~\eqref{eq:inclusion polar set 1}. Let us suppose that the following SDPs are upper bounded by 1:
			\begin{multicols}{2}
			\noindent
			\begin{align}\label{eq:SDP Additivity_individual 1}
				\sup_{\rho\in\Pos(A_0B_0)}&\Tr\left[\rho_{A_0B_0}V_1^\dagger\circ\Tr_{Y_0}^\dagger[\Gamma_{X_0Z_0}]\right]\notag\\
				&\qquad\qquad\suchthat \notag\\
				&\quad\qquad\rho_{A_0}=\psi_{A_0} 
			\end{align}
			\begin{align}\label{eq:SDP Additivity_individual 2}
				\sup_{\rho\in\Pos(A_1B_1)}&\Tr\left[\rho_{A_1B_1}V_2^\dagger\circ\Tr_{Y_1}^\dagger[\Gamma_{X_1Z_1}]\right] \notag,\\
				&\qquad\qquad\suchthat\notag\\
				&\qquad\quad\rho_{A_1}=\phi_{A_1}.
			\end{align}
		\end{multicols}
		\noindent Then, we only need to show the SDP
		\begin{align}\label{eq:SDP Additivity_joint}
		\sup_{\rho\in\Pos(A_0A_1B_0B_1)}&\Tr\left[\rho_{A_0A_1B_0B_1}V_1^\dagger\otimes V_2^\dagger\circ\Tr_{Y_0Y_1}^\dagger\left[\Gamma_{X_0Z_0}\otimes\Gamma_{X_1Z_1}\right]\right]\notag\\
		&\qquad\qquad\qquad\suchthat\notag\\
		&\qquad\quad\rho_{A_0A_1}=\psi_{A_0}\otimes\phi_{A_1}.
		\end{align}
		is bounded from above by 1.
	Following a similar reasoning as in Lemma~\ref{lemma:measured chain rule}, where 
	strong duality is established, we will focus on analyzing the dual problems. The duals corresponding to the SDPs in~\eqref{eq:SDP Additivity_individual 1} and~\eqref{eq:SDP Additivity_individual 2} are as follows:
		\begin{multicols}{2}
		\noindent
		\begin{align}\label{eq:SDP dual_Additivity_individual 1}
			&\inf_{\Lambda\in\operatorname{Herm}(A_0)}\Tr\left[\psi_{A_0}\Lambda_{A_0}\right] \notag\\
			&\qquad\qquad\qquad\suchthat \notag\\
			&\Lambda_{A_0}\otimes\id_{B_0}\geq V_1^\dagger[\Gamma_{X_0Z_0}\otimes\id_{Y_0}]
		\end{align}
		\begin{align}\label{eq:SDP dual_Additivity_individual 2}
			&\inf_{\Lambda\in\operatorname{Herm}(A_1)}\Tr\left[\phi_{A_1}\Lambda_{A_1}\right] \notag\\
			&\qquad\qquad\qquad\suchthat \notag\\
			&\Lambda_{A_1}\otimes\id_{B_1}\geq\! V_2^\dagger[\Gamma_{X_1Z_1}\otimes\id_{Y_1}],
		\end{align}
	\end{multicols}
	\noindent and the dual problem to the SDP in~\eqref{eq:SDP Additivity_joint} is
	\begin{align}
		\label{eq:SDP dual_Additivity_joint}
		&\inf_{\Lambda\in\operatorname{Herm}(A_0A_1)}\Tr\left[\left(\psi_{A_0}\otimes\phi_{A_1}\right)\Lambda_{A_0A_1}\right]\qquad\qquad\qquad\qquad\notag\\
		&\qquad\qquad\qquad\suchthat\notag\\
		&\Lambda_{A_0A_1}\otimes\id_{B_0B_1}\geq V_1^\dagger\left[\Gamma_{X_0Z_0}\otimes\id_{Y_0}\right]\otimes V_2^\dagger\left[\Gamma_{X_1Z_1}\otimes\id_{Y_1}\right].
	\end{align}
    Note that similar to the reasoning in Lemma~\ref{lemma:measured chain rule}, since $\Gamma_{X_0Z_0}\geq 0$, $\Gamma_{X_1Z_1}\geq 0$, and the adjoint of a CP map is also CP, the constraints in Eqs.~\eqref{eq:SDP dual_Additivity_individual 1},~\eqref{eq:SDP dual_Additivity_individual 2}, and~\eqref{eq:SDP dual_Additivity_joint} imply that $\Lambda_{A_0}\geq 0$, $\Lambda_{A_1}\geq 0$, and $\Lambda_{A_0A_1}\geq 0$. Hence without loss of generality, the optimization in Eqs.~\eqref{eq:SDP dual_Additivity_individual 1},~\eqref{eq:SDP dual_Additivity_individual 2}, and~\eqref{eq:SDP dual_Additivity_joint} can be restricted to $\Lambda\in\Pos(A_0)$, $\Lambda\in\Pos(A_1)$, and $\Lambda\in\Pos(A_0A_1)$, respectively.
    Also, we can again show the problems are strictly feasible and hence strong duality holds.
    
	Let $\Lambda^*_{A_0}$, and $\Lambda^*_{A_1}$ be arbitrary feasible points of the SDPs in~\eqref{eq:SDP dual_Additivity_individual 1} and~\eqref{eq:SDP dual_Additivity_individual 2}, respectively. 
	We can show that the product $\Lambda^*_{A_0}\otimes\Lambda^*_{A_1}$ is a feasible solution for the SDP given in~\eqref{eq:SDP dual_Additivity_joint}, as follows:
	\begin{align}
	&\Lambda_{A_0}^* \otimes \id_{B_0} 
	\geq V_1^\dagger[\Gamma_{X_0Z_0} \otimes \id_{Y_0}] \notag\\
	\implies\ 
	&\Lambda_{A_1}^*\otimes\id_{B_1}\otimes\Lambda_{A_0}^* \otimes \id_{B_0}
	\geq
	\Lambda_{A_1}^*\otimes\id_{B_1}\otimes  V_1^\dagger[\Gamma_{X_0Z_0} \otimes \id_{Y_0}]\notag\\
	\implies\ 
	&
	\Lambda_{A_0}^*\otimes\Lambda_{A_1}^*
	\otimes\id_{B_0B_1}\geq V_1^\dagger\left[\Gamma_{X_0Z_0}\otimes\id_{Y_0}\right]\otimes V_2^\dagger\left[\Gamma_{X_1Z_1}\otimes\id_{Y_1}\right],		
	\end{align}
	where the first line is the constraint of the SDP in Eq.~\eqref{eq:SDP dual_Additivity_individual 1}, the second line holds by noting that  $\Lambda_{A_1}^*\otimes\id_{B_1}\geq 0$ as mentioned above, 
	and last line follows from the constraint in Eq.~\eqref{eq:SDP dual_Additivity_individual 2} along with the fact that $\Gamma_{X_0Z_0} \geq 0$ and the adjoint of a CP map is also a CP map.
	Hence the relation in~\eqref{eq:measured additivity} holds; thus, the claim is proven. 
\end{proof}
\end{corollary}
\section{Marginal-constrained entropy accumulation theorem (MEAT)}
\label{sec:MEAT}
In this section, we will employ the chain rules that were established in the preceding section to derive an entropy accumulation bound that handles a specific form of constraint on the input state. 

\subsection{\texorpdfstring{$f$-weighted entropies}{f-weighted entropies}}
\label{subsec:f_entropy}

We begin by extending the result in Eq.~\eqref{eq:classmixHup} of Fact~\ref{fact:classmix} to the case where we have classical registers on both sides of the conditioning.
\begin{lemma}
	\label{lemma:twoclassmixHup}
	Let $\rho\in\dop{=}(\CS\CP QQ')$ be a state classical on $\CS\CP$. Then for any $\alpha\in (1,\infty]$, we have
	\begin{align}
		\label{eq:twoclassmixHup}
		&H_\alpha^{\uparrow}(Q\CS|\CP Q')_\rho \nonumber\\
		&=\frac{\alpha}{1-\alpha} \log \left( \sum_{\cP\in \supp(\bsym{\rho}_{\CP})}
		\rho(\cP)
        \inf\limits_{\sigma^{(\cP)}\in\dop{=}(Q')}
		\left[\sum_{\cS\in\supp(\bsym{\rho}_{\CS|\cP})} \rho(\cS|\cP)^\alpha\, 2^{(\alpha-1)  D_\alpha(\rho_{QQ'|\cS\cP}\Vert\id_Q\otimes\sigma_{Q'}^{(\cP)})} \right]^{\frac{1}{\alpha}} \right),
	\end{align}
	where $\bsym{\rho}_{\CS|\cP}=\left\{\rho\left(\cS|\cP\right)\big|\ \cS \in \alphCS
	\right\}$, and the $\alpha=\infty$ case is to be understood by taking the $\alpha\to\infty$ limit on both sides.\footnote{Some care is needed in taking the limit on the right-hand-side, as it 
    may contain expressions of the form $F(\alpha)^{\frac{1}{\alpha}}$ where $\lim_{\alpha\to\infty}F(\alpha)=0$; 
    however, we shall argue that the limit indeed exists.} 
    For $\alpha\in[\frac{1}{2},1)$, the above relation holds with the infima replaced by suprema. 
\end{lemma}

	\begin{proof}
    We focus on the $\alpha \in (1,\infty)$ regime, noting that we can take the $\alpha\to\infty$ limit at the end to obtain the result for $\alpha=\infty$ (recalling that the LHS of Eq.~\eqref{eq:twoclassmixHup} for $\alpha=\infty$ is given by that limit by definition, and hence if the equality holds for all $\alpha<\infty$, then it holds in the limit; in particular the limit on the RHS exists because the limit on the LHS exists). Moreover, given any such $\alpha$, it suffices to prove the equality for the case where $\rho$ has full support on $\CS\CP$, since it can then be extended to arbitrary $\rho$ by a continuity argument.\footnote{Essentially: given any $\rho_{Q\CS\CP Q'}$, we can construct a sequence of full-support classical states $\rho^{(n)}_{\CS\CP}$ converging to $\rho_{\CS\CP}$, and extend them to states on $Q\CS\CP Q'$ by setting $\rho^{(n)}_{QQ'|\cS\cP} = \rho_{QQ'|\cS\cP}$ (recall we define these as some arbitrary states if $\rho(\cS\cP)=0$). Given the premise that Eq.~\eqref{eq:twoclassmixHup} holds for all these $\rho^{(n)}$ (since they are full-support), it suffices to show that both sides of that equality converge to the corresponding expressions for $\rho$ as $n\to\infty$. For the LHS, observe we have $\lim_{n\to\infty} d\left(\rho^{(n)}_{Q\CS\CP Q'} , \rho_{Q\CS\CP Q'}\right) = 0$ and therefore $\lim_{n\to\infty} H_\alpha^{\uparrow}(Q\CS|\CP Q')_{\rho^{(n)}} = H_\alpha^{\uparrow}(Q\CS|\CP Q')_{\rho}$ by continuity of $H_\alpha^{\uparrow}$ (see e.g.~\cite{MD22,arx_BCG+24} or related works). For the RHS, first recall we chose $\rho^{(n)}_{QQ'|\cS\cP} = \rho_{QQ'|\cS\cP}$ for all $n$ and thus the $D_\alpha(\rho^{(n)}_{QQ'|\cS\cP}\Vert\id_Q\otimes\sigma_{Q'}^{(\cP)})$ terms are independent of $n$, and moreover for each $\cP$ the infimum over $\sigma_{Q'}^{(\cP)}$ can be restricted to the set $S'_{\cP}$ of all states $\sigma_{Q'}^{(\cP)}$ such that $\ker(\sigma_{Q'}^{(\cP)}) \supseteq \bigcap_{\cS} \ker(\rho_{Q'|\cS\cP})$ (since otherwise at least one divergence term would be infinite, given $\alpha>1$). With this, the expression on the RHS can be shown to converge in the desired manner as $\rho^{(n)}$ converges to $\rho$, as it is just a continuous function of the distribution on $\CS\CP$ (using the fact that the infima over $\sigma_{Q'}^{(\cP)}$ preserve continuity as they are taken over the compact sets $S'_{\cP}$). Note that for the $\alpha<1$ case later, this restriction to $S'_{\cP}$ would not be relevant since the conditions for $D_\alpha$ to be infinite are different in that regime.} With these restrictions in mind, we apply analogous techniques to those utilized in the proof of~\cite[Proposition~9]{MDS+13}. First we note that by definition we have
    \begin{align}\label{eq:twoclassmixHup_proof1}
            H_\alpha^{\uparrow}(Q\CS|\CP Q')_\rho&=\sup_{\sigma_{\CP Q'}\in\dop{=}(\CP Q')}-D_\alpha\left(\rho_{QQ'\CS\CP}\lVert\id_{Q\CS}\otimes\sigma_{\CP Q'}\right)\nonumber\\
            &=\log(d_{\CS}d_{Q})+\sup_{\sigma_{\CP Q'}\in\dop{=}(\CP Q')}-D_\alpha\left(\rho_{QQ'\CS\CP}\lVert\idnorm_{Q\CS}\otimes\sigma_{\CP Q'}\right),
        \end{align}
        where $\idnorm_{Q\CS}$ is the maximally mixed state on $Q\CS$.
        Then, by the data-processing property in Fact~\ref{fact:DPI}, we can restrict the optimization to be over only $\sigma$ classical on $\CP$ (by applying a pinching channel on $\CP$ in both $\rho$ and $\sigma$, which does not change $\rho$), i.e., $\sigma_{\CP Q'}=\sum_{\cP} \sigma(\cP)\ketbra{\cP}{\cP}\otimes\sigma_{Q'}^{(\cP)}$ for some normalized states $\sigma_{Q'}^{(\cP)}$.\footnote{Here we choose to denote these states as normalized states $\sigma_{Q'}^{(\cP)}$ indexed by $\cP$, rather than using the conditional-state notation $\sigma_{Q'|\cP}$ --- this is because in the rest of our analysis, they will only form ``dummy variables'' in the optimizations that may not always have straightforward interpretations as genuine conditional states.} Thus, from Eq.~\eqref{eq:classmixD} we have 
        \begin{align}\label{eq:twoclassmixHup_proof2}
            &H_\alpha^{\uparrow}(Q\CS|\CP Q')_\rho \nonumber\\
            &=\log(d_{\CS}d_{Q})+\sup_{\{\sigma(\cP)\}, \{\sigma_{Q'}^{(\cP)}\}}-\frac{1}{\alpha-1}\log\left(\sum_{\cS\cP}\rho(\cS\cP)^\alpha\left(\frac{\sigma(\cP)}{d_{\CS}}\right)^{1-\alpha}2^{(\alpha-1)D_\alpha(\rho_{QQ'|\cS\cP}\Vert\idnorm_Q\otimes\sigma_{Q'}^{(\cP)})}\right)\nonumber\\
            &=\log(d_{Q})+\sup_{\{\sigma(\cP)\}, \{\sigma_{Q'}^{(\cP)}\}}-\frac{1}{\alpha-1}\log\left(\sum_{\cS\cP}\rho(\cS\cP)^\alpha\left(\frac{\sigma(\cP)}{d_{Q}}\right)^{1-\alpha}2^{(\alpha-1)D_\alpha(\rho_{QQ'|\cS\cP}\Vert\id_Q\otimes\sigma_{Q'}^{(\cP)})}\right)\nonumber\\
            &=\sup_{\{\sigma(\cP)\}}\frac{1}{1-\alpha}\log\left(\sum_{\cP}\sigma(\cP)^{1-\alpha}\left[\inf\limits_{\sigma^{(\cP)}\in\dop{=}(Q')}\sum_{\cS}\rho(\cS\cP)^\alpha2^{(\alpha-1) D_\alpha(\rho_{QQ'|\cS\cP}\Vert\id_Q\otimes\sigma_{Q'}^{(\cP)})}\right]\right).
        \end{align}
        Let $r({\cS\cP})\defvar\rho(\cS\cP)2^{\frac{\alpha-1}{\alpha}D_\alpha(\rho_{QQ'|\cS\cP}\Vert\id_Q\otimes\sigma_{Q'}^{(\cP)})}$, then by employing Lagrange multiplier technique, it can be seen that the infimum is attained by
        \begin{align}
        \label{eq:twoclassmixHup_proof3}\sigma({\cP})=\frac{\left(\inf_{\sigma^{(\cP)}\in\dop{=}(Q')}\sum_{\cS}r(\cS\cP)^\alpha\right)^{1/\alpha}}{\sum_{\cP}\left(\inf_{\sigma^{(\cP)}\in\dop{=}(Q')}\sum_{\cS}r(\cS\cP)^\alpha\right)^{1/\alpha}}.
        \end{align}
        Substituting this solution into Eq.~\eqref{eq:twoclassmixHup_proof2}, followed by simple algebraic simplifications, yields the claimed result for $\alpha\in(1,\infty)$. The proof for $\alpha\in[\tfrac{1}{2},1)$ is identical except that the optimizations over $\sigma^{(\cP)}$ do not switch direction when brought inside the logarithm (since $\frac{1}{1-\alpha}>0$).
	\end{proof}

\begin{remark}
In the above statement, we restricted the summation domains to avoid potential ill-definedness of conditional probabilities $\rho(\cS|\cP)$ (or conditional states $\rho_{|\cS\cP}$, though in this work we have chosen the convention that such states can be arbitrarily defined). However, note that if we were to extend the summations to the full alphabets, then for any $\cP \notin \supp(\bsym{\rho}_{\CP})$, all terms involving conditional states $\rho_{|\cS\cP}$ or conditional probabilities $\rho(\cS|\cP)$ would be accompanied by a prefactor of the corresponding probability, i.e.~just $\rho(\cP)=0$. An analogous property holds for any $(\cS,\cP)$ such that $\cS \notin \supp(\bsym{\rho}_{\CS|\cP})$.
Hence in principle, one could instead sum over the full alphabets while assigning some arbitrary definition to these conditional probabilities or states.

Similar properties hold for many of our subsequent results as well, though we aim to explicitly specify the summation domains wherever possible to reduce ambiguity.
\end{remark}

We now introduce the concept of $f$-weighted entropies. The definition here slightly generalizes the definition in~\cite{inprep_weightentropy}, by allowing for a register $\CS$ on the left side of the conditioning.
\begin{definition}($f$-weighted {\Renyi} entropies)
	\label{def:QES}
	Let $\rho \in \dop{=}(\CS \CP Q Q')$ be a state where $\CS$ and $\CP$ are classical with alphabets $\alphCS$ and $\alphCP$ respectively. A \term{tradeoff function on $\CS \CP$} is simply a function $f:\alphCS \times \alphCP \to \mathbb{R}$; equivalently, we may denote it as a real-valued tuple $\mbf{f}
	\in \mathbb{R}^{|\alphCS \times \alphCP|}$ where each term in the tuple specifies the value $f(\cS \cP)$. Given a tradeoff function $f$ and a value $\alpha\in (1,\infty)$, the \term{$f$-weighted entropy of order $\alpha$} for $\rho$ is defined as\footnote{Building on this concept, \cite{AHT24} obtained various results for a slightly modified version of $f$-weighted entropies based on $H_\alpha^{\downsymb}$ instead of $H_\alpha^\uparrow$.
	}
	\begin{align}\label{eq:QESdefn}
		&H^{\uparrow ,f}_\alpha(Q\CS|\CP Q')_{\rho} \defvar\notag\\
        &\frac{\alpha}{1-\alpha} \log\!\left( \sum_{\cP\in \supp(\bsym{\rho}_{\CP})}
        \inf\limits_{\sigma^{(\cP)}\in\dop{=}(Q')}\left[\sum_{\cS\in\supp(\bsym{\rho}_{\CS|\cP})} \rho(\cS \cP)^\alpha\, 2^{(\alpha-1) \left(f(\cS \cP) + D_\alpha(\rho_{QQ'|\cS\cP}\Vert\id_Q\otimes\sigma_{Q'}^{(\cP)})\right)}\right]^{\frac{1}{\alpha}} \right) \nonumber\\
		&=\frac{\alpha}{1-\alpha} \log\!\left( \sum_{\cP\in \supp(\bsym{\rho}_{\CP})}\rho(\cP)
        \inf\limits_{\sigma^{(\cP)}\in\dop{=}(Q')}\left[\sum_{\cS\in\supp(\bsym{\rho}_{\CS|\cP})} \rho(\cS|\cP)^\alpha\, 2^{(\alpha-1) \left(f(\cS \cP)+D_\alpha(\rho_{QQ'|\cS\cP}\Vert\id_Q\otimes\sigma_{Q'}^{(\cP)})\right)}\right]^{\frac{1}{\alpha}} \right),
	\end{align}
	where 
	$\bsym{\rho}_{\CS|\cP}=\left\{\rho\left(\cS|\cP\right)\big|\ 
	\cS \in \alphCS
	\right\}$. 
	Moreover, we extend the definition to $\alpha=\infty$, by taking the limit $\alpha\rightarrow\infty$. Note that $f$-weighted entropies can also be represented by Schatten norms by the following construction: given $\rho$, $\sigma$, $\alpha$ and $f$, let us write $\mbf{x}^{\rho,\sigma,\alpha,f}_{|\cP}$ (for any $\cP\in \supp(\bsym{\rho}_{\CP})$) to denote a vector in $\mathbb{R}^{|\supp(\bsym{\rho}_{\CS|\cP})|}$ with components specified by
    \begin{align}\label{eq:normversionvector}
        x_{|\cP}^{\rho,\sigma,\alpha,f}(\cS)\defvar \rho(\cS|\cP)2^{\frac{\alpha-1}{\alpha}\left(f(\cS\cP)+D_\alpha(\rho_{QQ'|\cS\cP}\Vert\id_Q\otimes\sigma_{Q'}\right)}.
    \end{align}
    Then, Eq.~\eqref{eq:QESdefn} can be rewritten as
    \begin{align}
        H^{\uparrow ,f}_\alpha(Q\CS|\CP Q')_{\rho}&=\frac{\alpha}{1-\alpha} \log \left(\sum_{\cP\in \supp(\bsym{\rho}_{\CP})} \rho(\cP)
        \inf\limits_{\sigma^{(\cP)}\in\dop{=}(Q')}\norm{\mbf{x}_{|\cP}^{\rho,\sigma^{(\cP)},\alpha,f}}_{\alpha}\right) \nonumber\\
        &=\frac{\alpha}{1-\alpha} \log\mathbb{E}_{\bsym{\rho}_{\CP}}
        \inf\limits_{\sigma\in\dop{=}(Q')}\norm{\mbf{x}_{|\CP}^{\rho,\sigma,\alpha,f}}_{\alpha},
    \end{align}
    where in the last expression we have suppressed the index $\cP$ in the ``dummy variables'' $\sigma^{(\cP)}$, to avoid ambiguity in the random variable notation (with the understanding that the value of the infimum will be a function of the random variable $\CP$ in that expression). 

    For $\alpha\in[\frac{1}{2},1)$, we define $H^{\uparrow ,f}_\alpha(Q\CS|\CP Q')_{\rho}$ analogously, but with the infima over $\sigma^{(\cP)}$  replaced by suprema.
\end{definition}
Qualitatively, the registers $\CS$ and $\CP$ can be respectively viewed as ``secret'' and ``public'' classical registers, on which we define a function $f$ that captures some properties of the {\Renyi} entropy of the state, as we now describe.

Note that when $f(\cS\cP)=0$, $H_\alpha^{\uparrow ,f}(Q\CS|\CP Q')_\rho$ is exactly equal to the sandwiched {\Renyi} conditional entropy $H_\alpha^{\uparrow}(Q\CS|\CP Q')_\rho$. Furthermore, when the $\CS$ register is trivial it reduces to exactly the $f$-weighted entropy defined in~\cite{inprep_weightentropy}, 
where in this case for $\alpha=\infty$ we can use the convention $\frac{\alpha}{\alpha-1}=1$ without issues:
\begin{align}
	\label{eq:fweightedonlyH}
	H^{\uparrow ,f}_\alpha(Q|\CP Q')_{\rho} =
	\frac{\alpha}{1-\alpha} \log \left( \sum_{\cP} \rho(\cP) \, 2^{\left(\frac{1-\alpha}{\alpha}\right) \left(H^\uparrow_\alpha(Q|Q')_{\rho_{|\cP}} - f(\cP) \right) } \right) .
\end{align}
 
A similar intuition to that described in~\cite[Sec.~4.1]{AHT24} can also be developed here, through the concept of a ``log-mean-exponential" (also known as an exponential mean):
\begin{align}\label{eq:lme}
	\underset{P_X}{\operatorname{lme}_b} \left[g(X)\right] \defvar \log_b \left( \sum_{x} P_X(x) \, b^{g(x)} \right),
\end{align}
which allows us to rewrite~\eqref{eq:fweightedonlyH} (i.e.~the case where $\CP$ is trivial) as
\begin{align}\label{eq:fentropy_as_lme}
	H^{\uparrow ,f}_\alpha(Q|\CP Q')_{\rho} = \underset{\bsym{\rho}_{\CP}}{\operatorname{lme}_b} \left[H^\uparrow_\alpha(Q|Q')_{\rho_{|\CP}} - f(\CP) \right] , \quad\text{where } b = 2^{\frac{1-\alpha}{\alpha}} \in (0,\infty).
\end{align}
Thus we observe that having a tradeoff function $f(\CP)$ that lower bounds the conditional entropy $H^\uparrow_\alpha(Q|Q')_{\rho_{|\CP}}$ in a log-mean-exponential manner is equivalent to the non-negativity of $f$-weighted entropies, expressed as $H^{\uparrow ,f}_\alpha(Q|\CP Q')_{\rho}\geq 0$. 

We will now demonstrate that these entropies $H^{\uparrow ,f}_\alpha$ possess properties similar to those presented in~\cite{AHT24} for $H^{\downsymbcomma f}_\alpha$ (i.e.~the variant of $H^{\uparrow ,f}_\alpha$ based on $H^{\downsymb}_\alpha$; see~\cite[Definition~4.1]{AHT24}).
Specifically, in the subsequent lemma, we will show that, similar to Lemma~4.1 in~\cite{AHT24}, by introducing an extra register $D$ that encodes the value of $f(\CS\CP)$ via its entropy, $f$-weighted entropies can be related to the standard $H^\uparrow_\alpha$ conditional entropies.
\begin{lemma}
	\label{lemma:createD}
	Let $f$ be a tradeoff function on some classical registers $\CS\CP$, let $M\in\mathbb{R}$ be any real value such that $M-f(\cS\cP)>0$ for all $\cS\cP$, and take any $\alpha\in [\frac{1}{2},1)\cup(1,\infty]$. Consider any read-and-prepare channel $\mathcal{D}\in\CPTP(\CS\CP,\CS\CP D)$ such that the state it prepares on $D$ always satisfies
	\begin{align}\label{eq:D_entropy}
		H_\alpha(D)_{\rho_{|\cS \cP}} = H^\uparrow_\alpha(D)_{\rho_{|\cS \cP}} = M-f(\cS\cP).
	\end{align}
	(It is always possible to construct such a channel, and construct it such that $D$ is classical.)
	Then for any $\rho \in \dop{=}(\CS \CP Q Q')$ classical on $\CS \CP$, extending $\rho$ with this channel yields\footnote{There is no danger of ambiguity in having used $\rho$ to denote all states in this lemma, since a read-and-prepare channel always simply extends a state without ``disturbing'' any registers.} 
	\begin{align}\label{eq:createD}
		H_\alpha^{\uparrow}(DQ\CS|\CP Q')_{\rho} = M + H^{\uparrow ,f}_\alpha(Q\CS|\CP Q')_{\rho}.
	\end{align}
\begin{proof}
First we briefly verify that such channels can indeed be constructed: note that for any $\alpha>0$ and $h>0$, given a register $D$ with dimension at least $2^h$, one can straightforwardly construct a state on $D$ with $H_\alpha(D) = H^\uparrow_\alpha(D) = h$ (the first equality holds simply because $H_\alpha = H^\uparrow_\alpha$ when there is no conditioning system), e.g.~by taking a mixture between some pure state on $D$ and the maximally mixed state; furthermore this can clearly be achieved using a classical $D$. Therefore to obtain a read-and-prepare channel satisfying~\eqref{eq:D_entropy}, we simply need to it to read the classical value $\cS\cP$ and prepare a state on $D$ with the corresponding desired entropy value. 
	
	Then Eq.~\eqref{eq:createD} follows from the following calculation, focusing on $\alpha \in (1,\infty)$ (the $\alpha=\infty$ case follows by taking the limit at the end, and the $\alpha\in[\tfrac{1}{2},1)$ cases are analogous with the infima replaced by suprema):
	\begin{align}\label{eq:D_entropyproof}
		&H_\alpha^{\uparrow}(DQ\CS|\CP Q')_\rho=\notag\\
        &\frac{\alpha}{1-\alpha} \log \left( \sum_{\cP}\rho(\cP)\inf\limits_{\sigma^{(\cP)}\in\dop{=}(Q')}\left[\sum_{\cS} \rho(\cS|\cP)^\alpha\, 2^{(\alpha-1)  D_\alpha(\rho_{DQQ'|\cS\cP}\Vert\id_{DQ}\otimes\sigma_{Q'}^{(\cP)})}\right]^{\frac{1}{\alpha}} \right)\nonumber\\
		&=\frac{\alpha}{1-\alpha} \log \left( \sum_{\cP}\rho(\cP)\inf\limits_{\sigma^{(\cP)}\in\dop{=}(Q')}\left[\sum_{\cS} \rho(\cS|\cP)^\alpha\, 2^{(\alpha-1) \left(-H_\alpha^{\uparrow}(D)_{\rho|\cS\cP}+D_\alpha(\rho_{QQ'|\cS\cP}\Vert\id_Q\otimes\sigma_{Q'}^{(\cP)})\right)}\right]^{\frac{1}{\alpha}} \right)\nonumber\\
		&=\frac{\alpha}{1-\alpha} \log \left( \sum_{\cP}\rho(\cP)\inf\limits_{\sigma^{(\cP)}\in\dop{=}(Q')}\left[\sum_{\cS} \rho(\cS|\cP)^\alpha\, 2^{(\alpha-1) \left(f(\cS \cP) - M +D_\alpha(\rho_{QQ'|\cS\cP}\Vert\id_Q\otimes\sigma_{Q'}^{(\cP)})\right)}\right]^{\frac{1}{\alpha}} \right)\nonumber\\
		&=M+\frac{\alpha}{1-\alpha} \log \left( \sum_{\cP}\rho(\cP)\inf\limits_{\sigma^{(\cP)}\in\dop{=}(Q')}\left[\sum_{\cS} \rho(\cS|\cP)^\alpha\, 2^{(\alpha-1) \left(f(\cS \cP) + D_\alpha(\rho_{QQ'|\cS\cP}\Vert\id_Q\otimes\sigma_{Q'}^{(\cP)})\right)}\right]^{\frac{1}{\alpha}} \right)\nonumber\\
		&=M+H_\alpha^{\uparrow ,f}(Q\CS|\CP Q')_\rho,
	\end{align}
	where the second line holds since $\rho_{DQQ'|\cS\cP}=\rho_{D|\cS\cP}\otimes\rho_{QQ'|\cS\cP}$, and the third line holds by substituting from Eq.~\eqref{eq:createD}.
\end{proof}
\end{lemma}
We now present a relaxed version of the previous lemma, which is similar to Lemma~4.2 in~\cite{AHT24}, but again has $H^{\uparrow ,f}_\alpha$ rather than $H^f_\alpha$:
\begin{lemma}\label{lemma:createD_2}
	Let $f$ be a tradeoff function on some classical registers $\CS \CP$, let $M>0$ be any value such that $M - f(\cS \cP) > M/2 > 0$ for all $\cS \cP$.
	Consider any read-and-prepare channel $\mathcal{D}\in\CPTP(\CS \CP,\CS \CP D)$ such that the state it prepares on $D$ always satisfies
	\begin{align}\label{eq:D_entropy_2}
		\forall \alpha \in[0,\infty], \quad H_\alpha(D)_{\rho_{|\cS \cP}} = H^\uparrow_\alpha(D)_{\rho_{|\cS \cP}} \in
		\left[M-f(\cS\cP),M-f(\cS\cP)+2^{-\frac{M}{2}}\log e\right].
	\end{align}
	(It is always possible to construct such a channel, and construct it such that $D$ is classical.)
	Then for any $\rho \in \dop{=}(\CS \CP Q Q')$ classical on $\CS \CP$, extending $\rho$ with this channel yields
	\begin{align}\label{eq:createD_2}
		&\forall \alpha\in[\tfrac{1}{2},1)\cup(1,\infty], \notag\\ 
		&M + H^{\uparrow ,f}_\alpha(Q\CS|\CP Q')_{\rho} 
		\leq 
		H^{\uparrow}_\alpha(DQ\CS|\CP Q')_{\rho} 
		\leq 
		M + 2^{-\frac{M}{2}}\log e + H^{\uparrow ,f}_\alpha(Q\CS|\CP Q')_{\rho}.
	\end{align}
	\begin{proof}
		The proof is by the same argument as above, except we upper and lower bound the entropies on the $D$ register with~\eqref{eq:D_entropy_2} instead of~\eqref{eq:D_entropy}. (The existence of such a read-and-prepare channel is established by the same argument as in the proof of~\cite[Lemma~4.2]{AHT24}.)
	\end{proof}
\end{lemma}
Utilizing these lemmas, we ``transfer'' many properties of $H^{\uparrow}_\alpha$ to $H^{\uparrow ,f}_\alpha$, in very similar fashion to the results in~\cite{AHT24} that were instead based on $H^{\downsymb}_\alpha$. Many of these subsequent statements immediately include the $\alpha=\infty$ case without further elaboration, as the corresponding property is known to hold directly for $H^{\uparrow}_\infty$. The proofs of these lemmas follow exactly the same methodology as their counterparts in~\cite{AHT24} (simply use Lemmas~\ref{lemma:createD}--\ref{lemma:createD_2} to reformulate statements regarding $H^{\uparrow ,f}_\alpha$ in terms of statements regarding $H^{\uparrow}_\alpha$); therefore, we do not explicitly present them here except for Lemma~\ref{lemma:QES_up-to-down}, which did not have a full counterpart proven in~\cite{AHT24}. 
\begin{lemma}\label{lemma:DPI}
	(Data-processing) Let $\rho \in \dop{=}(\CS \CP Q Q')$ be classical on $\CS \CP$, let $f$ be a tradeoff function on $\CS \CP$, and take any $\alpha\in [\frac{1}{2},1)\cup(1,\infty]$. Then for any channel $\mathcal{E} \in \CPTP(Q',Q'')$, 
	\begin{align}
		H^{\uparrow ,f}_\alpha(Q\CS|\CP Q'')_{\mathcal{E}[\rho]} \geq H^{\uparrow ,f}_\alpha(Q\CS|\CP Q')_{\rho}.
	\end{align}
	If $\mathcal{E}$ is an isometry, then we have equality in the above bound.
\end{lemma}
\begin{lemma}\label{lemma:classmixQES}
	(Conditioning on classical registers) Let $\rho \in \dop{=}(\CS \CP Q Q' Z)$ be classical on $\CS \CP Z$, let $f$ be a tradeoff function on $\CS \CP$, and take any $\alpha\in[\tfrac{1}{2},1)\cup(1,\infty]$. Then (under the convention that $\frac{\alpha}{\alpha-1}=1$ when $\alpha=\infty$):
	\begin{align}\label{eq:classmixQES}
		H^{\uparrow ,f}_\alpha(Q \CS|\CP Q' Z)_{\rho} =
		\frac{\alpha}{1-\alpha} \log \left( \sum_{z} \rho(z) \, 2^{\left(\frac{1-\alpha}{\alpha}\right) H^{\uparrow ,f}_\alpha(Q \CS|\CP Q')_{\rho_{|z}}  } \right) .
	\end{align}
\end{lemma}

\begin{lemma}\label{lemma:Monotonicity}
	(Monotonicity in $\alpha$) Let $\rho \in \dop{=}(\CS \CP Q Q')$ be classical on $\CS \CP$, let $f$ be a tradeoff function on $\CS \CP$, and take any $\alpha, \alpha'\in [\frac{1}{2},1)\cup(1,\infty]$ such that $\alpha\geq\alpha'$. Then, 
	\begin{align}
		H^{\uparrow ,f}_\alpha(Q\CS|\CP Q')_{\rho} \leq H^{\uparrow ,f}_{\alpha'}(Q\CS|\CP Q')_{\rho}.
	\end{align}
\end{lemma}

\begin{lemma}
    \label{lemma:QES_up-to-down} (Relating $H_{\alpha}^{\uparrow ,f}$ and $H_{\alpha}^{\downsymbcomma f}$)
    Let $\rho\in\dop{=}(\CS\CP QQ')$ be classical on $\CS\CP$, and let $f$ be a tradeoff function on $\CS\CP$. Then,
    \begin{align}
    \begin{aligned}
        \forall \alpha\in[\tfrac{1}{2},1)\cup (1,\infty], \qquad & H_{\alpha}^{\uparrow ,f}(Q\CS|\CP Q')_\rho\geq  H_{\alpha}^{\downsymbcomma f}(Q\CS|\CP Q')_\rho ,\\
        \forall \alpha\in[\tfrac{1}{2},1)\cup (1,2], \qquad &   H_{\alpha}^{\downsymbcomma f}(Q\CS|\CP Q')_\rho\geq H_{\widehat{\alpha}}^{\uparrow ,f}(Q\CS|\CP Q')_\rho
    \end{aligned}
    \end{align}
    where $\widehat{\alpha}=\frac{1}{2-\alpha}$ (understood as $\infty$ when $\alpha=2$), and $H_\alpha^{\downsymbcomma f}$ is defined as in~\cite[Definition~4.1]{AHT24}. 
    \begin{proof}
        We show only the proof of the second inequality, as the proof of the first is analogous but simpler (since the {\Renyi} parameter remains unchanged).
        Let $M>0$ be any value such that $M-f(\cS\cP)>M/2>0$, and extend $\rho$ with a read-and-prepare channel as described in Lemma~\ref{lemma:createD_2}; note that this is the same as the channel in~\cite[Lemma~4.2]{AHT24}. Then we have
        \begin{align}
            H_\alpha^{\downsymbcomma f}(Q\CS|\CP Q')_\rho&\geq H_\alpha^{{\downsymb}}(DQ\CS|\CP Q')-M-2^{-M/2}\log e\notag\\
            &\geq H^{\uparrow}_{\widehat{\alpha}}(DQ\CS|\CP Q')-M-2^{-M/2}\log e\notag\\
            &\geq H_{\widehat{\alpha}}^{\uparrow ,f}(Q\CS|\CP Q')_\rho - 2^{-M/2}\log e,
        \end{align}
        where the first line is~\cite[Lemma~4.2]{AHT24}, the second line follows from~\cite[Corollary~4]{TBH14}, and the third line is Lemma~\ref{lemma:createD_2}. Since this inequality holds for arbitrary (sufficiently large) $M$, we can take the $M\rightarrow\infty$ limit to get the claimed inequality. 
    \end{proof}
\end{lemma}

\begin{lemma}\label{lemma:QES3Renyi}
	(Chain rules) Let $f$ be a tradeoff function on classical registers $\CS \CP$, take any $\alpha,\alpha',\alpha''\in (\frac{1}{2},1)\cup(1,\infty)$ such that $\frac{\alpha}{\alpha-1} = \frac{\alpha'}{\alpha'-1} + \frac{\alpha''}{\alpha''-1}$. Then we have for any $\rho \in \dop{=}(\CS \CP R Q Q')$:
	\begin{align}\label{eq:chain3Renyi2}
		\begin{gathered}
			H^{\uparrow ,f}_{\alpha}(R Q \CS|\CP Q')_{\rho} \leq H^{\uparrow ,f}_{\alpha'}(Q \CS | \CP Q' R)_{\rho} + H^{\uparrow}_{\alpha''}(R | \CP Q')_{\rho} \quad\text{ if } (\alpha-1)(\alpha'-1)(\alpha''-1) < 0 ,\\
			H^{\uparrow ,f}_{\alpha}(R Q \CS|\CP Q')_{\rho} \geq H^{\uparrow ,f}_{\alpha'}(Q \CS | \CP Q' R)_{\rho} + H^\uparrow_{\alpha''}(R | \CP Q')_{\rho} \quad\text{ if } (\alpha-1)(\alpha'-1)(\alpha''-1) > 0 .
		\end{gathered}
	\end{align}
	Furthermore, we also have for any $\rho \in \dop{=}(\CS \CP Q Q')$:
	\begin{align}\label{eq:chain3RenyiCS}
		\begin{gathered}
			H^{\uparrow ,f}_{\alpha}(Q \CS|\CP Q')_{\rho} \leq H^{\uparrow ,f}_{\alpha'}(Q | \CS \CP Q')_{\rho} + H^\uparrow_{\alpha''}(\CS|\CP Q')_{\rho} \quad\text{ if } (\alpha-1)(\alpha'-1)(\alpha''-1) < 0 ,\\
			H^{\uparrow ,f}_{\alpha}(Q \CS|\CP Q')_{\rho} \geq H^{\uparrow ,f}_{\alpha'}(Q | \CS \CP Q')_{\rho} + H^\uparrow_{\alpha''}(\CS|\CP Q')_{\rho} \quad\text{ if } (\alpha-1)(\alpha'-1)(\alpha''-1) > 0 ,
		\end{gathered}
	\end{align}
	where by $H^{\uparrow ,f}_{\alpha}(Q | \CS \CP Q')_{\rho}$ we mean
	\begin{align}\label{eq:CSinconditioning}
		H^{\uparrow ,f}_{\alpha}(Q | \CS \CP Q')_{\rho}= \frac{\alpha}{1-\alpha} \log \left( \sum_{\cS \cP} \rho(\cS \cP) 2^{\left(\frac{1-\alpha}{\alpha}\right) \left(-f(\cS \cP) + H^\uparrow_\alpha(Q|Q')_{\rho_{|\cS\cP}}\right) } \right).
	\end{align}
\end{lemma}

\begin{lemma}\label{lemma:normalize}
	(Normalization property) Let $\rho \in \dop{=}(\CS \CP Q Q')$ be classical on $\CS \CP$, let $f$ be a tradeoff function on $\CS \CP$, and take any $\alpha\in [\frac{1}{2},1)\cup(1,\infty]$. Then for any constant $\kappa\in\mathbb{R}$, we have
	\begin{align}
		H^{\uparrow ,f+\kappa}_\alpha(Q\CS|\CP Q')_{\rho} = H^{\uparrow ,f}_\alpha(Q\CS|\CP Q')_{\rho} - \kappa,
	\end{align}
	where $f+\kappa$ simply denotes the tradeoff function with values $f(\cS \cP) + \kappa$. 
\end{lemma}

\subsection{Simplified main results}
\label{subsec:simplified}

We are now in a position to present our initial key results. 
To establish an understanding, we will first present simplified versions of these results, postponing the most general statements and the proofs to Sec.~\ref{sec:generalresults}.

First, we demonstrate that tradeoff functions can accumulate to provide an estimation of the conditional entropy of a global state, provided that the state is generated by the application of a suitable sequence of channels (which would usually represent rounds of a protocol) to an arbitrary input state, as follows. As mentioned above, we shall postpone the proof to Sec.~\ref{sec:generalresults} below, where we in fact prove a slightly more general variant (Theorem~\ref{thrm:qes eat}).
\begin{manualtheorem}{4.1a}\label{thrm:qes eat_simp}
	For each $j\in\{1,2,\cdots,n\}$, take a state $\sigma^{(j-1)}\in\dop{=}(A_{j-1})$, and a channel $\mathcal{M}_j\in\CPTP(A_{j-1}E_{j-1},S_j\CP_jE_j)$, such that $\CP_j$ are classical. Let $\rho$ be a state of the form $\rho_{S_1^n\CP_1^nE_n}=\mathcal{M}_n\circ\cdots\circ\mathcal{M}_1[\omega_{A_0^{n-1}E_0}]$ for some $\omega\in\dop{=}(A_0^{n-1}E_0)$, such that $\omega_{A_0^{n-1}}=\sigma_{A_0}^{(0)}\otimes\cdots\otimes\sigma_{A_{n-1}}^{(n-1)}$. For each $j$, suppose that for every value $\cP_1^{j-1}$, we have a tradeoff function $f_{|\cP_1^{j-1}}$ on registers $\CP_j$. Define the following tradeoff function on $\CP_1^n$:
	\begin{align}
		\label{eq:full qes_simp}
		f_\mathrm{full}(\cP_1^n) \defvar \sum_{j=1}^n f_{|\cP_1^{j-1}}(\cP_j).
	\end{align}
	Then for any $\alpha\in (1,\infty]$ we have
	\begin{align}\label{eq:qes eat_simp}
		\begin{gathered}
			H_\alpha^{\uparrow,f_\mathrm{full}}(S_1^n|\CP_1^nE_n)_\rho\geq\sum_j\min_{\cP_1^{j-1}}\kappa_{\cP_1^{j-1}}\qquad
			\text{where}\quad \kappa_{\cP_1^{j-1}} \defvar \inf_{\nu\in\Sigma_j} H^{\uparrow, f_{|\cP_1^{j-1}}}_{\alpha}(S_j| \CP_j E_j \widetilde{E})_{\nu},
		\end{gathered}
	\end{align}
	defining 
    \begin{align}
        \Sigma_j \defvar \left\{
\EATchann_j\left[\omega_{A_{j-1}E_{j-1}\widetilde{E}}\right] 
\;\middle|\;
\omega \in \dop{=}(A_{j-1}E_{j-1}\widetilde{E}) \;\suchthat\; \omega_{A_{j-1}}=\sigma_{A_{j-1}}^{(j-1)}
\right\},
    \end{align} 
    with $\widetilde{E}$ being a register of large enough dimension to serve as a purifying register for any of the $A_{j-1}E_{j-1}$ registers.
	
	Consequently, if we instead define the following ``normalized'' tradeoff function on $\CP_1^n$:
	\begin{align}\label{eq:fullQESnorm_simp}
		\hat{f}_\mathrm{full}( \cP_1^n) \defvar \sum_{j=1}^n \hat{f}_{|\cP_1^{j-1}}(\cP_j), \quad\text{where}\quad \hat{f}_{|\cP_1^{j-1}}(\cP_j) \defvar f_{|\cP_1^{j-1}}(\cP_j) + \kappa_{\cP_1^{j-1}},
	\end{align}
	then
	\begin{align}\label{eq:chainQESnorm_simp}
		H^{\uparrow,\hat{f}_\mathrm{full}}_\alpha(S_1^n | \CP_1^n E_n)_\rho \geq 0.
	\end{align}
\end{manualtheorem}
Qualitatively, the above theorem tells us that for any choices of ``single-round'' tradeoff functions $f_{|\cP_1^{j-1}}$, we can ``accumulate'' them into an overall tradeoff function $\hat{f}_\mathrm{full}$ that yields a log-mean-exponential bound (as discussed in Eq.~\eqref{eq:fentropy_as_lme} earlier) on the {\Renyi} entropy of the final state. Furthermore, since the single-round optimizations preserve the marginal constraint on the input state, this makes it suitable for use with the source-replacement technique for PM protocols (see Sec.~\ref{sec:security} below). We defer to~\cite{AHT24,inprep_weightentropy} for a more detailed explanation of how to interpret and apply such bounds, instead first focusing on a particular application.

Specifically, in many scenarios, we may only be interested in the entropy conditioned on an event (e.g.~for QKD protocols that make an accept/abort decision and produce a key of fixed length if they accept).
Following similar reasoning to Sec.~5.1 in~\cite{AHT24}, the bound in Theorem~\ref{thrm:qes eat_simp} can be exploited to bound the entropy conditioned on the accept event. To describe the resulting bound, we first introduce the following terminology to describe the set of all mixtures of output states from some set of channels (with a marginal constraint).
\newcommand{\outQ}{Z}
\begin{definition}\label{def:range} 
(Marginal-constrained convex range) For $j\in\{1,2,\cdots,n\}$, consider channels $\mathcal{E}_j \in \CPTP(A_{j-1}\widetilde{A}_{j-1},\outQ_j)$ and states $\sigma^{(j-1)} \in\dop{=}(A_{j-1})$.  Suppose furthermore that $\outQ_j = \outQ'_j \outQ''_j \dots$ for some (finite) set of ``sub-registers'' $\outQ'_j \outQ''_j \dots$, and we have an embedding of all the output registers $\outQ'_j$ (resp.~$\outQ''_j , \dots$) in a common register $\outQ'$ (resp.~$\outQ'' , \dots$), in which case there is a well-defined notion of taking mixtures of states with distinct $\outQ_j$ (by embedding them in $\outQ \defvar \outQ'_j \outQ''_j \dots$). Then the \term{marginal-constrained convex range} $\Sigma_{\outQ}$ of\footnote{
Strictly speaking, the marginal-constrained convex range is not fully defined by only the channels $\mathcal{E}_j$ and states $\sigma^{(j-1)}$, since one also has to specify the embeddings. However, all of our subsequent results hold for any choice of embeddings, so we will usually not specify this explicitly.} the channels $\mathcal{E}_j$ and states $\sigma^{(j-1)}$ is defined as the set of all mixtures of states of the form
$\mathcal{E}_j\left[\omega_{A_{j-1}\widetilde{A}_{j-1}}\right]$ for some $j$ and some $\omega \in \dop{=}(A_{j-1}\widetilde{A}_{j-1})$ satisfying $ \omega_{A_{j-1}}=\sigma_{A_{j-1}}^{(j-1)}$. 
Slightly more formally, we can write
\begin{align}
\Sigma_{\outQ}
\defvar
\operatorname{conv}\left(
\bigcup_{j=1}^n
\left\{
\mathcal{E}_j\left[\omega_{A_{j-1}\widetilde{A}_{j-1}}\right] 
\;\middle|\;
\omega \in \dop{=}(A_{j-1}\widetilde{A}_{j-1}) \;\suchthat\; \omega_{A_{j-1}}=\sigma_{A_{j-1}}^{(j-1)}
\right\}
\right)
,
\end{align}
where $\operatorname{conv}(\mathcal{S})$ denotes the convex hull~\cite{BV04v8} of a set $\mathcal{S}$.
\end{definition}

If it happens that the channels $\mathcal{E}_j \in \CPTP(A_{j-1}\widetilde{A}_{j-1},\outQ_j)$ (resp.~states $\sigma^{(j-1)}\in\dop{=}(A_{j-1})$) are all isomorphic to a single channel $\mathcal{E} \in \CPTP(A \widetilde{A},\outQ)$ (resp.~state $\sigma \in \dop{=}(A)$), as was basically the case studied in~\cite{inprep_weightentropy}, one can simply take the marginal-constrained convex range to be the set
\begin{align}
\Sigma_{\outQ} = \left\{
\mathcal{E}\left[\omega_{A\widetilde{A}}\right]
\;\middle|\;
\omega \in \dop{=}(A\widetilde{A})
 \;\suchthat\;
\omega_{A}=\sigma_{A}
\right\}
,
\end{align}
since this set is already convex (therefore equal to its convex hull).

With this concept in mind, we can state the following result (again, we postpone its proof to Sec.~\ref{sec:generalresults} below, together with the statement of a variant that covers a slightly more general class of scenarios).
We stress once more that, in the following theorem, all registers are taken to be finite-dimensional.
\begin{manualtheorem}{4.2a} (Marginal-constrained entropy accumulation theorem, special case)
\label{thrm:MEAT_simp}
For each $j\in\{1,2,\cdots,n\}$, take a state $\sigma^{(j-1)}\in\dop{=}(A_{j-1})$, and a channel $\mathcal{M}_j\in\CPTP(A_{j-1}E_{j-1},S_j\CP_jE_j)$, such that $\CP_j$ are classical. Let $\rho$ be a state of the form $\rho_{S_1^n\CP_1^nE_n}=\mathcal{M}_n\circ\cdots\circ\mathcal{M}_1[\omega_{A_0^{n-1}E_0}]$ for some $\omega\in\dop{=}(A_0^{n-1}E_0)$, such that $\omega_{A_0^{n-1}}=\sigma_{A_0}^{(0)}\otimes\cdots\otimes\sigma_{A_{n-1}}^{(n-1)}$.

Suppose furthermore that $\rho = p_\Omega \rho_{|\Omega} + (1-p_\Omega) \rho_{|\overline{\Omega}}$ for some $p_\Omega \in (0,1]$ and normalized states $\rho_{|\Omega},\rho_{|\overline{\Omega}}$, and that all the $\CP_j$ registers are isomorphic to a single register $\CP$ with alphabet $\alphCP$. 
Let $S_\Omega$ be a convex set of probability distributions on the alphabet $\alphCP$, such that for all $\cP_1^n$ with nonzero probability in $\rho_{|\Omega}$, the frequency distribution $\freq_{\cP_1^n}$ lies in $S_\Omega$. 
Then, for any $\alpha\in(1,\infty]$, we have (under the convention that $\frac{\alpha}{\alpha-1}=1$ when $\alpha=\infty$):
\begin{align}\label{eq:fweightedREATworst_simp}
H^\uparrow_\alpha(S_1^n | \CP_1^n E_n)_{\rho_{|\Omega}} \geq  n  h^\uparrow_{\alpha}
- \frac{\alpha}{\alpha-1} \log\frac{1}{p_\Omega}, 
\end{align}
where (recalling $\bsym{\nu}_{\CP}$ denotes the distribution on $\CP$ induced by any state $\nu_{\CP}$)
\begin{align}
h^\uparrow_{\alpha} &= 
\inf_{\mbf{q} \in S_\Omega} \inf_{\nu\in\Sigma_{S\CP E\widetilde{E}}} \left( \frac{\alpha}{{\alpha}-1}D\left(\mbf{q} \middle\Vert \bsym{\nu}_{\CP}\right)+\sum_{\cP\in\supp(\bsym{\nu}_{\CP})}q(\cP)H^\uparrow_{{\alpha}}(S|E\widetilde{E})_{\nu_{|\cP}}  \right) 
,
\end{align}
with $\Sigma_{S\CP E \widetilde{E}}$ being the marginal-constrained convex range (Definition~\ref{def:range}) of the channels $\EATchann_j \otimes \idmap_{\widetilde{E}}$ and states $\sigma^{(j-1)}$, where $\widetilde{E}$ is a register of large enough dimension to serve as a purifying register for any of the $A_{j-1}E_{j-1}$ registers. 
\end{manualtheorem}
The above theorem allows us to bound the {\Renyi} entropy of the state conditioned on the event~$\Omega$ (which in most applications would be the accept event in some protocol), in terms of an optimization problem $h^\uparrow_{\alpha}$ that only involves single rounds. We refer the interested reader to~\cite{AHT24} for an explanation of some intuitive interpretations of this optimization. Note however that the optimization in the form presented above is not convex with respect to the states $\nu$, since the {\Renyi} entropy $H^\uparrow_\alpha$ is not a convex function of quantum states. Reformulating it as a convex optimization involves a slightly elaborate construction if the states $\sigma^{(j-1)}$ are different across rounds; therefore, we defer such a reformulation to Sec.~\ref{sec:generalresults}, where it serves a critical role in the proof of Theorem~\ref{thrm:MEAT}. More broadly though, for some applications, it may be more practical to instead relax the optimization to some yet larger set that can be evaluated using convex optimization techniques.

Our results are similar to those that were obtained in~\cite{inprep_weightentropy}.\footnote{Our results and theirs both differ from the GEAT by allowing marginal constraints to be included when evaluating the single-round values such as $\kappa_{\cP_1^{j-1}}$, at the cost of not allowing for a ``secret'' memory register to be passed between channels in different rounds --- see Sec.~\ref{sec:conclusion}.} However, our results generalize it in a few important ways: 
\begin{itemize}
\item Both Theorem~\ref{thrm:qes eat_simp} and Theorem~\ref{thrm:MEAT_simp} involve sequences of channels, rather than IID tensor products of channels as in~\cite{inprep_weightentropy}, hence covering a wider range of protocols as compared to that work (we discuss this further in Sec.~\ref{sec:security}).
Furthermore, they allow a different marginal state $\sigma_{A_{j}}^{(j)}$ to be imposed as a constraint in each round, rather than requiring the same state in each round. 
\item Theorem~\ref{thrm:qes eat_simp} allows choosing \emph{different} tradeoff functions in different rounds, and each choice can depend on the classical ``public'' values $\cP_1^{j-1}$ from preceding rounds, reflected by the notation $f_{|\cP_1^{j-1}}$. As discussed in~\cite{ZFK20} (see also~\cite[Sec.~4.3]{AHT24} for an explanation more tailored to this notation), this allows for ``fully adaptive'' procedures that update the tradeoff function\footnote{The corresponding concept in~\cite{ZFK20} is \term{quantum estimation factors}; in~\cite{AHT24} it is \term{quantum estimation score-systems}.} choices over the course of a protocol, which was not possible in~\cite{inprep_weightentropy}. In fact, our result in this context is more general even compared to~\cite{FHKR25}, because the result in that work allows for choosing a different tradeoff function $f_j$ in each round, but the choice cannot depend on the observed values $\cP_1^{j-1}$ from preceding rounds.\footnote{Essentially, this arises because~\cite{FHKR25} proved a result equivalent to Corollary~\ref{cor:channel cond ent additivity}, but not Theorem~\ref{thrm:channel cond ent chain rule}, which is more general as it allows channels that act on the ``side-information'' registers. On the other hand, we again highlight that~\cite{FHKR25} obtained other chain rules that are ``incomparable'' to Theorem~\ref{thrm:channel cond ent chain rule}; the main limitation of those chain rules in the context of PM-QKD is that they do not preserve a marginal constraint.}
\item The variants we present below (Theorems~\ref{thrm:qes eat} and~\ref{thrm:MEAT}) also have the above properties, but additionally handle a slightly more general class of scenarios, where additional ``secret'' classical registers $\CS_j$ can be produced in each round.
\end{itemize}

Similar to the point noted above regarding $h^\uparrow_\alpha$, the optimization in the definition of $\kappa_{\cP_1^{j-1}}$ is generally not a convex optimization. Nevertheless, we demonstrate that this optimization for each individual round can be reinterpreted as an equivalent convex optimization problem, by using purifying functions (Definition~\ref{def:purify}), which simplifies calculations of $\kappa_{\cP_1^{j-1}}$. 
\begin{lemma}\label{lemma:convexity}
	Let $\EATchann\in\CPTP(Q,SE\CS\CP)$ where the output is always classical on $\CS \CP$, and let $f$ be a tradeoff function on $\CS \CP$. Let $\widetilde{E}$ be a register such that $\dim(Q) \leq \dim(\widetilde{E})$, and let $\pf$ be a purifying function for $Q$ onto $\widetilde{E}$ (Definition~\ref{def:purify}). Then for any state $\omega_{\inQ \widetilde{E}}$ and $\alpha\in [\frac{1}{2},1)\cup(1,\infty]$,
	\begin{align}\label{eq:purification_proccesing}
		H^{\uparrow ,f}_\alpha(S \CS | \CP E \widetilde{E})_{\EATchann\left[\omega_{\inQ \widetilde{E}}\right]} \geq H^{\uparrow ,f}_\alpha(S \CS | \CP E \widetilde{E})_{\EATchann\left[ \pf\left(\omega_{\inQ}\right) \right]},
	\end{align}
	with equality if $\omega_{\inQ \widetilde{E}}$ is pure, and furthermore for $\alpha\in[1,\infty]$, the right-hand-side is a convex function of $\omega_{\inQ}$.
	\begin{proof}
		The proof follows from the same steps as in Lemma~4.8 in~\cite{AHT24}.
	\end{proof}
\end{lemma}
The above convexity property will also be relevant in our proof of Theorem~\ref{thrm:MEAT_simp} (and Theorem~\ref{thrm:MEAT} later), which we now turn to presenting. (The challenge in reformulating $h^\uparrow_\alpha$ in a similar fashion is that it needs to ``account for'' all the rounds, as opposed to $\kappa_{\cP_1^{j-1}}$ which only involves a single round.)

\subsection{More general results and proofs}
\label{sec:generalresults}
We now present and prove our most general formulation of $f$-weighted entropy accumulation, in which (as compared to Theorem~\ref{thrm:qes eat_simp}) we allow the channels to generate an additional ``secret'' classical register $\CS_j$ in each round.
Throughout almost all of this section, we have constructed the statements and proof steps such that they directly hold for $\alpha=\infty$, keeping in mind the conventions for how  $H^{\uparrow ,f}_\alpha$, $H^{\uparrow}_\alpha$ and $\frac{\alpha}{\alpha-1}$ are defined in that case. We do so in order to avoid technical subtleties regarding whether taking the $\alpha\to\infty$ limit commutes with various suprema/infima in the computations.

\begin{manualtheorem}{4.1b}\label{thrm:qes eat}
	For each $j\in\{1,2,\cdots,n\}$, take a state $\sigma^{(j-1)}\in\dop{=}(A_{j-1})$, and a channel $\mathcal{M}_j\in\CPTP(A_{j-1}E_{j-1},S_j\CS_j\CP_jE_j)$, such that $\CS_j\CP_j$ are classical. Let $\rho$ be a state of the form $\rho_{S_1^n\CS_1^n\CP_1^nE_n}=\mathcal{M}_n\circ\cdots\circ\mathcal{M}_1[\omega_{A_0^{n-1}E_0}]$ for some $\omega\in\dop{=}(A_0^{n-1}E_0)$, such that $\omega_{A_0^{n-1}}=\sigma_{A_0}^{(0)}\otimes\cdots\otimes\sigma_{A_{n-1}}^{(n-1)}$. For each $j$, suppose that for every value $\cP_1^{j-1}$, we have a tradeoff function $f_{|\cP_1^{j-1}}$ on registers $\CS_j\CP_j$. Define the following tradeoff function on $\CS_1^n\CP_1^n$:
	\begin{align}
		\label{eq:full qes}
		f_\mathrm{full}(\cS_1^n\cP_1^n) \defvar \sum_{j=1}^n f_{|\cP_1^{j-1}}(\cS_j\cP_j).
	\end{align}
	Then for any $\alpha\in (1,\infty]$ we have
	\begin{align}\label{eq:qes eat}
		\begin{gathered}	H_\alpha^{\uparrow,f_\mathrm{full}}(S_1^n\CS_1^n|\CP_1^nE_n)_\rho\geq\sum_j\min_{\cP_1^{j-1}}\kappa_{\cP_1^{j-1}}\qquad
			\text{where}\quad \kappa_{\cP_1^{j-1}} \defvar \inf_{\nu\in\Sigma_j} H^{\uparrow, f_{|\cP_1^{j-1}}}_{\alpha}(S_j \CS_j| \CP_j E_j \widetilde{E})_{\nu},
		\end{gathered}
	\end{align}
	defining
    \begin{align}\label{eq:Sigmajdefn}
        \Sigma_j \defvar \left\{
\EATchann_j\left[\omega_{A_{j-1}E_{j-1}\widetilde{E}}\right] 
\;\middle|\;
\omega \in \dop{=}(A_{j-1}E_{j-1}\widetilde{E}) \;\suchthat\; \omega_{A_{j-1}}=\sigma_{A_{j-1}}^{(j-1)}
\right\},
    \end{align}  
    with $\widetilde{E}$ being a register of large enough dimension to serve as a purifying register for any of the $A_{j-1}E_{j-1}$ registers.
	
	Consequently, if we instead define the following ``normalized'' tradeoff function on $\CS_1^n\CP_1^n$:
	\begin{align}\label{eq:fullQESnorm}
		\hat{f}_\mathrm{full}( \cS_1^n\cP_1^n) \defvar \sum_{j=1}^n \hat{f}_{|\cP_1^{j-1}}(\cS_j\cP_j), \quad\text{where}\quad \hat{f}_{|\cP_1^{j-1}}(\cS_j\cP_j) \defvar f_{|\cP_1^{j-1}}(\cS_j\cP_j) + \kappa_{\cP_1^{j-1}},
	\end{align}
	then
	\begin{align}\label{eq:chainQESnorm}
		H^{\uparrow,\hat{f}_\mathrm{full}}_\alpha(S_1^n \CS_1^n| \CP_1^n E_n)_\rho \geq 0.
	\end{align}
\end{manualtheorem}

We highlight a subtle difference between the above result and its analogue in~\cite[Theorem~4.1]{AHT24} (which was based on the GEAT). Namely: in~\cite[Theorem~4.1]{AHT24}, the choices of tradeoff functions could depend on \emph{both} the ``secret'' and ``public'' classical values $\cS_1^{j-1}$ and $\cP_1^{j-1}$ of previous rounds, whereas in our result, they can only depend on the latter (as reflected via the notation $f_{|\cP_1^{j-1}}$). This is because in the proof we present below, we were unable to accommodate the more general dependency, due to the lack of a ``secret'' memory register as compared to the GEAT model. We comment further on this at the relevant point in the proof. 
	
\begin{proof}
	The proof is nearly identical to that of~\cite[Theorem~4.1]{AHT24}, except we invoke Theorem~\ref{thrm:channel cond ent chain rule} in place of the chain rule used in that proof. Note that all steps of the following analysis are valid even for $\alpha=\infty$, keeping in mind the conventions for how  $H^{\uparrow ,f}_\alpha$, $H^{\uparrow}_\alpha$ and $\frac{\alpha}{\alpha-1}$ are defined in that case.
    
	First, let $M>0$ be any value such that $M - f_{|\cP_1^{j-1}}(\cS_j \cP_j) > 0$ for all the $f_{|\cP_1^{j-1}}(\cS_j \cP_j)$ in the theorem statement (for all $j$). 
	Now for each $j$, define a read-and-prepare channel $\mathcal{D}_j\in\CPTP(\CS_j \CP_1^j,\CS_j \CP_1^j D_j)$ of the form described in Lemma~\ref{lemma:createD}, so that the state it prepares on $D_j$ satisfies
	\begin{align}\label{eq:Dj_entropy}
		H_\alpha(D)_{\rho_{|\cP_1^j\cS_j}} = H^\uparrow_\alpha(D)_{\rho_{|\cP_1^j\cS_j}} = M-f_{|\cP_1^{j-1}}(\cS_j\cP_j).
	\end{align}
	
	Now let $\mathcal{N}_j\in\CPTP(A_{j-1} E_{j-1}  \CP_1^{j-1},D_j S_j \CS_j E_j \CP_1^j)$ denote a channel that does the following 
	\begin{enumerate}
		\item Apply $\EATchann_j \otimes \mathcal{P}_j$, where $\mathcal{P}_j$ is a pinching channel on $ \CP_1^{j-1}$ in its classical basis, hence making the $ \CP_1^{j-1}$ registers classical.
		\item Generate a $D_j$ register by applying the above read-and-prepare channel $\mathcal{D}_j$ on $\CS_j \CP_1^j$.
	\end{enumerate}
	Note that here, in each round $j$ the tradeoff function only depends on previous public announcements $\cP_1^{j-1}$, as opposed to $\cS_1^{j-1}\cP_1^{j-1}$ in~\cite{AHT24}.\footnote{One would need to keep a copy of the registers $\CS_1^{j-1}$ in the output registers of $\mathcal{N}_j$ for the tradeoff function to be able to depend on $\cS_1^{j-1}$ as well. This was handled in~\cite{AHT24} by keeping that copy in a ``secret'' memory register; however, that work was under the framework of the GEAT~\cite{MFSR24}, which allowed having a ``secret'' memory register such that there is no signalling from it to the adversary's register (see the discussion in Sec.~\ref{sec:conclusion} later). In order to accommodate a memory register in this framework, one would need to construct a chain rule similar to the one in Theorem~\ref{thrm:channel cond ent chain rule}, but generalized to allow handling a memory register under a non-signalling condition to the adversary. Such a chain rule seems highly non-trivial to obtain, and we leave the analysis to future work.} 
	
	Observing that $\mathcal{N}_n\circ\cdots\circ\mathcal{N}_1[\omega]$ is a valid extension of the state $\rho$ in the theorem, we can use $\rho$ to refer to this state as well without any ambiguity. We now apply Theorem~\ref{thrm:channel cond ent chain rule} iteratively to this state. Specifically, by identifying the channels and registers in that theorem with those in this scenario (for each $j$) as:
	\begin{itemize}
		\item $\mathcal{E}_1 \in\CPTP(A_0Y_0,X_1Y_1) \;\leftrightarrow\; \mathcal{N}_{j-1} \circ \dots \circ \mathcal{N}_1 \in \CPTP(A_1^{j-2} E_{0},D_1^{j-1} S_1^{j-1} \CS_1^{j-1} E_{j-1} \CP_1^{j-1})$
		\item $A_0 \leftrightarrow A_0^{j-2}$
		\item $Y_0 \leftrightarrow E_0$
		\item $X_1\leftrightarrow D_1^{j-1}S_1^{j-1}\CS_1^{j-1}$,
		\item $Y_1 \leftrightarrow E_{j-1}\CP_1^{j-1}$
		\item $\mathcal{E}_2 \in\CPTP(A_1Y_1,X_2Y_2) \;\leftrightarrow\; \mathcal{N}_j \in \CPTP(A_{j-1} E_{j-1}  \CP_1^{j-1},D_j S_j \CS_j E_j \CP_1^j)$
		\item $A_1 \leftrightarrow A_{j-1}$
		\item $X_2\leftrightarrow D_jS_j\CS_j$ 
		\item $Y_2 \leftrightarrow E_j\CP_1^j$
		,
	\end{itemize}
	we have:
	\begin{align}\label{eq:GEATnotest}
		H^{\uparrow}_\alpha(D_1^n S_1^n \CS_1^n | \CP_1^n E_n)_\rho &\geq 
		H^{\uparrow}_\alpha\left(\mathcal{N}_n\circ\cdots\circ\mathcal{N}_1,D_1^n S_1^n \CS_1^n,[\omega_{A_0^{n-1}}]\right)\nonumber\\
		&\geq 				H^{\uparrow}_\alpha\left(\mathcal{N}_n,D_n S_n \CS_n,[\omega_{A_{n-1}}]\right)\nonumber\\
		&\quad + H^{\uparrow}_\alpha\left(\mathcal{N}_{n-1}\circ\cdots\circ\mathcal{N}_1,D_1^{n-1} S_1^{n-1} \CS_1^{n-1},[\omega_{A_0^{n-2}}]\right)\nonumber\\
		&\quad\vdots \nonumber\\
		&\geq \sum_j H^{\uparrow}_\alpha\left(\mathcal{N}_j,D_j S_j \CS_j,[\omega_{A_{j-1}}]\right)\nonumber\\
		&=\sum_j \inf_{\nu'\in\Sigma'_j} H_\alpha^{\uparrow}(D_j S_j \CS_j | \CP_1^j E_j \widetilde{E})_{\nu'},
	\end{align}
	where $\Sigma'_j$ denotes the set of all states that could be produced by $\mathcal{N}_j$ acting on some initial state $\omega' \in \dop{=}(A_{j-1} E_{j-1} \CS_{j-1} \CP_1^{j-1} \widetilde{E})$\footnote{In this step, let us take $\widetilde{E}$ to be of large enough dimension to be a purifying register for the input registers in the $\mathcal{N}_j$ scenario as well; this can be achieved without loss of generality by expanding its dimension as necessary.}, such that $\omega'_{A_{j-1}}=\sigma_{A_{j-1}}^{(j-1)}$.
	
	The remainder of the proof is identical to the~\cite[Theorem~4.1]{AHT24} proof, i.e.~by applying calculations similar to Eq.~\eqref{eq:D_entropyproof} above, we show that
	\begin{align}
	    H^{\uparrow}_\alpha(D_1^n S_1^n \CS_1^n | \CP_1^n E_n)_\rho &= nM + H_\alpha^{\uparrow,f_\mathrm{full}}(S_1^n\CS_1^n|\CP_1^nE_n)_\rho, \label{eq:finalfullbnd} \\ 
	    \inf_{\nu'\in\Sigma'_j} H_\alpha^{\uparrow}(D_j S_j \CS_j | \CP_1^j E_j \widetilde{E})_{\nu'} 
	    &= M + \min_{\cP_1^{j-1}} \inf_{\nu\in\Sigma_j} H^{\uparrow, f_{|\cP_1^{j-1}}}_{\alpha}(S_j \CS_j| \CP_j E_j \widetilde{E})_{\nu} \label{eq:1rndreduction},
	\end{align}
	then substitute these relations into Eq.~\eqref{eq:GEATnotest}, which yields the claimed result by cancelling $nM$ from both sides.
\end{proof}

We now turn to proving Theorem~\ref{thrm:MEAT_simp} and a slight variant of it.
We start with the definition of (Legendre-Fenchel) convex conjugates~\cite{BV04v8}:
\begin{definition}\label{def:conjugate}
	(Convex conjugate) For a function $F:D\to\mathbb{R}\cup\{-\infty,+\infty\}$ where $D$ is a convex subset of $\mathbb{R}^k$, its \term{convex conjugate} is the function $F^*:\mathbb{R}^k\to\mathbb{R}\cup\{-\infty,+\infty\}$ given by
	\begin{align}
		F^*(\mbf{y}) \defvar \sup_{\mbf{x}\in D} \left(\mbf{x}\cdot\mbf{y} - F(\mbf{x})\right).
	\end{align}
\end{definition}

Under this definition, we have the following result.
\begin{lemma}\label{lemma:Legendre_conjugate}
	Consider any $\alpha\in (1,\infty]$ and any $\rho \in \dop{=}(QQ'\CP)$ 
	such that $\CP$ is classical with alphabet $\alphCP$; let $\mathbb{P}_{\alphCP}$ denote the set of probability distributions on $\alphCP$. Then the pair of functions $\left(G_{\alpha,\rho}
	,G_{\alpha,\rho}^*
	\right)$ on $\mathbb{R}^{|\alphCP|}$ defined as follows (under the convention that $\frac{\alpha}{\alpha-1}=1$ when $\alpha=\infty$)
	\begin{align}\label{eq:GandGstar_f}
		\begin{aligned}
			&G_{\alpha,\rho}({\mbf{f}}) \coloneqq - H^{\uparrow ,f}_\alpha(Q|\CP Q')_{\rho}, \\
			&G_{\alpha,\rho}^*(\bsym{\lambda}) \coloneqq 
			\begin{cases}
				\frac{\alpha}{\alpha-1}D\left(\bsym{\lambda} \middle\Vert \bsym{\rho}_{\CP}\right)+\sum\limits_{\cP\in\supp(\bsym{\rho}_{\CP})}\lambda(\cP)H_\alpha^\uparrow(Q|Q')_{\rho_{|\cP}} &
				\bsym{\lambda} \in \mathbb{P}_{\alphCP}
				, \\
				+\infty  & \text{else,}
			\end{cases}
		\end{aligned}
	\end{align}
	are convex functions that are convex conjugates to each other, where in the $G_{\alpha,\rho}$ formula we recall that any tuple $\mbf{f}\in\mathbb{R}^{| \alphCP|}$ is equivalent to a tradeoff function on $\CP$, and in the $G^*_{\alpha,\rho}$ formula we recall that $\bsym{\rho}_{\CP}$ denotes the distribution induced by $\rho$ on $\CP$.
\end{lemma}
\begin{proof}
Note that all steps of the following analysis are valid even for $\alpha=\infty$, keeping in mind the conventions for how  $H^{\uparrow ,f}_\alpha$, $H^{\uparrow}_\alpha$ and $\frac{\alpha}{\alpha-1}$ are defined in that case.
	First, we establish the convexity of $G_{\alpha,\rho}({\mbf{f}})$. To achieve this, let $\tilde{\mbf{f}},\widetilde{\bsym{\lambda}},\widetilde{\bsym{\rho}}_{\CP}$ denote the restrictions of ${\mbf{f}},{\bsym{\lambda}},\bsym{\rho}_{\CP}$, respectively, to the support of $\bsym{\rho}_{\CP}$, i.e., $\supp(\bsym{\rho}_{\CP})$. Furthermore, let $\widetilde{\mbf{H}}_\alpha$ denote the tuple of values $\{H_\alpha^\uparrow(Q|Q')_{\rho_{|\cP}}\}_{\cP\in\supp(\bsym{\rho}_{\CP})}$. Similar to the proof of~\cite[Lemma~5.3]{AHT24} we exploit the notion of log-sum-exponential function to rewrite $G_{\alpha,\rho}({\mbf{f}})$ as follows:
    \begin{align}\label{eq:G_to_logsumexp_up}
G_{\alpha,\rho}({\mbf{f}}) &= \frac{\alpha}{(\alpha-1)\ln2} \ln \left( \sum_{ \cP \in \supp(\bsym{\rho}_{\CP})} \rho(\cP) e^{\left(\frac{\alpha-1}{\alpha}\right)(\ln2) \left({f}(\cP) - H_\alpha^\uparrow(Q|Q')_{\rho_{|\cP}}\right) } \right) \nonumber\\
&= \frac{\alpha}{(\alpha-1)\ln2} \operatorname{logsumexp} \left( 
\ln\left(\widetilde{\bsym{\rho}}_{\CP}\right) + \left(\frac{\alpha-1}{\alpha}\right)(\ln2) \left(\tilde{\mbf{f}} - \widetilde{\mbf{H}}_\alpha \right)
\right)
,
\end{align} 
where $\ln\left(\widetilde{\bsym{\rho}}_{\CP}\right)$ is the elementwise logarithm of $\widetilde{\bsym{\rho}}_{\CP}$. Convexity follows by noting that $\alpha > 1$, the log-sum-exponential function is convex~\cite{BV04v8}, 
and $G_{\alpha,\rho}(\mbf{f})$ is 
constant with respect to any $\mbf{f}$ components outside of $\tilde{\mbf{f}}$.
	
	Since $G_{\alpha,\rho}$ is convex and finite over the entire set $\mathbb{R}^{|\alphCP|}$, taking the convex conjugate twice returns the original function~\cite{BV04v8}, so to prove the claimed result, it suffices to show its convex conjugate has the formula in~\eqref{eq:GandGstar_f}.
	The convex conjugate of $G_{\alpha,\rho}(\mbf{f})$ is, by Definition~\ref{def:conjugate},
	\begin{align}
		\label{eq:conjugate_G}
		G_{\alpha,\rho}^*(\bsym{\lambda})=\sup_{\mbf{f}\in\mathbb{R}^{| \alphCP|}}\left(\bsym{\lambda}\cdot\mbf{f}-G_{\alpha,\rho}(\mbf{f})\right).
	\end{align}
	Note that if $\bsym{\lambda}$ is such that $\supp(\bsym{\lambda})\nsubseteq\supp(\bsym{\rho}_{\CP})$, the optimization is $+\infty$. This is because for those values of $\cP$ such that $\lambda(\cP)\neq 0$
	and $\rho(\cP)= 0$,
	$G_{\alpha,\rho}(\mbf{f})$ becomes independent of the corresponding $f(\cP)$, and therefore, depending on the sign of $\lambda(\cP)$ we either set $f(\cP)\to\infty$ or $f(\cP)\to -\infty$ to make the objective diverge to $+\infty$.
	This behavior matches the formula given in Eq.~\eqref{eq:GandGstar_f}, since for those $\bsym{\lambda}$ that $\supp(\bsym{\lambda})\nsubseteq\supp(\bsym{\rho}_{\CP})$, either $\bsym{\lambda}\notin\mathbb{P}_{\CP}$, 
	in which case the optimization yields $+\infty$, which is reflected in the formula in~\eqref{eq:GandGstar_f}, or $\bsym{\lambda}\in\mathbb{P}_{\CP}$, in which case $D(\bsym{\lambda}\Vert\bsym{\rho}_{\CP})$ becomes $+\infty$. 
	
	Thus, we only focus on the case where $\supp(\bsym{\lambda})\subseteq\supp(\bsym{\rho}_{\CP})$. Hence, for such $\bsym{\lambda}$, the optimization for $G_{\alpha,\rho}^*$ reduces to
	\begin{align}\label{eq:simplified_conjugate_f}
		G_{\alpha,\rho}^*(\bsym{\lambda}) &= \sup_{\mbf{f}\in\mathbb{R}^{|\alphCP|}}
		\left(\widetilde{\bsym{\lambda}}\cdot\tilde{\mbf{f}}-G_{\alpha,\rho}(\tilde{\mbf{f}})\right) \nonumber\\
		&= \sup_{\mbf{g}\in\mathbb{R}^{\left|\supp(\bsym{\rho}_{\CP})\right|}}\left(\widetilde{\bsym{\lambda}}\cdot\mbf{g}-G_{\alpha,\rho}(\mbf{g})\right),
	\end{align}
	
	Since the above optimization is concave, any stationary point would be a global maximum. We differentiate the objective to find that such a stationary point would have to satisfy:
	\begin{align}\label{eq:stationary_f}
		\forall \cP\in\supp(\bsym{\rho}_{\CP}), \quad 
		\lambda(\cP) &= 
		\frac{\rho(\cP)e^{\left(\frac{\alpha-1}{\alpha}\right)(\ln 2)\left(g(\cP)-H_\alpha^{\uparrow}(Q|Q')_{\rho_{|\cP}}\right)}}{ \sum_{\cP\in\supp(\bsym{\rho}_{\cP})}\rho(\cP)e^{\left(\frac{\alpha-1}{\alpha}\right)(\ln 2)\left(g(\cP)-H_\alpha^{\uparrow}(Q|Q')_{\rho_{|\cP}}\right)}}.
	\end{align}
	We first note that the above equations can only have a solution if $\sum_{\cP}\lambda(\cP)=1$ (this can be seen by summing the equations over $\cP$). Therefore, let us first focus on the case where $\bsym{\lambda}\in\mathbb{P}_{\CP}$ and has support equal to $\supp(\bsym{\rho}_{\CP})$. In this case,
	$\log{\lambda(\cP)}$ and $\log\rho(\cP)$ are finite for all $\cP\in\supp(\bsym{\rho}_{\CP})$, which allows us to take the log on both sides of Eq.~\eqref{eq:stationary_f} and show that a valid solution exists.
	Evaluating the objective function at that solution yields
	\begin{align}
		\label{eq:conjugate_function_f}
		G_{\alpha,\rho}^*(\bsym{\lambda})=\frac{\alpha}{\alpha-1}D\left(\bsym{\lambda} \middle\Vert \bsym{\rho}_{\CP}\right)+\sum_{\cP\in\supp(\bsym{\rho}_{\CP})}\lambda(\cP)H_\alpha^{\uparrow}(Q|Q')_{\rho_{|\cP}},
	\end{align}
	where the first term is the KL divergence.
	Other remaining cases 
	can be handled similarly to the proof of~\cite[Lemma~5.3]{AHT24}. 
\end{proof}
We use this to derive the following convex duality result.
\newcommand{\margdom}{\Theta}
\begin{lemma}\label{lemma:convduality}
Let $\EATchann\in\CPTP(Q,SE\CP)$ where the output is always classical on $\CP$, let $\widetilde{E}$ be a register such that $\dim(Q) \leq \dim(\widetilde{E})$, and let $\pf$ be a purifying function for $Q$ onto $\widetilde{E}$ (Definition~\ref{def:purify}). 
Let $\mathbb{P}_{\CP}$ denote the set of probability distributions on $\CP$, and consider any $S_\Omega \subseteq \mathbb{P}_{\CP}$ and any $\margdom \subseteq \dop{=}(Q)$. For any $\omega\in\dop{=}(Q)$, define a corresponding state $\nu^\omega \defvar \EATchann\left[ \pf\left(\omega\right)\right]$. Then for any $\alpha\in(1,\infty]$,
the optimization 
\begin{align}\label{eq:dualf}
\sup_{\mbf{f}} \inf_{\mbf{q} \in S_\Omega}
\inf_{\omega\in\margdom}
\left(H^{\uparrow, f}_{\alpha}(S | \CP E\widetilde{E})_{\nu^\omega} + \mbf{f}\cdot\mbf{q} \right),
\end{align}
is the Lagrange dual problem\footnote{In this lemma, Lagrange dual problems are defined in the sense described in e.g.~\cite{BV04v8}. Since the constraints are equality constraints, there is an arbitrary sign convention to pick when defining the dual variables; see Eq.~\eqref{eq:dualfrewrite} for the sign convention we used.} of the following constrained optimization (under the convention that $\frac{\alpha}{\alpha-1}=1$ when $\alpha=\infty$):
\begin{align}\label{eq:primalomega}
\begin{gathered}
\inf_{\mbf{q} \in S_\Omega} \inf_{\bsym{\lambda} \in \mathbb{P}_{\CP}} \inf_{\omega\in\margdom}
\left( \frac{\alpha}{{\alpha}-1}D\left(\bsym{\lambda} \middle\Vert \bsym{\nu}^\omega_{\CP}\right)+\sum_{\cP\in\supp(\bsym{\nu}^\omega_{\CP})}\lambda(\cP)H^\uparrow_{{\alpha}}(S|E\widetilde{E})_{\nu^\omega_{|\cP}}  \right) \\
\suchthat\quad \mbf{q}-\bsym{\lambda} = \mbf{0}
\end{gathered}
,
\end{align}
in which the objective function is jointly convex in $\bsym{\lambda}$ and $\omega$ (and $\mbf{q}$, trivially).

Furthermore, if $S_\Omega$ is convex and $\margdom$ is convex and compact, the values of~\eqref{eq:dualf} and~\eqref{eq:primalomega} are equal, i.e.~strong duality holds.
\end{lemma}
\begin{proof}
Note that all steps of the following analysis are valid even for $\alpha=\infty$, keeping in mind the conventions for how  $H^{\uparrow ,f}_\alpha$, $H^{\uparrow}_\alpha$ and $\frac{\alpha}{\alpha-1}$ are defined in that case.
The proof is identical to the proof of~\cite[Lemma~5.4]{AHT24}, except with the convex-conjugate result in Lemma~\ref{lemma:Legendre_conjugate} instead: basically, for $G_{\alpha,\nu^\omega}$ as defined in Lemma~\ref{lemma:Legendre_conjugate}, one rewrites~\eqref{eq:dualf} as  
\begin{align}\label{eq:dualfrewrite}
\sup_{\mbf{f}} \inf_{\mbf{q} \in S_\Omega} \inf_{\bsym{\lambda} \in \mathbb{P}_{\CP}} \inf_{\omega \in \margdom}  \left(G^*_{\alpha,\nu^\omega}(\bsym{\lambda}) + \left(\mbf{q}-\bsym{\lambda}\right)\cdot{\mbf{f}} \right),
\end{align}
which (after substituting the formula for $G^*_{\alpha,\nu^\omega}$) is exactly the dual problem of~\eqref{eq:primalomega}, by the definition of Lagrange duals~\cite{BV04v8}. Strong duality is then proven using the Clark-Duffin condition~\cite{Duff78}; note that for the proof to carry through the same way, one would use the fact that Lemma~\ref{lemma:convexity} ensures $H^{\uparrow, f}_{\alpha}(S | \CP E\widetilde{E})_{\nu^\omega} $ is convex in $\omega$, as well as the convexity and/or compactness conditions on $S_\Omega$ and $\margdom$. {In fact, using the background assumption in this work that all systems are finite-dimensional, it may not be strictly necessary to impose the condition that $\margdom$ is compact --- the proof of~\cite[Lemma~5.4]{AHT24} involves a continuity argument in which $S_\Omega$ is replaced with its closure, and a similar argument should apply to $\margdom$ if it is not compact, as long as the space is finite-dimensional. We leave this technical point to be resolved in future work if necessary.}
\end{proof}

Aided by the above lemmas, we can obtain the following bound on {\Renyi} conditional entropy conditioned on an event. 
We stress again that, in the following theorem, all registers are taken to be finite-dimensional.
\begin{manualtheorem}{4.2b}
(Marginal-constrained entropy accumulation theorem)
\label{thrm:MEAT}
For each $j\in\{1,2,\cdots,n\}$, take a state $\sigma^{(j-1)}\in\dop{=}(A_{j-1})$, and a channel $\mathcal{M}_j\in\CPTP(A_{j-1}E_{j-1},S_j\CS_j\CP_jE_j)$, such that $\CS_j\CP_j$ are classical. Let $\rho$ be a state of the form $\rho_{S_1^n\CS_1^n\CP_1^nE_n}=\mathcal{M}_n\circ\cdots\circ\mathcal{M}_1[\omega_{A_0^{n-1}E_0}]$ for some $\omega\in\dop{=}(A_0^{n-1}E_0)$, such that $\omega_{A_0^{n-1}}=\sigma_{A_0}^{(0)}\otimes\cdots\otimes\sigma_{A_{n-1}}^{(n-1)}$. 

Suppose furthermore that $\rho = p_\Omega \rho_{|\Omega} + (1-p_\Omega) \rho_{|\overline{\Omega}}$ for some $p_\Omega \in (0,1]$ and normalized states $\rho_{|\Omega},\rho_{|\overline{\Omega}}$, and that all the $\CS_j$ (resp.~$\CP_j$) registers are isomorphic to a single register $\CS$ (resp.~$\CP$) with alphabet $\alphCS$ (resp.~$\alphCP$).
Let $S_\Omega$ be a convex set of probability distributions on the alphabet $\alphCS \times \alphCP$, such that for all $\cS_1^n \cP_1^n$ with nonzero probability in $\rho_{|\Omega}$, the frequency distribution $\freq_{\cS_1^n \cP_1^n}$ lies in $S_\Omega$. 
Then, for any $\alpha\in(1,\infty]$, we have (under the convention that $\frac{\alpha}{\alpha-1}=1$ and $\frac{1}{\alpha-1}=0$ when $\alpha=\infty$):
\begin{align}\label{eq:fweightedREATworst}
H^\uparrow_\alpha(S_1^n \CS_1^n | \CP_1^n E_n)_{\rho_{|\Omega}} \geq  n  h_{\alpha}
- \frac{\alpha}{\alpha-1} \log\frac{1}{p_\Omega}, 
\end{align}
where (recalling $\bsym{\nu}_{\CS\CP}$ denotes the distribution on $\CS\CP$ induced by any state $\nu_{\CS\CP}$)
\begin{align}\label{eq:halphaMEAT}
\quad h_{\alpha} &= 
\inf_{\mbf{q} \in S_\Omega} \inf_{\nu\in\Sigma_{S\CS \CP E\widetilde{E}}} \left( \frac{1}{{\alpha}-1}D\left(\mbf{q} \middle\Vert \bsym{\nu}_{\CS\CP}\right)-\sum_{\cS\cP\in \supp(\bsym{\nu}_{\CS\CP})}q(\cS\cP)D_{\alpha}\left(\nu_{SE\widetilde{E} \land \cS\cP} \middle\Vert \id_S\otimes\nu_{E\widetilde{E} \land \cP}\right)\right) 
,
\end{align}
with $\Sigma_{S\CS \CP E\widetilde{E}}$ being the marginal-constrained convex range (Definition~\ref{def:range}) of the channels $\EATchann_j \otimes \idmap_{\widetilde{E}}$ and states $\sigma^{(j-1)}$, 
where $\widetilde{E}$ is a register of large enough dimension to serve as a purifying register for any of the $A_{j-1}E_{j-1}$ registers. 
\end{manualtheorem}

We now present the proofs of Theorem~\ref{thrm:MEAT_simp} and Theorem~\ref{thrm:MEAT} together, as their structures are extremely similar. Inconveniently, the former is not literally a special case of the latter, because our proof of the latter currently involves a relaxation from $H^{\uparrow,f}_\alpha$ to $H^f_\alpha$ that is not present in the former. Still, given that the proof structures are nearly identical, we choose to present them together in this fashion.

\begin{proof}
Note that all steps of the following analysis are valid even for $\alpha=\infty$ unless otherwise stated, keeping in mind the conventions for how  $H^{\uparrow ,f}_\alpha$, $H^{\uparrow}_\alpha$ and $\frac{\alpha}{\alpha-1}$ are defined in that case.
The proofs are basically the same as in the proof of \cite[Theorem~5.1]{AHT24}. 
To begin, consider the same channels $\mathcal{N}_j$ and state as in the Theorem~\ref{thrm:qes eat} proof above. Then  we have an analogous result to~\cite[Corollary~4.1]{AHT24} (letting $\widetilde{\Omega}$ be the set of all values $\cS_1^n \cP_1^n$ with nonzero probability in $\rho_{|\Omega}$):
\begin{align}\label{eq:bndfromaccept}
H^{\uparrow}_\alpha(S_1^n \CS_1^n | \CP_1^n E_n)_{\rho_{|\Omega}} &\geq H^{\uparrow}_\alpha(D_1^n S_1^n \CS_1^n | \CP_1^n E_n)_{\rho_{|\Omega}} - \max_{\cS_1^n \cP_1^n \in \widetilde{\Omega}} H_\alpha(D_1^n)_{\rho_{|\cS_1^n \cP_1^n}} \nonumber\\
&\geq H^{\uparrow}_\alpha(D_1^n S_1^n \CS_1^n | \CP_1^n E_n)_{\rho} - \max_{\cS_1^n \cP_1^n \in \widetilde{\Omega}} H_\alpha(D_1^n)_{\rho_{|\cS_1^n \cP_1^n}} - \frac{\alpha}{\alpha-1} \log\frac{1}{p_\Omega} \nonumber\\
&= H^{\uparrow}_\alpha(D_1^n S_1^n \CS_1^n | \CP_1^n E_n)_{\rho} - nM + \min_{\cS_1^n \cP_1^n \in \widetilde{\Omega}} f_\mathrm{full}(\cS_1^n\cP_1^n) - \frac{\alpha}{\alpha-1} \log\frac{1}{p_\Omega}\nonumber\\
&\geq 
\left(\sum_j  \inf_{\nu\in\Sigma_j} H^{\uparrow, f_{|\cP_1^{j-1}}}_{\alpha}(S_j \CS_j| \CP_j E_j \widetilde{E})_{\nu}\right) + \min_{\cS_1^n \cP_1^n \in \widetilde{\Omega}} f_\mathrm{full}(\cS_1^n\cP_1^n) - \frac{\alpha}{\alpha-1} \log\frac{1}{p_\Omega}
,
\end{align}
where the first line is~\cite
[Lemma~4.9]
{AHT24}, the second line is~\cite[Lemma~B.5]{DFR20}, the third line is~\eqref{eq:D_entropy}, and the fourth line is by~\eqref{eq:GEATnotest} and~\eqref{eq:1rndreduction}. (While the former two results were not explicitly stated for $\alpha=\infty$, they can be immediately seen to hold for that case by taking limits on each expression, noting that the limit exists for each individual $H_\alpha(D_1^n)_{\rho_{|\cS_1^n \cP_1^n}}$ term in the maximum.)

We now focus on presenting the proof of Theorem~\ref{thrm:MEAT_simp} first. In that theorem, there are no $\CS_j$ registers and thus we set those registers to be trivial in Eq.~\eqref{eq:bndfromaccept} above. 
Now consider an arbitrary tradeoff function $f$ on $\CP$ (we will optimize over this choice of $f$ at the end). By setting all the tradeoff functions $f_{|\cP_1^{j-1}}$ be equal to $f$, the Eq.~\eqref{eq:bndfromaccept} bound yields
\begin{align}
H^{\uparrow}_\alpha(S_1^n | \CP_1^n E_n)_{\rho_{|\Omega}} 
&\geq 
\left(\sum_j  \inf_{\nu\in\Sigma_j} H^{\uparrow, f}_{\alpha}(S_j | \CP_j E_j \widetilde{E})_{\nu}\right) + \inf_{\mbf{q} \in S_\Omega} \mbf{f}\cdot\mbf{q} \, n 
- \frac{\alpha}{\alpha-1} \log\frac{1}{p_\Omega} \nonumber\\
&\geq 
\inf_{\mbf{q} \in S_\Omega}
\inf_{\nu\in\Sigma_{S\CP E\widetilde{E}}}
\left(H^{\uparrow, f}_{\alpha}(S | \CP E\widetilde{E})_{\nu} + \mbf{f}\cdot\mbf{q} \right) n 
- \frac{\alpha}{\alpha-1} \log\frac{1}{p_\Omega}
, \label{eq:samefbound}
\end{align}
where the first line rearranges the summation in $f_\mathrm{full}$ the same way as the~\cite[Corollary~5.1]{AHT24} proof, and the second line holds since the marginal-constrained convex range $\Sigma_{S\CP E\widetilde{E}}$ contains every state $\nu\in\Sigma_j$. 

\newcommand{\margdomAEJ}{\Theta_{AEJ}}
We now reparametrize the optimization over $\Sigma_{S\CP E\widetilde{E}}$ in terms of a specifically constructed channel $\EATchann$. Embed all the registers $A_{j-1}$ (resp.~$ E_{j-1}$) in some register $A$ (resp.~$ E$) of sufficiently large dimension, and extend all the channels $\EATchann_j$ to act on $\dop{=}(AE)$ (rather than the original domains $\dop{=}(A_{j-1}E_{j-1})$) in some arbitrary fashion. Now let $J$ be a register of dimension $n$, and define $\EATchann \in \CPTP(AEJ,S\CP E)$ to be a channel that does the following:\footnote{A slight variant of this construction would be to instead define $AE$ to be the direct sum of all the $A_{j-1}E_{j-1}$ registers, and instead define $\EATchann \in \CPTP(AE,S\CP E)$ to be a channel that first performs a measurement with projectors $\{P_j\}$ where $P_j$ is the projector onto the $A_{j-1}E_{j-1}$ subspace --- this approach was used in~\cite{AHT24} at some points. However, here we find the explanation is simpler if we introduce an explicit $J$ register to ``label'' the blocks.}
\begin{enumerate}
\item Measure $J$ in some basis $\{\ket{j}\}$, which we shall henceforth refer to as its classical basis (even for states that are not classical on $J$).
\item Conditioned on outcome value $j$, implement the channel $\EATchann_j$ on the registers $AE$, producing an output state on $S\CP E$ (via the embedding chosen in the definition of $\Sigma_{S\CP E\widetilde{E}}$).
\end{enumerate}
Furthermore, we define the following set, which we note is convex and compact: 
\begin{align}
\margdomAEJ \defvar \left\{\omega \in \dop{=}(AEJ)
\;\middle|\;
\omega \text{ is classical on $J$, with $\omega_{AE|j} \in \dop{=}(A_{j-1}E_{j-1})$ and $ \omega_{A|j} = \sigma^{(j-1)}_{A_{j-1}}$
}
\right\},
\end{align}
where the conditions on $\omega_{|j}$ are interpreted with respect to the embeddings of $A_{j-1}E_{j-1}$ in $AE$. 

This channel and set have the following critical property:
\begin{align}\label{eq:domainrewrite}
\Sigma_{S\CP E\widetilde{E}} = \left\{
\EATchann\left[\omega_{AEJ \widetilde{E}}\right]
\;\middle|\;
\omega \in \dop{=}(AEJ\widetilde{E})
\;\suchthat\;
\omega_{AEJ} \in \margdomAEJ
\right\},
\end{align}
in other words, $\Sigma_{S\CP E\widetilde{E}}$ is \emph{exactly equal to} the set of all states produced by $\EATchann$ acting on some extension (onto $\widetilde{E}$) of a state in $\margdomAEJ$. To see this, we first prove the $\subseteq$ direction, which is straightforward: notice every state in $\Sigma_{S\CP E\widetilde{E}}$ can be produced by having $\EATchann$ act on a state of the form $\omega_{AEJ\widetilde{E}} = \sum_j p_j \omega_{A_{j-1}E_{j-1}\widetilde{E}|j} \otimes \pure{j}_J$ for some probabilities $p_j$ and states satisfying $ {\omega}_{A|j} = \sigma^{(j-1)}_{A_{j-1}}$, so this input state indeed satisfies $\omega_{AEJ} \in \margdomAEJ$. As for the $\supseteq$ direction, note that the first operation performed by $\EATchann$ is to measure $J$ in its classical basis, which will collapse \emph{any} input state $\omega_{AEJ\widetilde{E}}$ into a mixture of the form $\overline{\omega}_{AEJ\widetilde{E}} = \sum_j p_j \overline{\omega}_{AE\widetilde{E}|j} \otimes \pure{j}_J$ for some probabilities $p_j$ and normalized conditional states $\overline{\omega}_{AE\widetilde{E}|j}$. Furthermore, given that the input state satisfies $\omega_{AEJ} \in \margdomAEJ$, by definition this means\footnote{Note that we are not claiming that the \emph{overall} input state $\omega_{AEJ\widetilde{E}}$ is classical in $J$, as that may not be the case.} $\omega_{AEJ}$ is classical on $J$, thus it is unchanged by the measurement and we have $\overline{\omega}_{AEJ} = \omega_{AEJ} \in \margdomAEJ$. This implies $\overline{\omega}_{AE|j} \in \dop{=}(A_{j-1}E_{j-1})$ and $ \overline{\omega}_{A|j} = \sigma^{(j-1)}_{A_{j-1}}$, thus the remaining action of $\EATchann$ (applying $\EATchann_j$ conditioned on $j$) produces a mixture of states $\EATchann_j\left[\overline{\omega}_{A_{j-1}E_{j-1}\widetilde{E}|j}\right]$ for some $\overline{\omega}_{A_{j-1}E_{j-1}\widetilde{E}|j}$ satisfying the marginal constraints, as desired. 

This lets us \emph{exactly} reparametrize any optimization over $\Sigma_{S\CP E\widetilde{E}}$ into an optimization defined via $\EATchann$ and $\margdomAEJ$. In particular, the optimization in~\eqref{eq:samefbound} can be rewritten as follows:
\begin{align}
\inf_{\mbf{q} \in S_\Omega}
\inf_{\nu\in\Sigma_{S\CP E\widetilde{E}}}
\left(H^{\uparrow, f}_{\alpha}(S  | \CP E\widetilde{E})_{\nu} + \mbf{f}\cdot\mbf{q} \right) &= 
\inf_{\mbf{q} \in S_\Omega}
\inf_{\substack{\omega\in\dop{=}(AEJ\widetilde{E})\\ \suchthat\ \omega_{AEJ}\in\margdomAEJ}}
\left(H^{\uparrow, f}_{\alpha}(S  | \CP E\widetilde{E})_{\EATchann[\omega_{AEJ\widetilde{E}}]} + \mbf{f}\cdot\mbf{q} \right)
 \nonumber\\
&=\inf_{\mbf{q} \in S_\Omega}
\inf_{\omega\in\margdomAEJ}
\left(H^{\uparrow, f}_{\alpha}(S  | \CP E\widetilde{E})_{\nu^\omega} + \mbf{f}\cdot\mbf{q} \right),
\label{eq:optrewrite}
\end{align}
introducing the notation $\nu^\omega \defvar \EATchann\left[ \pf\left(\omega_{AEJ}\right)\right]$ for some arbitrary choice of purifying function $\pf$ (Definition~\ref{def:purify}) for $AEJ$ onto $\widetilde{E}$, expanding the dimension of $\widetilde{E}$ if necessary to purify $AEJ$.
In the above, the first line holds by~\eqref{eq:domainrewrite}, and the second line holds since $H^{\uparrow, f}_{\alpha}(S  | \CP E\widetilde{E})$ satisfies data-processing on $\widetilde{E}$ (Lemma~\ref{lemma:DPI}).
Note also that by Lemma~\ref{lemma:convexity},  the objective function in the last line is jointly convex in $\mbf{q}$ and $\omega$ (since $\alpha>1$).

Finally, recall that the above bound holds for \emph{any} choice of the tradeoff function $\mbf{f}$, and therefore the bound can be optimized by maximizing over $\mbf{f}$. However, maximizing over $\mbf{f}$ simply gives exactly the optimization~\eqref{eq:dualf} in Lemma~\ref{lemma:convduality} (setting $\margdom=\margdomAEJ$), which according to that lemma statement has the same value as~\eqref{eq:primalomega}, since the lemma conditions are satisfied.
From this, our claimed result in Theorem~\ref{thrm:MEAT_simp} follows immediately by observing the value of~\eqref{eq:primalomega} is equal to $h^\uparrow_{\alpha}$, since we can substitute the $\mbf{q}-\bsym{\lambda} = \mbf{0}$ constraint into the objective, and use the same reasoning as in~\eqref{eq:optrewrite} to again reparametrize the optimization over input states $\omega\in\margdomAEJ$ into an optimization over $\Sigma_{S\CP E\widetilde{E}}$ (noting that $H^\uparrow_{{\alpha}}(S|E\widetilde{E})$ satisfies data-processing on $\widetilde{E}$).  

We now turn to Theorem~\ref{thrm:MEAT}. Here however, we first restrict to $\alpha\in(1,\infty)$ because some properties of $H_\infty$ are slightly less convenient than $H^{\uparrow}_\infty$, along the lines of Remark~\ref{rem:classmixlimits}.
We start by using Lemma~\ref{lemma:QES_up-to-down} to slightly relax the last line of the Eq.~\eqref{eq:bndfromaccept} bound from $H^{\uparrow,f}_\alpha$ to $H^f_\alpha$: 
\begin{align}
H^{\uparrow}_\alpha(S_1^n \CS_1^n | \CP_1^n E_n)_{\rho_{|\Omega}} &\geq 
\left(\sum_j  \inf_{\nu\in\Sigma_j} H^{f_{|\cP_1^{j-1}}}_{\alpha}(S_j \CS_j| \CP_j E_j \widetilde{E})_{\nu}\right) + \min_{\cS_1^n \cP_1^n \in \widetilde{\Omega}} f_\mathrm{full}(\cS_1^n\cP_1^n) - \frac{\alpha}{\alpha-1} \log\frac{1}{p_\Omega}
.
\end{align}
The remainder of the proof then proceeds in exactly analogous fashion, except based on the convexity and duality properties of $H^f_{\alpha}$ that were established in~\cite{AHT24}, rather than those of $H^{\uparrow,f}_\alpha$ here --- for instance, use~\cite[Lemma~5.3]{AHT24} in place of Lemma~\ref{lemma:Legendre_conjugate} here. The main obstruction to making this proof work directly with $H^{\uparrow,f}_\alpha$ is that Lemma~\ref{lemma:Legendre_conjugate} does not seem straightforward to generalize to the case where the $\CS_j$ registers are nontrivial, as it seems unclear how to show $-H^{\uparrow ,f}_\alpha(Q \CS|\CP Q')_{\rho}$ is convex in $\mbf{f}$ (the infima in the definition are in the ``wrong direction'' to preserve convexity).
Finally, to extend to $\alpha=\infty$, we observe that $h_\alpha$ is a nonincreasing function of $\alpha$, therefore
\begin{align}
\lim_{\alpha\to\infty} h_\alpha = \inf_{\alpha\in(1,\infty)} h_\alpha,
\end{align}
and this infimum over $\alpha$ commutes with the various infima in~\eqref{eq:halphaMEAT}, yielding the claimed result. (In principle we could also have used such a limiting argument in the proof of Theorem~\ref{thrm:MEAT_simp}; however, we find it potentially useful to know that the intermediate steps also directly hold for $\alpha=\infty$.)
\end{proof}

\begin{remark}
Observe that from the use of Lemma~\ref{lemma:convduality} in the above proof, one obtains the fact that the objective function in~\eqref{eq:primalomega} is jointly convex in $(\mbf{q},\bsym{\lambda},\omega)$ (therefore also jointly convex in just $(\mbf{q},\omega)$ after substituting the constraint). 
Hence in principle this may provide a way to compute $h^\uparrow_{\alpha}$ via a convex optimization. However, the channel $\EATchann$ constructed here is somewhat elaborate, so in practice it may be easier to instead find a different convex reformulation of the optimization for $h^\uparrow_{\alpha}$. Note that any relaxation of that optimization from $\Sigma_{S\CP E\widetilde{E}}$ to some larger set would of course also yield a valid lower bound, hence any suitable such relaxation could be used if it is easier to analyze.

Furthermore, one also obtains the fact that the optimal solution of the dual problem~\eqref{eq:dualf} is precisely the optimal choice of $\mbf{f}$ to use in~\eqref{eq:samefbound}. Hence this should also provide a method to find choices of $\mbf{f}$ that yield good keyrates when applying Theorems~\ref{thrm:qes eat_simp} and \ref{thrm:qes eat}, by extracting the dual solution when solving~\eqref{eq:primalomega}. If that optimization is difficult to solve directly, as mentioned above, one could instead solve some suitable reformulation or relaxation of it, which would presumably find an approximately optimal $\mbf{f}$. We highlight that failing to find the optimal $\mbf{f}$ in such a procedure would only mean the resulting keyrates are suboptimal, rather than obtaining an ``insecure'' result (as long as the $\kappa_{\cP_1^{j-1}}$ values are securely computed or lower-bounded) --- this is because Theorems~\ref{thrm:qes eat_simp} and \ref{thrm:qes eat} are valid for \emph{any} choice of tradeoff functions.
\end{remark}

For the sake of compatibility with previous work, we also present here a statement explicitly formulated to be similar to the EAT version in~\cite{LLR+21}. In particular, it implies that any bounds derived under the previous EAT or GEAT models (which were of the form shown in Eq.~\eqref{eq:previousbound} below, or worse) are also valid under the model we consider in this work.

\newcommand{\fmin}{f_\mathrm{min}}
\newcommand{\CvsQ}{\zeta}
\newcommand{\Max}{\operatorname{Max}}
\newcommand{\Min}{\operatorname{Min}}

\begin{corollary}\label{cor:oldbound}
Consider the same conditions and notation as in Theorem~\ref{thrm:MEAT}. Let $\fmin$ be\footnote{This function was usually referred to as a \term{min-tradeoff function} in previous work; however, it should not be confused with the notion of tradeoff functions used throughout the rest of this work.} an affine function of probability distributions on $\CS\CP$, with the following property for all $j$:
\begin{align}
\forall \nu \in \Sigma_j , \quad H(S_j\CS_j | \CP_jE_j \widetilde{E})_\nu \geq \fmin(\bsym{\nu}_{\CS_j\CP_j}),
\end{align}
where $\Sigma_j$ is defined as in~\eqref{eq:Sigmajdefn} (i.e.~the set of output states of $\EATchann_j$ under a marginal constraint and with a purifying register on the input state). 
Let  
$h_\mathrm{vN}$ be a value satisfying $\fmin(\freq_{\cS_1^n \cP_1^n}) \geq h_\mathrm{vN}$ for all $\cS_1^n \cP_1^n$ with nonzero probability in $\rho_{|\Omega}$.
Then for $\alpha \in (1,2)$ and $\widehat{\alpha}=1/(2-\alpha)$, we have
\begin{align}\label{eq:previousbound}
\begin{aligned}
H^\uparrow_\alpha(S_1^n \CS_1^n | \CP_1^n E_n)_{\rho_{|\Omega}} &\geq  n h_\mathrm{vN} + n T_\alpha(\fmin) - n (\alpha-1)^2 K_\alpha(\fmin) - \frac{\alpha}{\alpha-1} \log\frac{1}{p_\Omega}, \\
H^\uparrow_\alpha(S_1^n \CS_1^n | \CP_1^n E_n)_{\rho_{|\Omega}} &\geq  n h_\mathrm{vN} + n T_{\widehat{\alpha}}(\fmin) - n (\widehat{\alpha}-1)^2 K_{\widehat{\alpha}}(\fmin) - \frac{\alpha}{\alpha-1} \log\frac{1}{p_\Omega},
\end{aligned}
\end{align}
where (letting $\CvsQ \defvar 1$ if $S$ is always classical and $\CvsQ \defvar 2$ otherwise, and letting $\delta_{\cS\cP}$ denote the distribution with all its weight on the value $\cS\cP$):
\begin{align}
\begin{aligned}
T_\alpha(\fmin) &\defvar \min_j \inf_{\nu\in\Sigma_j}\left( 
H(S_j\CS_j | \CP_jE_j \widetilde{E})_\nu - \fmin(\bsym{\nu}_{\CS\CP}) - (\alpha-1) V(\fmin,\bsym{\nu}_{\CS\CP})
\right), \\ 
V(\fmin,\mbf{p}) &\defvar \frac{\ln 2}{2} \left(\log\left(1+2 d_{SC}^{\CvsQ}\right) + \sqrt{2+\operatorname{Var}(\fmin,\mbf{p})} \right)^2 , \text{ where } d_{SC}\defvar \max_j \dim(S_j \CS_j), \\
\operatorname{Var}(\fmin,\mbf{p}) &\defvar \sum_{\cS\cP} p(\cS\cP) \left(\fmin(\delta_{\cS\cP}) - \sum_{\cS'\cP'} p(\cS'\cP') \fmin(\delta_{\cS'\cP'}) \right)^2 , \\
K_\alpha(\fmin) &\defvar
\frac{2^{(\alpha-1)(\CvsQ \log d_{SC} + \Max(\fmin)-\Min_{\mathcal{Q}}(\fmin))} }{6(2-\alpha)^3\ln2}
\ln^3\left(2^{\CvsQ\log d_{SC} + \Max(\fmin)-\Min_{\mathcal{Q}}(\fmin)} + e^2\right), \\
\Max(\fmin) &\defvar \sup_{\mbf{p}\in\mathbb{P}_{\alphCS\alphCP}} \fmin(\mbf{p}), \qquad
\Min_{\mathcal{Q}}(\fmin) \defvar \min_j \inf_{\nu\in\Sigma_j} \fmin(\bsym{\nu}_{\CS\CP})
.
\end{aligned}
\end{align}
\end{corollary}
\begin{proof}
The proof is again along similar lines. Let $\mathcal{D}_j \in \CPTP(\CS_j\CP_j , \CS_j\CP_j D_j)$ be read-and-prepare channels as defined in the proof of~\cite[Proposition~V.3]{DF19} (and also the proof of~\cite[Supplement, Theorem~2]{LLR+21}), in which the possible entropies of the $D_j$ register depend on $\fmin$ in a quite similar fashion to how they depend on the tradeoff-function values in Lemma~\ref{lemma:createD}. Defining $\mathcal{N}_j \defvar \mathcal{D}_j \circ \EATchann_j$, we can again apply Theorem~\ref{thrm:channel cond ent chain rule} iteratively to obtain
\begin{align}
	H^{\uparrow}_\alpha(D_1^n S_1^n \CS_1^n | \CP_1^n E_n)_{\mathcal{N}_n\circ\cdots\circ\mathcal{N}_1[\omega]} &= H^{\uparrow}_\alpha(D_1^n S_1^n \CS_1^n | \CP_1^n E_n)_{\left(\mathcal{N}_n \otimes \idmap_{\CP_1^{n-1}}\right)\circ\cdots\circ\mathcal{N}_1[\omega]} \nonumber\\ 
	&\geq\sum_j H^{\uparrow}_\alpha\left(\mathcal{N}_j \otimes \idmap_{\CP_1^{j-1}} , D_j S_j \CS_j,[\omega_{A_{j-1}}]\right) \nonumber\\
	&=\sum_j H^{\uparrow}_\alpha\left(\mathcal{N}_j , D_j S_j \CS_j,[\omega_{A_{j-1}}]\right) \nonumber\\
	&= \sum_j \inf_{\nu'\in\Sigma'_j} H_\alpha^{\uparrow}(D_j S_j \CS_j | \CP_j E_j \widetilde{E})_{\nu'},
\end{align}
where $\Sigma'_j$ denotes the set of all states that could be produced by $\mathcal{N}_j$ acting on some initial state $\omega' \in \dop{=}(A_{j-1} E_{j-1} \widetilde{E})$ such that $\omega'_{A_{j-1}}=\sigma_{A_{j-1}}^{(j-1)}$.

The remainder of the proof then proceeds identically to the proof of~\cite[Supplement, Theorem~2]{LLR+21}: to summarize the key steps, first they show that for any $\nu' \in \Sigma'_j$, we have (given $\alpha\in(1,2)$):
\begin{align}\label{eq:EATlocalcont}
H^{\uparrow}_\alpha(D_j S_j \CS_j | \CP_j E_j \widetilde{E})_{\nu'} \geq H(D_j S_j \CS_j | \CP_j E_j \widetilde{E})_{\nu'} - (\alpha-1) V(\fmin,\bsym{\nu}'_{\CS\CP})
- (\alpha-1)^2 K_\alpha(\fmin) .
\end{align}
By chain rules for von Neumann entropy, we have
\begin{align}
H(D_j S_j \CS_j | \CP_j E_j \widetilde{E})_{\nu'}
&= H(S_j \CS_j | \CP_j E_j \widetilde{E})_{\nu'} + H(D_j| S_j \CS_j \CP_j E_j \widetilde{E})_{\nu'} \nonumber\\
&= H(S_j \CS_j | \CP_j E_j \widetilde{E})_{\nu'} + H(D_j| \CS_j \CP_j)_{\nu'} ,
\end{align}
where the last equality holds because we have a Markov chain structure $D_j \leftrightarrow \CS_j \CP_j \leftrightarrow S_j E_j \widetilde{E}$. The proof is then completed by relating $H^\uparrow_\alpha(D_1^n S_1^n \CS_1^n | \CP_1^n E_n)_{\mathcal{N}_n\circ\cdots\circ\mathcal{N}_1[\omega]}$ to $H^\uparrow_\alpha(S_1^n \CS_1^n | \CP_1^n E_n)_{\rho_{|\Omega}}$ in a similar way to the proofs of Theorems~\ref{thrm:qes eat}--\ref{thrm:MEAT} in this work, and relating $H(D_j| \CS_j \CP_j)_{\nu'}$ to the $\fmin$ values in a similar way to Lemma~\ref{lemma:createD} in this work. (For the second line in~\eqref{eq:previousbound}, simply lower bound $H^{\uparrow}_\alpha(D_j S_j \CS_j | \CP_j E_j \widetilde{E})_{\nu'}$ with $H^{\uparrow}_{\widehat{\alpha}}(D_j S_j \CS_j | \CP_j E_j \widetilde{E})_{\nu'}$ before applying the bound used to obtain~\eqref{eq:EATlocalcont}.)
\end{proof}

For convenience in future applications, we also explicitly state here the bounds obtained by specializing our above results to tensor products of channels. (For many security proofs, the $I_j$ registers would correspond to public announcements, as we describe further in Sec.~\ref{sec:security}.)
\begin{corollary}\label{cor:tensorcase}
For each $j\in\{1,2,\cdots,n\}$, take a state $\sigma^{(j-1)}\in\dop{=}(A_{j-1})$, and a channel $\widetilde{\EATchann}_j \in \CPTP(A_{j-1} B_{j-1} , S_j \CS_j \CP_j I_j)$, such that $\CS_j\CP_j$ are classical.
Let $\rho$ be a state of the form 
$\rho_{S_1^n \CS_1^n \CP_1^n I_1^n \widehat{E}} = \widetilde{\EATchann}_n \otimes \dots \otimes \widetilde{\EATchann}_1 [\omega^0]$ for some $\omega\in\dop{=}(A_0^{n-1}B_0^{n-1}\widehat{E})$  such that $\omega_{A_0^{n-1}}=\sigma_{A_0}^{(0)}\otimes\cdots\otimes\sigma_{A_{n-1}}^{(n-1)}$. Then the bounds in Theorems~\ref{thrm:qes eat_simp}--\ref{thrm:MEAT} and Corollary~\ref{cor:oldbound} all hold with suitable register substitutions; for instance, letting $\Sigma_j$ denote the set of all states of the form $\widetilde{\EATchann}_j\left[\omega_{A_{j-1}B_{j-1}\widetilde{E}}\right]$ for some $\omega\in\dop{=}(A_{j-1}B_{j-1}\widetilde{E})$ such that $\omega_{A_{j-1}}=\sigma_{A_{j-1}}^{(j-1)}$, in Theorem~\ref{thrm:qes eat} we instead have
\begin{align}
\begin{aligned}
H_\alpha^{\uparrow,f_\mathrm{full}}(S_1^n\CS_1^n|\CP_1^nI_1^n \widehat{E})_\rho &\geq\sum_j\min_{\cP_1^{j-1}}\kappa_{\cP_1^{j-1}}, \quad\text{where}\quad \kappa_{\cP_1^{j-1}} \defvar \inf_{\nu\in\Sigma_j} H^{\uparrow, f_{|\cP_1^{j-1}}}_{\alpha}(S_j \CS_j| \CP_j I_j \widetilde{E})_{\nu}, \\
H^{\uparrow,\hat{f}_\mathrm{full}}_\alpha(S_1^n \CS_1^n| \CP_1^n I_1^n \widehat{E})_\rho &\geq 0, \quad\text{where}\quad 
\hat{f}_\mathrm{full}( \cS_1^n\cP_1^n) \defvar \sum_{j=1}^n \left(f_{|\cP_1^{j-1}}(\cS_j\cP_j) + \kappa_{\cP_1^{j-1}}\right),
\end{aligned}
\end{align}
in Theorem~\ref{thrm:MEAT_simp} we instead have (interpreting the $\CS$ registers to be trivial)
\begin{align}
\begin{gathered}
H^\uparrow_\alpha(S_1^n | \CP_1^n I_1^n \widehat{E})_{\rho_{|\Omega}} \geq  n  h^\uparrow_{\alpha}
- \frac{\alpha}{\alpha-1} \log\frac{1}{p_\Omega}, \quad \text{where}\\
h^\uparrow_{\alpha} =
\inf_{\mbf{q} \in S_\Omega} \inf_{\nu\in\Sigma_{S \CP I E}} \left( \frac{\alpha}{{\alpha}-1}D\left(\mbf{q} \middle\Vert \bsym{\nu}_{\CP}\right)+\sum_{\cP\in\supp(\bsym{\nu}_{\CP})}q(\cP)H^\uparrow_{{\alpha}}(S|I\widetilde{E})_{\nu_{|\cP}}  \right) 
,
\end{gathered}
\end{align}
in Theorem~\ref{thrm:MEAT} we instead have
\begin{align}
\begin{gathered}
H^\uparrow_\alpha(S_1^n \CS_1^n | \CP_1^n I_1^n \widehat{E})_{\rho_{|\Omega}} \geq  n  h_{\alpha}
- \frac{\alpha}{\alpha-1} \log\frac{1}{p_\Omega}, \quad \text{where}\\
\quad h_{\alpha} = 
\inf_{\mbf{q} \in S_\Omega} \inf_{\nu\in\Sigma_{S\CS \CP I \widetilde{E}}} \left( \frac{1}{{\alpha}-1}D\left(\mbf{q} \middle\Vert \bsym{\nu}_{\CS\CP}\right)-\sum_{\cS\cP\in \supp(\bsym{\nu}_{\CS\CP})}q(\cS\cP)D_{\alpha}\left(\nu_{SI\widetilde{E} \land \cS\cP} \middle\Vert \id_S\otimes\nu_{I\widetilde{E} \land \cP}\right)\right) 
,
\end{gathered}
\end{align}
and in Corollary~\ref{cor:oldbound} we instead have
\begin{align}
\begin{aligned}
H^\uparrow_\alpha(S_1^n \CS_1^n | \CP_1^n I_1^n \widehat{E})_{\rho_{|\Omega}} &\geq  n h_\mathrm{vN} + n T_\alpha(\fmin) - n (\alpha-1)^2 K_\alpha(\fmin) - \frac{\alpha}{\alpha-1} \log\frac{1}{p_\Omega}, \\
H^\uparrow_\alpha(S_1^n \CS_1^n | \CP_1^n I_1^n \widehat{E})_{\rho_{|\Omega}} &\geq  n h_\mathrm{vN} + n T_{\widehat{\alpha}}(\fmin) - n (\widehat{\alpha}-1)^2 K_{\widehat{\alpha}}(\fmin) - \frac{\alpha}{\alpha-1} \log\frac{1}{p_\Omega},
\end{aligned}
\end{align}
where $\fmin$ is an affine function satisfying $ H(S_j\CS_j | \CP_jI_j \widetilde{E})_\nu \geq \fmin(\bsym{\nu}_{\CS_j\CP_j}) \;\forall\; \nu \in \Sigma_j $, and so on.
\end{corollary}
\begin{proof}
We simply identify the $E_j$ registers in the above results with the registers $B_j^{n-1} I_1^j \widehat{E}$ in this scenario. With this identification, we can then apply the above results to the channels $\EATchann_j \in \CPTP(A_{j-1} E_{j-1}, S_j \CS_j \CP_j E_j) \equiv \CPTP(A_{j-1} B_{j-1}^{n-1} I_1^{j-1} \widehat{E}, S_j \CS_j \CP_j B_{j}^{n-1} I_1^j \widehat{E})$
defined by $\EATchann_j \defvar \widetilde{\EATchann}_j \otimes \idmap_{B_{j}^{n-1} I_1^{j-1} \widehat{E}}$. Then observe that it is not necessary to explicitly account for the $\idmap_{B_{j}^{n-1} I_1^{j-1} \widehat{E}}$ terms in these channels when computing the single-round quantities, because all those computations are stabilized by the purifying register $\widetilde{E}$, which can be used to absorb the optimization over the input registers to the identity channel $\idmap_{B_{j}^{n-1} I_1^{j-1} \widehat{E}}$.
\end{proof}

\section{Security of PM-QKD protocols via the MEAT}
\label{sec:security}

In this section, we state how one can use the result of the previous section to obtain a bound on the amount of secure extractable key from a PM-QKD protocol. 

\begin{remark}
Unfortunately, there is an ``off-by-one'' discrepancy between the most natural indexings of Alice's registers in the QKD protocol (i.e.~$A_1^n$) and our earlier theorems (i.e.~$A_0^{n-1}$). We highlight this as a point to keep in mind to avoid confusion when applying our earlier theorems.
\end{remark}

We first note that if the protocol is such that all the measurements are completed before any public announcements are made, this can be achieved by a rather standard argument. Namely, the source-replacement technique~\cite{CLL04,FL12} is used to write the state produced in the protocol (before error correction and privacy amplification; see below) in the form $\widetilde{\EATchann}_n \otimes \dots \otimes \widetilde{\EATchann}_1 [\omega^0]$, for some measurement channels\footnote{In this notation, the $S_j \CS_j \CP_j$ registers are interpreted in the same way as our above results, while the $I_j$ registers represent single-round public announcements as described in Protocol~\ref{Prot:PM Protocol} below.} $\widetilde{\EATchann}_j \in \CPTP(A_j B_j , S_j \CS_j \CP_j I_j)$ acting on an Alice-Bob-Eve entangled state $\omega^0 \in \dop{=}(A_1^n B_1^n \widehat{E})$ satisfying a marginal constraint on $A_1^n$. (Note that \emph{no repetition-rate restrictions are required} to perform this conversion, unlike the existing GEAT-based security proofs.) With this, we can then apply Corollary~\ref{cor:tensorcase} to bound the relevant entropies for a security analysis.\footnote{As discussed in the previous section, our results are more general than~\cite{inprep_weightentropy,FHKR25} even in such a scenario, because the tradeoff function choices $f_{|\cP_1^{j-1}}$ can depend on the values $\cP_1^{j-1}$ from preceding rounds, allowing ``fully adaptive'' updating of these choices in the manner described in~\cite{ZFK20,AHT24}. (It is also more general than~\cite{inprep_weightentropy} by allowing for different channels and marginal states in different rounds, though this was also achieved in~\cite{FHKR25}.)} Note also that in our theorems, the single-round computations all incorporate the marginal constraints on Alice's single-round registers, which is often necessary to obtain nontrivial keyrates for PM-QKD under the source-replacement technique.

However, one may be interested in protocols where some announcements are made before the measurements are completed, sometimes referred to as ``on-the-fly'' announcements. This requires a somewhat more sophisticated analysis, which we now describe. We begin by describing the form of the protocols we can analyze in our framework, as Protocol~\ref{Prot:PM Protocol} below.

\begin{prot}{Prepare-and-measure protocol.}\label{Prot:PM Protocol} 

\noindent \textbf{Parameters:}

\begin{tabularx}{0.9\linewidth}{r c X}
\(n \in \mathbb{N}_0\) 			    &:& 	Total number of rounds \\
\(\{\rho^{(j)}_{X_jQ_j}\}_{j=1}^n\) 	&:& 	Classical-quantum states prepared by Alice \\
\(\{t_j^A,t_j^B,t^{\text{ann}}_j\}_{j=1}^n\) 						&:& 	Timings of various steps (see below)
\end{tabularx}

\noindent We require the times listed above to satisfy the following conditions:\footnote{
Note that these conditions are extremely minimal: the first condition is simply the fundamental physical requirement that Alice sends the state in each round before Bob receives it (and the announcements involving that round only take place afterwards, to prevent Eve attacking the state based on the basis choices), while the second condition is simply a basic time-ordering condition on the sequences of preparations, measurements, and announcements \emph{individually}, without otherwise constraining them with respect to each other. In particular, these conditions do \emph{not} limit the repetition rate as in the existing GEAT-based proofs; they are extremely minimal conditions that should hold in practice.}
\begin{itemize}
\item For each $j \in \{1,2,\dots,n\}$, we have $t^A_j<t^B_j<t^{\text{ann}}_j$. 
\item Each of the sequences $\{t^A_j\}_{j=1}^n$, $\{t^B_j\}_{j=1}^n$, and $\{t^{\text{ann}}_j\}_{j=1}^n$ is monotonically increasing.
\end{itemize}
We do not impose \emph{any} other restrictions on those timings; for instance, the sequences $\{t^A_j\}_{j=1}^n$, $\{t^B_j\}_{j=1}^n$, and $\{t^{\text{ann}}_j\}_{j=1}^n$ can be ``interleaved'' with each other in an arbitrary fashion as long as the first condition is satisfied.

\noindent \textbf{Protocol steps:}
\begin{enumerate}
\item Alice and Bob perform the following steps for every $j \in \{1,2,\dots,n\}$. \label{step:singlerounds}
\begin{enumerate}
\item \textbf{State preparation and transmission:} At time \(t_j^A\), independently for each round, Alice prepares some state \(\rho^{(j)}_{
X_jQ_j}\) where \(X_j\) is a classical register. 
She then sends out \({Q_j}\) via a public quantum channel, after which Eve may interact freely with it.

\item \textbf{Measurements:} 
At time $t^B_j$, Bob receives some quantum register from Eve and performs some POVM\footnote{Here, we shall use the formalism that if e.g.~Bob chooses randomly from multiple possible bases to measure in, all those choices are incorporated into a single POVM, with the basis choice being encoded into the POVM outcomes.} on it, storing the outcome in a register \(Y_j\). 

\item \label{step:announce} \textbf{Public announcement:} At time \(t_j^\text{ann}\), Alice and Bob publicly announce and/or communicate some information based only on the values $X_j Y_j$. We denote all such public information as a classical register $I_j$, and will suppose that they contain a ``public test data'' register $\CP_j$ for use in later steps.

\item \label{step:sift} \textbf{Raw data processing:} 
At any time after \(t_j^\text{ann}\), Alice computes a classical value \(S_j\) based only on her raw data \(X_j\) and the public announcements \(I_j\).\footnote{Examples of such processing include ``sifting'', in which Alice erases rounds where she used a different basis from Bob, or (for optical protocols) where Bob did not receive a detection.} This value \(S_j\) will later be used to generate her final key via privacy amplification. If desired, at this point Alice can also compute some ``secret test data'' register $\CS_j$ that is not announced.

\end{enumerate}
\item \label{step:classical_processing} 
Alice and Bob apply various further classical procedures such as an acceptance test or variable-length decision\footnote{In earlier work such as~\cite{rennerthesis}, these steps were usually referred to as ``parameter estimation''.} (depending on whether it is a fixed-length or variable-length protocol~\cite{PR21}), error correction, and privacy amplification; we defer details to e.g.~\cite{rennerthesis,inprep_weightentropy,AHT24} as they are not relevant to our discussion here.
\end{enumerate}
\end{prot}	

For a security proof, we would be interested in analyzing the state at the end of Step~\ref{step:singlerounds}; more specifically, bounding its $f$-weighted entropy, or the {\Renyi} entropy conditioned on various later steps accepting.
However, the structure of the above protocol does not exactly match the theorem requirements in Sec.~\ref{sec:MEAT}, so we first need to reduce the analysis to another scenario that is compatible with those theorems. 

We first consider a modification to the above protocol, where the only change compared to the original is that the steps take place at new times $\tilde{t}^A_j$, $\tilde{t}^B_j$, and $\tilde{t}^{\text{ann}}_j$ satisfying
\begin{align}
	\label{eq:intermediate-time order}
	\tilde{t}^A_1<\cdots <\tilde{t}^A_n<\tilde{t}^B_1<\tilde{t}^{\text{ann}}_1<\tilde{t}^B_2<\tilde{t}^{\text{ann}}_2<\cdots <\tilde{t}^{\text{ann}}_n,
\end{align}
which is a stricter condition than the originally listed timing conditions.
To do so, we preserve the timings of Bob's measurements (i.e.~set $\tilde{t}^B_j=t^B_j$), but move forward the timings of all of Alice's preparations to before Bob's first measurement timing $\tilde{t}^B_1$, and move forward the timings of all the announcements so that $\tilde{t}^{\text{ann}}_j$ is immediately after\footnote{More precisely, it lies between $\tilde{t}^B_j$ and $\tilde{t}^B_{j+1}$. We also note that physically speaking, the public communication might have to take place over some nonzero time interval, especially if it involves two-way interaction. However, in this modified protocol we shall consider a ``virtual'' model in which all public communication in each round happens ``instantaneously'' at time $\tilde{t}^{\text{ann}}_j$. (While this is not physically possible, this modified protocol should be considered as merely a virtual scenario such that proving security of this modified protocol implies security of the actual protocol.)} $\tilde{t}^B_j$. 
So in this protocol, at time $\tilde{t}^A_1$, Alice prepares and transmits the first state, continuing this process until $\tilde{t}^A_n$, when she sends the $n$-th state. After all states have been sent, Bob begins receiving and measuring states, starting with the first one at $\tilde{t}^B_1$. Immediately
after this measurement, at $\tilde{t}^\text{ann}_1$, Alice and Bob announce some information $I_1$ about $X_1 Y_1$.
Bob then proceeds to receive and measure a second state at $\tilde{t}^B_2$, followed by another announcement at $\tilde{t}^\text{ann}_2$, and this pattern continues until all measurements and announcements are completed according to the specified order of times in~\eqref{eq:intermediate-time order}. 

Note that we do not lose any generality by considering this modified protocol, since moving forward the timings of Alice's preparations and the public announcements does not reduce Eve's capabilities --- everything Eve could have done in Protocol~\ref{Prot:PM Protocol} remains possible in this modified version, since the only change is that she gets access to various registers earlier, in which case she can just ignore them until the time where she would have interacted with them in the original protocol.

Since the security of the modified protocol ensures the security of Protocol~\ref{Prot:PM Protocol}, we can now focus on analyzing the modified version.
In this formulation, Alice prepares all quantum states before Bob begins his measurements. 
With this in mind, we apply the source-replacement technique~\cite{CLL04,FL12}, which recasts Alice's preparation of the classical-quantum state $\bigotimes_{j=1}^n \rho^{(j)}_{X_jQ_j}$ as an exactly equivalent process where she first prepares a pure state $\bigotimes_{j=1}^n \ket{\psi^{(j)}}_{A_j Q_j}$, then performs a suitable measurement on the $A_j$ registers\footnote{These registers would include ``shield systems'' (on which Alice's measurement acts trivially) if necessary to describe the states that Alice prepares.} to produce the $X_j$ registers. 
Crucially, we now observe that in this picture, Alice's measurement on $A_j$ commutes with all other operations except the public announcements involving $X_j$, meaning it can be deferred to any time up to that point, without affecting the state produced at the end of the protocol in any way. In particular, we can suppose she simply measures her registers at the same time as Bob. With this in mind, we consider the following virtual entanglement-based protocol, which produces \emph{exactly} the same final state as the modified prepare-and-measure protocol we described above (interpreting the times to be the same as in that modified protocol).

\begin{prot}{Virtual entanglement-based protocol.}\label{Prot:PM virtual_Protocol} 

\noindent \textbf{Parameters:} 

\begin{tabularx}{0.9\linewidth}{r c X}
\(n \in \mathbb{N}_0\) 			    &:& 	Total number of rounds \\
\(\left\{\ket{\psi^{(j)}}_{A_j Q_j}\right\}_{j=1}^n\) 	&:& 	Quantum states prepared by Alice \\
\(\{\tilde{t}_j^B,\tilde{t}^{\text{ann}}_j\}_{j=1}^n\) 						&:& 	Timings of various steps (see below)
\end{tabularx}
\vspace{20pt}

\noindent \textbf{Protocol steps:}
\begin{enumerate}
\setcounter{enumi}{-1}
\item \label{step:prep} \textbf{State preparation and transmission:} 
Alice prepares a pure state $\bigotimes_{j=1}^n \ket{\psi^{(j)}}_{A_j Q_j}$. 
She then sends out all the \({Q_1^n}\) registers via a public quantum channel, after which Eve may interact freely with them.
\item For each round $j \in \{1,2,\dots,n\}$,
Alice and Bob perform the following steps.
\begin{enumerate}
\item \label{step:measurement} \textbf{Measurements:} 
At time $\tilde{t}^B_j$, Bob receives some quantum register from Eve and performs some POVM on it, storing the outcome in a register \(Y_j\). At the same time, Alice measures $A_j$ and stores the outcome in a register $X_j$.

\item \textbf{Public announcement:} At time \(\tilde{t}_j^\text{ann}\), Alice and Bob publicly announce and/or communicate some information based only on the values $X_j Y_j$. We denote all such public information as a classical register $I_j$, and will suppose that they contain a ``public test data'' register $\CP_j$ for use in later steps.

\item \textbf{Raw data processing:} 
At time\footnote{Here we have also slightly changed the timing of this step as compared to the original protocol; however, this change is rather trivial since the sifting and/or key mapping again commutes with all operations after \(\tilde{t}_j^\text{ann}\).} \(\tilde{t}_j^\text{ann}\), Alice computes a classical value \(S_j\) based only on her raw data \(X_j\) and the public announcements \(I_j\). This value \(S_j\) will later be used to generate her final key via privacy amplification. If desired, at this point Alice can also compute some ``secret test data'' register $\CS_j$ that is not announced.

\end{enumerate}
\item
Alice and Bob apply various further classical procedures such as an acceptance test or variable-length decision (depending on whether it is a fixed-length or variable-length protocol~\cite{PR21}), error correction, and privacy amplification; we defer details to e.g.~\cite{rennerthesis,inprep_weightentropy,AHT24} as they are not relevant to our discussion here.
\end{enumerate}
\end{prot}	
\begin{figure}
    \centering
    \includegraphics[width=0.9\linewidth]{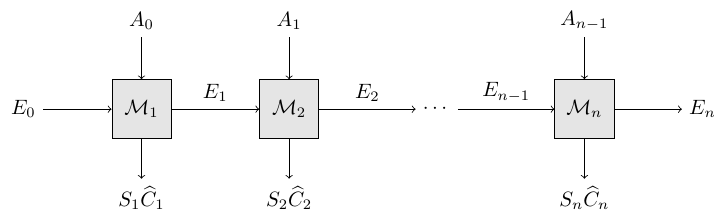}
    \caption{\markup{The setting for the marginal-constrained entropy accumulation theorem. For application in PM-QKD protocols, the critical feature to note is that each $A_j$ register (i.e.~Alice's marginal state in the source-replaced picture) can be prepared at \emph{any} time prior to being supplied to the corresponding $\mathcal{M}_j$ channel; phrased another way, the preparation of these registers commutes with the $\mathcal{M}_j$ channels. In particular, it is not necessary to delay the preparation of $A_{j}$ until after the $\mathcal{M}_{j-1}$ channel has completed, which was basically the limitation of the GEAT channel structure.}}
    \label{fig:MEAT}
\end{figure}

With this, we can now model the state produced in Protocol~\ref{Prot:PM virtual_Protocol} (at time $\tilde{t}_n^\text{ann}$) via a sequence of channels $\mathcal{M}_j$ that are compatible with the theorem conditions in Sec.~\ref{sec:MEAT}. 
Namely, let us identify the initial registers $A_1^n$ (resp.~$Q_1^n$) in the above protocol with the initial registers $A_0^{n-1}$ (resp.~$E_0$) in those theorems, in which case this is indeed an initial state satisfying a suitable marginal constraint.
We can then take each channel $\mathcal{M}_j$ as performing all the physical processes that happen in the protocol during the time interval $\left(\tilde{t}_{j-1}^\text{ann},\tilde{t}_j^\text{ann} \right]$. 
In other words, it includes Eve performing any operations on her side-information based on $I_{j-1}$ and preparing a state to send to Bob, followed by Alice and Bob performing measurements on their $j^\text{th}$ systems, announcing the public information $I_j$ (which Eve immediately incorporates into her side-information), and generating Alice's value $S_j$.
This channel structure fits the conditions of our above theorems, and they can hence be used to analyze this state.\footnote{Strictly speaking, for full compatibility, each of these channels should trace out protocol registers such as $Y_j$ that cannot be incorporated into the output registers $S_j\CS_j\CP_jE_j$ in the theorem statements; this does not cause any issues since a QKD security proof typically only needs to bound the entropy of Alice's registers $S_1^n$.} We again emphasize, however, that the original protocol was not subject to any repetition-rate restrictions; what we have achieved in the above argument is to reduce its security analysis to a virtual scenario which could be analyzed using the sequential structure in our previous theorems.

\markup{
This sequential virtual representation is illustrated in Fig.~\ref{fig:MEAT}. As discussed above, we highlight that the channel structure in the diagram only reflects the channel decomposition used to apply MEAT, rather than imposing an additional round-by-round sequential timing constraint on the protocol. This is because Alice's state preparation and transmission steps are not contained within the MEAT channels. In fact, all the transmitted systems can already be available to Eve before the sequential channels are applied (see Protocol~\ref{Prot:PM virtual_Protocol}). Therefore, the preparation and transmission of a signal state need not wait for the completion of the preceding round, and hence there is no restriction on the repetition rate of the protocol.}

\section{Conclusion}
\label{sec:conclusion}
Overall, in this work we derived a chain rule for channel conditional entropies defined through sandwiched {\Renyi} entropies, for the region $\alpha\geq 1$. Furthermore, we showed that these channel conditional entropies are strongly additive across tensor product of channels. Motivated by cryptography, we combined this chain rule with techniques developed in~\cite{AHT24,inprep_weightentropy} to prove the MEAT, which is a new entropy accumulation result that is compatible with having some form of marginal constraint on the input. Finally, we explained how one can use the analysis in this work to get a security statement for prepare-and-measure protocols, without imposing any of the repetition-rate constraints that were required in the earlier GEAT approaches, and covering a wider range of scenarios than~\cite{inprep_weightentropy}. 

Going forward, several directions can be explored. Notably, as briefly mentioned in the introduction and Sec.~\ref{sec:MEAT}, one limitation of this framework compared to GEAT is its handling of a ``secret'' memory register. In particular, the GEAT (in the notation of~\cite{AHT24}) considers channels of the form  
\begin{align}
\mathcal{M}_j\in\CPTP(R_{j-1}E_{j-1}, S_jR_jE_j\CS_j\CP_j),
\end{align}
where the $R_j$ registers serve as internal memory registers that can be updated in each round and passed to the next channel. (The remaining registers play roles analogous to those in this work.) 
Crucially, these registers are ``secret'' in the sense that the single-round entropies analyzed in the GEAT have the form $H_\alpha(S_j\CS_j | \CP_jE_j \widetilde{E})$, \emph{without} the $R_j$ registers appearing, as long as the channels satisfy a technical ``non-signalling'' condition from $R_{j-1}$ to $E_j$ (see~\cite{MFSR24} for details of this condition).\footnote{We note that without such a non-signalling assumption, the memory register would, in fact, 
affect the entropies
by appearing in the conditioning register of the single-round conditional entropy.}
The MEAT does not accommodate such a property, since \emph{all} the output registers of each channel appear somewhere in the single-round entropies in our results. 
Conversely, while the GEAT channel structure can in principle incorporate the MEAT channel structure\footnote{ 
By setting the $R_j$ registers in the GEAT channels to be trivial (in which case the NS condition automatically holds) and identifying each $E_j$ register with the registers $A_{j}^{n-1} E_j$ in the MEAT channel structure, taking the $j^\text{th}$ GEAT channel to act as identity on the registers $A_{j}^{n-1}$. Note that this means the registers $A_{j}^{n-1}$ would appear in the conditioning systems for the $j^\text{th}$-round entropy term in the GEAT; however, in the MEAT they were effectively also present as part of the stabilizing register, in some sense. 
}, this comes at the cost that the GEAT does not impose any marginal constraints on the input states, which makes it difficult to use for analyzing PM-QKD in the manner we described in Sec.~\ref{sec:security}.

Certainly, a natural question is whether the GEAT and MEAT frameworks can be unified. The GEAT was obtained by deriving a chain rule similar to our Theorem~\ref{thrm:channel cond ent chain rule}, but where each channel can output registers $R_j$ that do not appear in the conditioning, via a series of arguments based on the NS condition.
Unfortunately, the series of arguments used in~\cite{MFSR24} does not directly translate to the MEAT framework. For instance, it is unclear how to adapt the statement of Lemma~3.3 in~\cite{MFSR24} to one that aligns with the structure of the MEAT framework. This remains an important question for future work. However, we caution that there are counterexamples to the most ``straightforward'' result one might attempt to obtain when unifying the approaches. We describe this in Appendix~\ref{app:counterexample}.

Another potential direction for future work is to explore whether a similar entropy accumulation statement can be formulated with different marginal constraints. For example, instead of requiring the global marginal state in Theorems~\ref{thrm:qes eat_simp}--\ref{thrm:MEAT} to be a tensor product of individual states, one could consider a less restrictive class, such as certain forms of separable states that are not necessarily tensor products. In a completely different direction, it might be valuable to investigate whether the techniques used in this work to prove the strong additivity of channel conditional entropy could provide insights into additivity problems in other areas of quantum information theory.

\section*{Acknowledgements}
We thank Peter Brown, Renato Renner, Roberto Rubboli, Martin Sandfuchs, and Marco Tomamichel for helpful discussions and feedback. We also thank the authors of~\cite{FHKR25} (Omar Fawzi, Jan Kochanowski, Cambyze Rouz\'{e}, and Thomas Van Himbeeck) for coordinating a joint release of our results.
This research was conducted at the Institute for Quantum Computing, at the University of Waterloo, which is supported by Innovation, Science, and Economic Development Canada. Support was also provided by NSERC under the Discovery Grants Program, Grant No.~341495.

\section*{Author contributions}
The authors contributed equally to this work.

\printbibliography

\appendix

\section{Counterexamples to avoid}
\label{app:counterexample}

\noindent (We thank Renato Renner and Martin Sandfuchs for presenting this construction to us.)

To incorporate marginal constraints into the GEAT, one might attempt the following. Consider channels $\mathcal{M}_j\in\CPTP(R_{j-1}E_{j-1}, S_jR_jE_j)$ satisfying the relevant NS condition (see~\cite{MFSR24}) from $R_{j-1}$ to $E_j$.\footnote{For this discussion we take the $\CS_j \CP_j$ registers to be trivial, since a chain rule that held while including such registers would also include the version here as a special case. Also, a chain rule of the form here would conversely suffice as a ``building block'' to obtain the desired results involving $\CS_j \CP_j$, via similar arguments to the Sec.~\ref{sec:MEAT} proofs.} Furthermore, suppose $R_j$ and/or $E_j$ have ``sub-register'' structure in the sense $R_j = R'_j R''_j$ and/or $E_j = E'_j E''_j$ respectively. 
One might then attempt to prove a chain rule of the following form, for some {\Renyi} parameter values $\alpha,\widehat{\alpha} \in (1, \infty]$:
\begin{align}\label{eq:falseGEAT_Rcon}
\mathbb{H}_\alpha(S_1^n | E_n)_{\EATchann_n \circ \dots \circ \EATchann_1 [\omega^0]} \stackrel{?}{\geq} \sum_j \inf_{\substack{\omega\in\dop{=}(R_{j-1} E_{j-1} \widetilde{E})\\ \suchthat\ \tr[R''_{j-1} E_{j-1} \widetilde{E}]{\omega} = \sigma^{(j-1)}_{R'_{j-1}} }}  \mathbb{H}_{\widehat{\alpha}}(S_j | E_j \widetilde{E})_{\EATchann_j[\omega]} ,
\end{align}
or
\begin{align}\label{eq:falseGEAT_Econ}
\mathbb{H}_\alpha(S_1^n | E_n)_{\EATchann_n \circ \dots \circ \EATchann_1 [\omega^0]} \stackrel{?}{\geq} \sum_j \inf_{\substack{\omega\in\dop{=}(R_{j-1} E_{j-1} \widetilde{E})\\ \suchthat\ \tr[R_{j-1} E''_{j-1} \widetilde{E}]{\omega} = \sigma^{(j-1)}_{E'_{j-1}} }}  \mathbb{H}_{\widehat{\alpha}}(S_j | E_j \widetilde{E})_{\EATchann_j[\omega]} ,
\end{align}
where $\mathbb{H}_\alpha$ represents either $H^\uparrow_\alpha$ or $H_\alpha$ (or even their Petz counterparts; the same analysis holds in suitable {\Renyi} parameter regimes), and $\sigma^{(j-1)} = \EATchann_{j-1} \circ \dots \circ \EATchann_1 [\omega^0]$. In other words, the domain in each minimization is restricted by constraining some marginal of the input state to match that of the actual state generated from $\omega^0$ by the earlier channels.

Unfortunately, there are counterexamples to the above bounds (except if $\mathbb{H}_{\widehat{\alpha}} = H_\infty$). 
First note that if they were true, they would also hold for the special case where $R''_j$ and $E''_j$ are trivial, in which case we just have $R_j = R'_j$ and $E_j = E'_j$, i.e.~the marginal constraints in the above formulas hold on the entirety of the $R_j$ or $E_j$ registers.
Now consider a scenario where the initial state is simply the classical state $\omega^0_{R_0 E_0} = \frac{1}{2} (\pure{00} + \pure{11})$, i.e.~$R_0 E_0$ are copies of a uniform coinflip. Suppose each $\EATchann_j$ first measures $R_{j-1}E_{j-1}$ (in their classical bases) and then simply writes the outcomes into $R_jE_j$, so essentially the initial value on $R_0E_0$ gets passed along unchanged throughout the sequence of channels.
It also generates a uniformly random bit on $S_j$ (independent of all other registers) if the outcome from measuring $R_{j-1}$ was $0$, and sets $S_j$ to some deterministic value $\perp$ otherwise. These channels can be straightforwardly shown to satisfy the NS condition. Furthermore, note that by applying Fact~\ref{fact:classmix} to the resulting state on $S_1^n E_n$, we have for any $\alpha>1$
\begin{align}
H_\alpha(S_1^n | E_n) \leq H^\uparrow_\alpha(S_1^n | E_n) &= \frac{\alpha}{1-\alpha}\log\left(\frac{1}{2}\, 2^{\frac{1-\alpha}{\alpha}n} + \frac{1}{2}\, 2^{0}\right) 
\nonumber\\
&\leq \frac{\alpha}{1-\alpha}\log\left(\frac{1}{2}\, (0) + \frac{1}{2}\, (1)\right) = \frac{\alpha}{\alpha-1},
\end{align}
i.e.~$\mathbb{H}_\alpha(S_1^n | E_n)$ is upper bounded by a \emph{constant}. However, observe that for any $\omega_{R_{j-1} E_{j-1} \widetilde{E}}$ such that $\omega_{R_{j-1}} = \frac{1}{2} (\pure{0} + \pure{1})$, the state produced from it by $\EATchann_j$ satisfies (by applying data-processing and Fact~\ref{fact:classmix} to the classical register $R_j$):
\begin{align}
H^\uparrow_{\widehat{\alpha}}(S_j | E_j \widetilde{E})_{\EATchann_j[\omega]} 
\geq H^\uparrow_{\widehat{\alpha}}(S_j | R_j E_j \widetilde{E})_{\EATchann_j[\omega]}
&= \frac{\widehat{\alpha}}{1-\widehat{\alpha}}\log\left(\frac{1}{2}\, 2^{\frac{1-\widehat{\alpha}}{\widehat{\alpha}}} + \frac{1}{2}\, 2^{0}\right) \nonumber\\
&> \frac{\widehat{\alpha}}{1-\widehat{\alpha}}\log\left(\frac{1}{2}\, (1) + \frac{1}{2}\, (1)\right) = 0, \label{eq:lbnd_Rcon}
\end{align}
i.e.~$H^\uparrow_{\widehat{\alpha}}(S_j | E_j \widetilde{E})_{\EATchann_j[\omega]}$ is lower bounded by a \emph{strictly positive constant}; similarly, we can obtain an analogous bound for any $H_{\widehat{\alpha}}(S_j | E_j \widetilde{E})_{\EATchann_j[\omega]}$ with $\widehat{\alpha}\neq\infty$.\footnote{For $H_\infty$ we in fact have $H_{\infty}(S_j | R_j E_j \widetilde{E})_{\EATchann_j[\omega]} = 0$ by Fact~\ref{fact:classmix}. However, we shall not concern ourselves too much with that somewhat extreme scenario, since in any case the bounds in~\eqref{eq:falseGEAT_Rcon}--\eqref{eq:falseGEAT_Econ} would then simply reduce to a special case of~\cite[Proposition~5.5]{Tom16}, which holds even without minimizing over input states to the channels.}
Hence the RHS of~\eqref{eq:falseGEAT_Rcon} would be $\Omega(n)$, resulting in a contradiction. 

Similarly, if we just slightly modify this construction by having the channels set the value of $S_j$ based on the outcome from measuring $E_{j-1}$ instead of $R_{j-1}$, they would again still satisfy the NS condition, and for any $\omega_{R_{j-1} E_{j-1} \widetilde{E}}$ such that $\omega_{E_{j-1}} = \frac{1}{2} (\pure{0} + \pure{1})$, we would again have that $\mathbb{H}_{\widehat{\alpha}}(S_j | E_j \widetilde{E})_{\EATchann_j[\omega]}$ is lower bounded by a strictly positive constant, as long as $\mathbb{H}_{\widehat{\alpha}} \neq H_\infty$. Hence this would provide a counterexample to~\eqref{eq:falseGEAT_Econ}.

To work around this counterexample, it would be necessary to impose additional structure or conditions on the channels, in such a way that they do not allow the bounds~\eqref{eq:falseGEAT_Rcon}--\eqref{eq:falseGEAT_Econ} as a special case. One would also need to ensure that some simple modification of the above construction does not provide another counterexample. For instance, one might instead consider channels $\mathcal{M}_j\in\CPTP(R_{j-1}E_{j-1}, S_jE_j)$ acting on an initial state of the form $\omega^0_{R_0^{n-1}E_0}$, i.e.~each channel acts on a separate initial $R_{j-1}$ register rather than ``passing memory registers between them''. However, this alone would still not be sufficient to obtain bounds such as~\eqref{eq:falseGEAT_Rcon}, because it is still subject to the same form of counterexample described above, by initializing all the $R_{j-1}$ registers to be classical \emph{copies} of a uniformly random bit.
Note that while this channel structure already basically matches the MEAT\footnote{While we do have a match here in channel structure, when it comes to the bounds, there is a potential discrepancy regarding whether the $R_j^{n-1}$ registers are to appear in the conditioning, under suitable NS conditions (i.e.~the main contribution of the GEAT as compared to previous chain rules). However, this does not affect our overall point here, since the lower bound in~\eqref{eq:lbnd_Rcon} holds even conditioned on $R_j^{n-1}$ in this scenario.} (identifying $R_{j-1} \leftrightarrow A_{j-1}$), it appears that in the MEAT we essentially rule out this counterexample via the further restriction that the state $\omega^0_{A_0^{n-1}}$ has a tensor-product form.\footnote{From the perspective of our proofs, if we were to attempt a similar approach to prove either of the bounds~\eqref{eq:falseGEAT_Rcon}--\eqref{eq:falseGEAT_Econ}, the critical step that appears not to carry over is the analysis of the dual SDPs~\eqref{eq:SDP dual individual 1}--\eqref{eq:SDP dual joint} (or~\eqref{eq:SDP Additivity_individual 1}--\eqref{eq:SDP Additivity_joint}): in such an attempt, the resulting dual SDPs do not seem to have the property that the tensor product of feasible points for the smaller SDPs gives a feasible point of the larger SDP.} 

Another method to handle marginal constraints in the EAT or GEAT was proposed in~\cite{BGW+24}, based on introducing some ``tomography'' rounds that aim to characterize the marginal states. It appears to avoid the above counterexample in several respects: for instance, it aims to certify that the marginal state is IID (similar to the above discussion of the MEAT initial state conditions), and furthermore it considers the state \emph{conditioned} on the tomography procedure accepting, which may ``disrupt'' the state enough to avoid the above counterexample.

\end{document}